\documentclass{aastex63}

\usepackage{bm}
\usepackage{xcolor}
\usepackage{multibib}
\newcites{main}{References}
\newcites{appendix}{References (Appendix)}
\usepackage{subfigure}
\usepackage{soul}
\usepackage{ulem}
\usepackage{tcolorbox}
\usepackage{amsmath}
\usepackage{verbatim}

\usepackage{threeparttable} 
\usepackage{longtable}

\newcommand{\beq}{\begin{equation}}
\newcommand{\eeq}{\end{equation}}
\newcommand{\bseq}{\begin{subequations}}
\newcommand{\eseq}{\end{subequations}}
\newcommand{\bary}{\begin{eqnarray}}
\newcommand{\eary}{\end{eqnarray}}
\newcommand{\bwt}{\begin{widetext}}
\newcommand{\ewt}{\end{widetext}}

\begin{document}
\title{Deciphering the VHE gamma-rays spectra of Extreme High-Frequency peaked BL Lac Objects through a Photohadronic Model
}

%

\author[0000-0003-0038-5548]{Sarira Sahu}
\email{sarira@nucleares.unam.mx}
\affiliation{Instituto de Ciencias Nucleares, Universidad Nacional Aut\'onoma de M\'exico,\\ Circuito Exterior S/N, C.U., A. Postal 70-543, CDMX 04510, México.}

\author[0009-0001-2172-4556]{D. I. Páez-Sánchez}
\email{diana.paez@correo.nucleares.unam.mx}
\affiliation{Instituto de Ciencias Nucleares, Universidad Nacional Aut\'onoma de M\'exico,\\ Circuito Exterior S/N, C.U., A. Postal 70-543, CDMX 04510, México.}

\author[0009-0006-4454-7165]{A. U. Puga Oliveros}
\email{angel.puga@correo.nucleares.unam.mx}
\affiliation{Instituto de Ciencias Nucleares, Universidad Nacional Aut\'onoma de M\'exico,\\ Circuito Exterior S/N, C.U., A. Postal 70-543, CDMX 04510, México.}

\author[0000-0003-1405-5099]{M. E. Iglesias Martínez}
\email{miigmar@upv.es}
\affiliation{Instituto Universitario de Matemática Pura y Aplicada,\\ Universitat Politècnica de València, Camino de Vera s/n, 46022 Valencia, Spain
}
\author[0000-0002-0477-7998]{Jose Guerra Carmenate}
\email{jguecar@doctor.upv.es}
\affiliation{Instituto Universitario de Matemática Pura y Aplicada,\\ Universitat Politècnica de València, Camino de Vera s/n, 46022 Valencia, Spain
}

\author{R.de J. Pacheco-Aké}
\email{rodrigo.pacheco@cinvestav.mx}
\affiliation{Instituto de Física, Circuito de la Investigación Científica, C. U. , Coyoacán, C.P. 04510, Ciudad de México
}

\author[0000-0002-0291-2412]{G. Sánchez-Colón}
\email{gabriel.sanchez@cinvestav.mx}
\affiliation{Departamento de Física Aplicada, Centro de Investigación y de Estudios Avanzados del IPN,\\ Unidad Mérida. A. Postal 73, Cordemex, Mérida, Yucatán 97310, México
}

\author[0000-0003-1967-7217]{Subhash Rajpoot}
\email{Subhash.Rajpoot@csulb.edu}
\affiliation{Department of Physics and Astronomy, California State University,\\
1250 Bellflower Boulevard, Long Beach, CA 90840, USA.
}

\author[0000-0002-0347-7280]{P. Fernández de Córdoba}
\email{pfernandez@mat.upv.es}
\affiliation{Instituto Universitario de Matemática Pura y Aplicada,\\ Universitat Politècnica de València, Camino de Vera s/n, 46022 Valencia, Spain
}

\author[0000-0001-7574-2330]{Gaetano Lambiase}
\email{lambiase@sa.infn.it}
\affiliation{Dipartimento di Fisica “E.R. Caianiello”, Universit`a di Salerno,
Via Giovanni Paolo II, I-84084 Fisciano (SA), Italy}
\affiliation{INFN, Gruppo collegato di Salerno,
Via Giovanni Paolo II, I-84084 Fisciano (SA), Italy}

\begin{abstract}
 Extreme high-frequency peaked BL Lac objects (EHBLs) have their synchrotron peak above $\sim 1$ keV and their inverse Compton peak extends up to several TeVs. Due to low luminosity of EHBLs, these objects are often undetected, and only a small number of them are thus known. As a result, they are poorly understood. Using a photohadronic model and including extragalactic background light correction we analyzed 22 very high-energy (VHE) gamma-ray spectra of 15 EHBL sources observed in different periods. We find that the photohadronic model used can adequately describe the observed VHE spectral energy distribution
and furthermore, we classify them according to the value of the photon spectral index, $\delta$. The redshift of the EHBL HESS J1943+213 is also tightly constrained using the same photohadronic model. Within the sample, a plurality of flaring epochs fall in the high-emission state which is followed by the low-emission state and the very-high-emission state respectively. These fractions of flaring epochs characterize only the present sample, not the EHBL population at large. Due to the anomalous behavior of the VHE gamma-ray spectra of the EHBL H 2356-309, we suggest it to be a new subclass of EHBL having hard and very soft intrinsic spectra during different flaring epochs. 

 
\end{abstract}


\section{Introduction}
Extreme high-frequency peaked BL Lac objects (EHBLs) are an emerging subclass of BL Lac objects with extreme properties.  By definition, their synchrotron peak frequency, $\nu^p_s$, is above $10^{17}$ Hz ($\sim 1$ keV)~\citep{2001A&A...371..512C,2002A&A...384...56C}, and their inverse Compton (IC) peak extends up to several TeVs \citep{2019MNRAS.486.1741F, 2018MNRAS.477.4257C}.The EHBLs exhibit the following characteristics: a hard TeV spectrum (intrinsic spectral index $<2$ in Very High Energies; for which these objects are also called "hard-TeV blazars") \citep{2018MNRAS.477.4257C,2019MNRAS.486.4233A}, have low luminosity, and the absorption of very high energy (VHE $>\,100$ GeV) gamma-rays from these sources by the extragalactic background light (EBL) causes the sources to be faint and often undetected \citep{2019MNRAS.486.1741F}.In VHE gamma-rays, EHBLs consistently display an almost stable spectrum, unlike the high-frequency peaked BL Lac objects (HBLs), which display rapid flux variability spanning timescales hours to few minutes. These unique features of EHBLs pose a significant challenge in modeling the spectra ~\citep{2018MNRAS.477.4257C,2020ApJS..247...16A}. The leptonic modeling of the spectral energy distribution (SED) of such sources requires an unusually low magnetic field and a very high electron Lorentz factor~\citep{2011MNRAS.414.3566T}. It is observed that during VHE flaring, some HBLs might exhibit temporary EHBL (tEHBL)-like behavior
by temporarily shifting their synchrotron peak towards higher energies~\citep{2015A&A...578A..22A,2018A&A...620A.181A,2020A&A...638A..14M,Biteau_2020}. 
Recently it is observed that this transient behavior is not limited to just a few blazars. In this sense, tEHBLs have been identified as a new subclass of EHBLs, characterized by a different spectral behavior in the VHE regime~\citep{2020ApJ...901..132S,2021ApJ...906...91S,2021ApJ...914..120S,2022MNRAS.515.5235S,Sahu_2026}. Therefore, identifying such transient states and discarding them from the list of stable EHBL sources is crucial for a complete understanding of these sources \citep{2020ApJ...891..170V,2024A&A...685A.117M}. The exact number of stable/permanent EHBLs is difficult to determine due to the faintness of their spectrum and contamination by tEHBLs, only a limited number of these sources are known~\citep {Biteau_2020,2022MNRAS.512..137N,2025arXiv250702718D}.

Among all the EHBLs observed so far, the archetypal EHBL 1ES 0229+200 has the highest synchrotron peak~\citep{2011A&A...534A.130K}. Also, this source has been observed in multiwavelength several times. Several other EHBLs such as: 1ES 0347-121 \citep{2007AA...473L..25A}, RGB J0710+591 \citep{2010ApJ...715L..49A}, and 1ES 1101-232 \citep{2007AA...470..475A}
have also shown hard intrinsic spectra like  1ES 0229+200 making them ideal probes for testing models of EBL~\citep{Franceschini:2008tp,2011MNRAS.410.2556D,2014MNRAS.438.3255T,2015arXiv151205080T,2021MNRAS.507.5144S,2012MNRAS.422.3189G}. The propagation of gamma rays above tens of GeV are attenuated by EBL by producing electron-positron pairs~\citep{1992ApJ...390L..49S,2012Sci...338.1190A}. Also, this attenuation factor depends on the photon energy and the redshift of the source. Thus, observation of TeV photons from these sources will help in the analysis of the EBL features, particularly in the long infrared wavelength regime.

The electromagnetic cascade produced by photons above 1 TeV is sensitive to the intergalactic magnetic field and leaves its imprint on the reprocessed gamma rays~\citep{2009PhRvD..80b3010E}. As a consequence, excess photons in the GeV energy range can be detected by Fermi-LAT. Thus, intergalactic magnetic field can be constrained using observed VHE photons from EHBLs~\citep{2023ApJ...950L..16A}.

EHBLs have the highest Doppler factors within the blazar sequence~\citep{2017MNRAS.469..255G}, which clearly suggests that EHBLs could be possible sources of extreme particle accelerators, producing ultrahigh energy cosmic rays (mostly protons) and neutrinos. In this process, the neutrino carries $\sim 5\%$ of the proton energy~\citep{2022icrc.confE..30O}. 
The correlation of the IceCube neutrino event (IC170922A) with the flaring blazar TXS 0506+056 establishes a potential connection between blazars and high-energy neutrino production sites~\citep{2018Sci...361.1378I,2018Sci...361..147I}. In the individual level, the EHBL 3HSP J095507.9+355101 was found to be within the error region of the IceCube neutrino event IC200107A and the source had its synchrotron peak above 10 keV at the time of neutrino detection~\citep{2020A&A...640L...4G}. 
Similarly, several other IceCube neutrino events were within the 90\% uncertainty region of many EHBLs, such as: IC 111216A with 3HSP J023248.6+201717, IC 170506A with 3HSP J144656.8-265658 and 3HSP J094620.2+010452 with IC 190819 respectively~\citep{2020MNRAS.497..865G,Aguilar-Ruiz:2023aqt}. However, the lack of adequate monitoring before and after the neutrino events was a limitation in tracing emission activities and morphological changes in all the sources. Thus, it is important to study these sources and their flaring mechanisms. 

The blazar-neutrino connection is investigated through several population-based 
neutrino counterpart search without compelling evidence~\citep{2020PhRvD.101j3015L,2021A&A...650A..83H,2022PhRvD.106h3024L,2022ApJ...934..180K,2023ApJ...955L..32B,2024JCAP...05..133P,2024MNRAS.527L..26S}. Moreover, spatio-temporal analyses on the connection between blazar flares and IceCube neutrinos also
do not find any significant correlation~\citep{Kouch_2024,2026A&A...708A.383K}.
The maximum contribution of blazars in the 2nd Fermi-LAT AGN catalog (2LAC) to the observed astrophysical neutrino flux can be 27\% or less for neutrinos in the energy range 10 TeV to 2 PeV~\citep{2017ApJ...835...45A}.

A variety of models have been proposed to explain the hard VHE spectra of EHBLs and simultaneously avoiding unrealistic parameters needed by the SSC model. On such model is by ~\cite{2011ApJ...740...64L},
wherein a time-dependent SSC model, the dominance of adiabatic losses over synchrotron losses or a Maxwellian-like electron distribution is considered. Also, by proposing that the 
momentum distribution of the most energetic electrons in the BL Lac jet is anisotropic,
the hard VHE spectrum can be explained under quipartition of energy density between the magnetic field and the relativistic electrons~\citep{2020MNRAS.491.2198T}. The proton synchrotron, synchrotron emission from $p\gamma$ induced cascades (lepto-hadronic) models are also proposed to account for the hard VHE spectra of EHBLs~\citep{2015MNRAS.448..910C,2016A&A...585A...8Z}.
A lepto-hadronic two-zone model is proposed by~\cite{2022MNRAS.512.1557A} to explain the observed SED of EHBLs. In this model, the photons above TeV energies are produced from the $p\gamma$ interaction, where the photons are the annihilation line photons from the plasma in the inner blob.

Previously, it was shown that the VHE gamma-ray spectra of several HBLs could be explained very well using the photohadronic model of~\cite{2013PhRvD..87j3015S,2020MNRAS.492.2261S} and depending on the photon spectral index, the VHE flaring epochs were classified into three categories~\citep{2019ApJ...884L..17S}. Also,  by extending this photohadronic model to two-zones, VHE gamma-ray spectra of several tEHBL are explained successfully, and it is shown that, tEHBLs are in fact a different subclass of EHBLs~\citep{2020ApJ...901..132S,2021ApJ...906...91S,2021ApJ...914..120S,2022MNRAS.515.5235S,Sahu_2026}. In this work, our goal is to extend  this photohadronic model to study the VHE gamma-ray spectra of EHBLs and to classify them according to their photon spectral index. Finally, we compare and contrast the EHBLs with the HBLs.

The paper is organized as follows: In section 2,  our version of the photohadronic model and its kinematics condition are briefly discussed. In section 3, the selection criteria for the EHBL sources and VHE flaring epochs from 15 EHBL sources are analyzed in details by including different EBL models to the photohadronic model mentioned in section 2. We have discussed different flaring epochs of H 2356-309 in the context of photohadronic model and compare with other sources in section 4 and finally a brief discussion is given in section 5. 

\section{The Photohadronic Model}\label{Section2}

The one-zone leptonic models predict a softer SSC spectrum in the Klein-Nishina regime, thus it is difficult to explain the VHE spectrum of an EHBL. However, by adopting extreme parameters, such as very low magnetic field, very large bulk Lorentz factor and electron Lorentz factor it is possible to explain the spectra~\citep{2018A&A...620A.181A,2020A&A...638A..14M}. Additionally, there are several alternative models which include, two-zone leptonic model, the inverse Compton scattering of relativistic electrons with the cosmic microwave background, the spine-layer structured jet model, and different hadronic models to explain these spectral features~\citep{2018A&A...620A.181A,2008ApJ...679L...9B,2019MNRAS.490.2284M,2022MNRAS.512.1557A}.

In the context of this photohadronic scenario, the VHE gamma-rays are produced from the interactions of high energy protons with the background seed photons in the blazar jet~\citep{Sahu:2019lwj,2019ApJ...884L..17S}. During the VHE flaring of the source, a double jet structure is proposed. Such a double jet configuration has also been proposed in earlier works~\citep{2008MNRAS.387.1669G,2010MNRAS.402.1649G}. In the present scenario, a compact and narrow jet of size $R'_f$, is formed within the jet of size $R'_b$, $R'_f < R'_b$ (the quantities in the commoving frame are indicated by primes) and both lie along the same axis \citep{2016EPJC...76..127S,2018EPJC...78..484S}. In the inner region of the jet, the photon density, $n'_{\gamma,f}$ is unknown, but is significantly greater than the photon density in the outer jet region, $n'_\gamma$ ($n'_{\gamma,f} \gg n'_\gamma$). Also, it is to be noted that the present photohadronic model is based on the conventional interpretation of the first two peaks in the SED using the leptonic model. In the leptonic scenario, the first peak is formed from the synchrotron radiation emitted by relativistic electrons within the jet environment and the second peak is formed from the well known synchrotron self-Compton (SSC) process. During the flaring process, the inner jet 
has a bulk Lorentz factor $\Gamma_{\rm int}$, slightly higher than the bulk Lorentz factor $\Gamma_{\rm ext}$, in the outer jet. However, for simplicity, we assume $\Gamma_{\rm int} \simeq \Gamma_{\rm ext}\equiv \Gamma$ and both having a common Doppler factor $\mathcal{D}$~\citep{Ghisellini:1998it,Krawczynski:2003fq}.
Also, for blazars, we have $\Gamma\simeq {\cal D}$.

Acoording to \cite{Sahu:2019lwj}, the protons in the inner jet region are accelerated to very high energies with a power law differential spectrum given by 
\beq
\frac{dN_p}{dE_p} \propto E^{-\alpha}_p, 
\label{eq:difspect}
\eeq
where $E_p$ is the proton energy and the spectral index $\alpha \ge 2$~\citep{1993ApJ...416..458D}. These protons interact with the tail region of the SSC photons in the inner jet region through the process
$p+\gamma\rightarrow \Delta^+$. As the photon density in the inner jet is unknown, we assume a scaling behavior of the photon densities at two different energies $\epsilon_{\gamma_1}$ and $\epsilon_{\gamma_2}$ in the inner and the outer regions as \citep{2013PhRvD..87j3015S,2019ApJ...884L..17S}
\beq
    \frac{n'_{\gamma,f}(\epsilon_{\gamma_1})}{n'_{\gamma,f}(\epsilon_{\gamma_2})} 
    \approx 
    \frac{n'_\gamma(\epsilon_{\gamma_1})}{n'_\gamma(\epsilon_{\gamma_2})}.
\label{eq:denscal}
\eeq
The right side is known from the observation, but the left side is unknown. Thus, one can express the unknown photon density in the inner region in terms of the known photon density in the outer jet region. In a normal blazar jet there is no inner jet region and the photon density is low, thus, the production of $\Delta$-resonance is inefficient and needs super-Eddington luminosity for protons 
\citep{2017MNRAS.465.3506P}. However, by shifting the $\Delta$ production to the inner region we could avoid excessive proton luminosity. Furthermore, in the inner region the $\Delta$-resonance decays to $\pi^0$ and then to gamma-rays, or to $\pi^+$ and then to neutrinos. In the photohadronic process, the nonresonant pion production and also multi-pion production take place. However, these are subdominant processes, so we do not include them.

The kinematical condition to produce gamma-rays from the photohadronic process is given by 
\beq
E_\gamma \epsilon_\gamma= 0.032\, \frac{\Gamma^2}{(1+z)^{2}}\, \text{GeV}^2,
\label{eq:Epegmaa}
\eeq
where $E_\gamma=0.1\,E_p$, $\epsilon_\gamma$ and $z$ are the observed VHE photon energy, background seed photon energy, and the redshift of the source, respectively. The intrinsic VHE gamma-ray flux $F_{in}$ is proportional to the proton flux 
$F_p\equiv E^2_p\,dN/dE_p$ and the background photon density $n'_{\gamma ,f}$ in the low energy tail region of the SSC flux, $\Phi_{SSC}$. 
This gives $F_{in}\propto F_{p}\,n'_{\gamma ,f}  $.
Thus the intrinsic gamma-ray flux from the $\pi^0$ decay is 
\beq
F_{in}(E_{\gamma}) \equiv E^2_{\gamma} \frac{dN(E_\gamma)}{dE_\gamma} 
\propto  E^2_p \frac{dN(E_p)}{dE_p} n'_{\gamma,f} .
\eeq
Again, using Eq. (\ref{eq:denscal}) we can express  $n'_{\gamma,f}$ in terms of $n'_{\gamma}$, where we have
\beq
n'_{\gamma}(\epsilon_{\gamma})=\eta \left ( \frac{d_L}{R'_b} \right
)^2 \frac{1}{(1+z)} \frac{\Phi_{SSC}(\epsilon_{\gamma})}{{\cal
    D}^{2+\kappa}\, \epsilon_{\gamma}}.
\label{photondensity}
\eeq
In the above equation, $d_L$ is the
luminosity distance to the source, $\eta$ is the efficiency of the SSC process and $\kappa=0(1)$ 
corresponds to continuous (discrete) blazar jet. We take $\eta=1$ for 100\% efficiency. 

The VHE gamma rays coming from the extragalactic sources get attenuated by interacting with the EBL through the process $\gamma\gamma\rightarrow e^+e^-$. This attenuation depends on $E_{\gamma}$ and $z$ and the shape of the observed spectrum changes at very high energies. Several EBL models are developed to correct the observed spectrum~\citep{Franceschini:2008tp,2011MNRAS.410.2556D,2014MNRAS.438.3255T,2015arXiv151205080T,2021MNRAS.507.5144S,2012MNRAS.422.3189G}.
Expressing $\epsilon_{\gamma}$ in terms of $E_{\gamma}$ by using Eq.(\ref{eq:Epegmaa}), and taking EBL correction into account, the observed VHE gamma-ray flux can be written as
\beq
F_{\gamma}(E_{\gamma})=F_{in}(E_\gamma)\,e^{-\tau_{\gamma\gamma}}
=F_0\left (\frac{E_{\gamma}}{TeV}\right )^{-\delta+3}e^{-\tau_{\gamma\gamma}},
\label{eq:flux}
\eeq
where $\tau_{\gamma \gamma}$ is the optical depth of the $e^+e^-$ pair production process and the exponential is the survival probability of the VHE photons. For an observed VHE SED, the flux normalization factor $F_0$ can be fixed. The free parameter $\delta=\alpha+\beta$ is the VHE photon spectral index and the component $\beta$ is the spectral index of the background seed photon flux in the low energy tail region of the SSC flux expressed as $\Phi_{SSC}\propto \epsilon^{\beta}_{\gamma}$~\citep{Sahu:2019lwj}. Henceforth, we shall refer to the above model as "the photohadronic model".

The broadband SED of most of the EHBLs have low-energy tail region of the SSC spectrum above $10^{20}$ Hz ($>\, 0.4$ MeV) \citep{2011MNRAS.414.3566T, acciari2020new} and this is the region where $\Delta$-resonance is produced and also VHE photons are produced from the $\pi^0$ decay. We assume a mild efficiency of the $\Delta$-resonance production by taking the optical depth $\tau_{p\gamma}=n'_{\gamma,f} \sigma_{\Delta} R'_f\, < \,1$, where $\sigma_{\Delta}\simeq 5\times 10^{-28}\,\mathrm{cm^2}$ \citep{1999PASA...16..160M}. In this region, the high energy electrons and positrons interact with the seed photons and lose energy. So, its corresponding optical depth $\tau_{e\gamma}$ must satisfy $\tau_{e\gamma} \, > 1$. By taking the inner jet size $R'_f\simeq 10^{16}\,\mathrm{cm}$ \citep{2015MNRAS.448..910C}, the  $n'_{\gamma,f} \sigma_{\Delta} R'_f\, < \,1$ condition gives $n'_{\gamma,f} \, < \, 2\times 10^{11}\,\mathrm{cm^{-3}}$. Again, from $\tau_{e\gamma} \, > 1$ we get $n'_{\gamma,f} \, < \, 1.5\times 10^{8}\,\mathrm{cm^{-3}}$. Thus we can take
 $n'_{\gamma,f} \, \simeq \, 10^{9}\,\mathrm{cm^{-3}}$, which gives $\tau_{p\gamma}\simeq 5\times 10^{-3}$.

The produced VHE photons from the $\pi^0$ decay interact with the seed photons, with energy  $\epsilon_{\gamma} \gtrsim 0.4$ MeV to produce $e^+e^-$ pair. However, the $e^+e^-$ pair production efficiency depends on the VHE photon energy $E_{\gamma}$, the seed photon energy $\epsilon_{\gamma}$ and its density $n'_{\gamma,f}$. In the inner jet region the pair production cross section is $\sigma_{\gamma\gamma}\, \lesssim 10^{-28}\, \mathrm{cm^2}$ which corresponds to the mean free path $\lambda_{\gamma\gamma}= (n'_{\gamma,f}\,\sigma_{\gamma\gamma})^{-1}\simeq 10^{19}\,\mathrm{cm}\gg R'_f$. It shows that the $\gamma\gamma\rightarrow e^+e^-$ process is inefficient in the low energy tail region of the SSC spectrum in the inner jet. Moreover, in the outer jet region, the photon density $n'_{\gamma}$ is very low, hence the VHE photons stream out from the jet without attenuation. Thus, the EBL is solely responsible for the attenuation of the propagating VHE photons.

The pair production by the Bethe Heitler (BH) process is responsible for the proton cooling in the jet. In the magnetic field this lepton pair emit synchrotron photons and the maximum energy of these photons will be in the low energy tail region of the SSC spectrum~\citep{Petropoulou:2014rla,Petropoulou:2015upa}. The photohadronic process works well
for $E_{\gamma}>\,100$ GeV and below this energy the leptonic models have the dominant contribution to the SED. However, around 100 GeV photon energy, we expect contribution from the SSC process also.

A distinctive feature of an EHBL is its intrinsic differential spectrum, given as $dN/dE \propto E^{-\Gamma_{in}}_{\gamma}$ and its intrinsic flux as $F_{in}\propto E^{2-\Gamma_{in}}_{\gamma}$. The spectrum is defined as hard for $\Gamma_{in} < 2$~\citep{2007AA...475L...9A} and soft for $\Gamma_{in} > 2$~\citep{2001A&A...371..512C}. However, in the photohadronic model
we have $F_{in}\propto E^{-\delta+3}_{\gamma}$. By comparing both, we get
$\delta=1+\Gamma_{in}$ and consequently in the photohadronic model, the spectrum is hard for $\delta < 3.0$ and it is soft and flat for $\delta=3.0$. Additionally, we define the spectrum as very soft for $\delta > 3.0$.

In a previous study, ~\cite{2019ApJ...884L..17S} classified the VHE emission from HBLs  depending on the spectral index $\delta$. The emission state is called low when $\delta=3.0$. The high-emission state is observed for $2.6 < \delta < 3.0$ and the very high-emission state corresponds to $2.5\leq \delta \leq 2.6$. It is observed that the EHBLs spectra can also be fitted very well with the spectral index in the range $2.5\leq \delta \leq 3.0$. Thus, we stick to the same classification scheme that was given for HBLs and the spectra of the EHBLs are analyzed individually below. 

\begin{figure}[!tb]
  \begin{center}
    \includegraphics[width=0.4\textwidth]{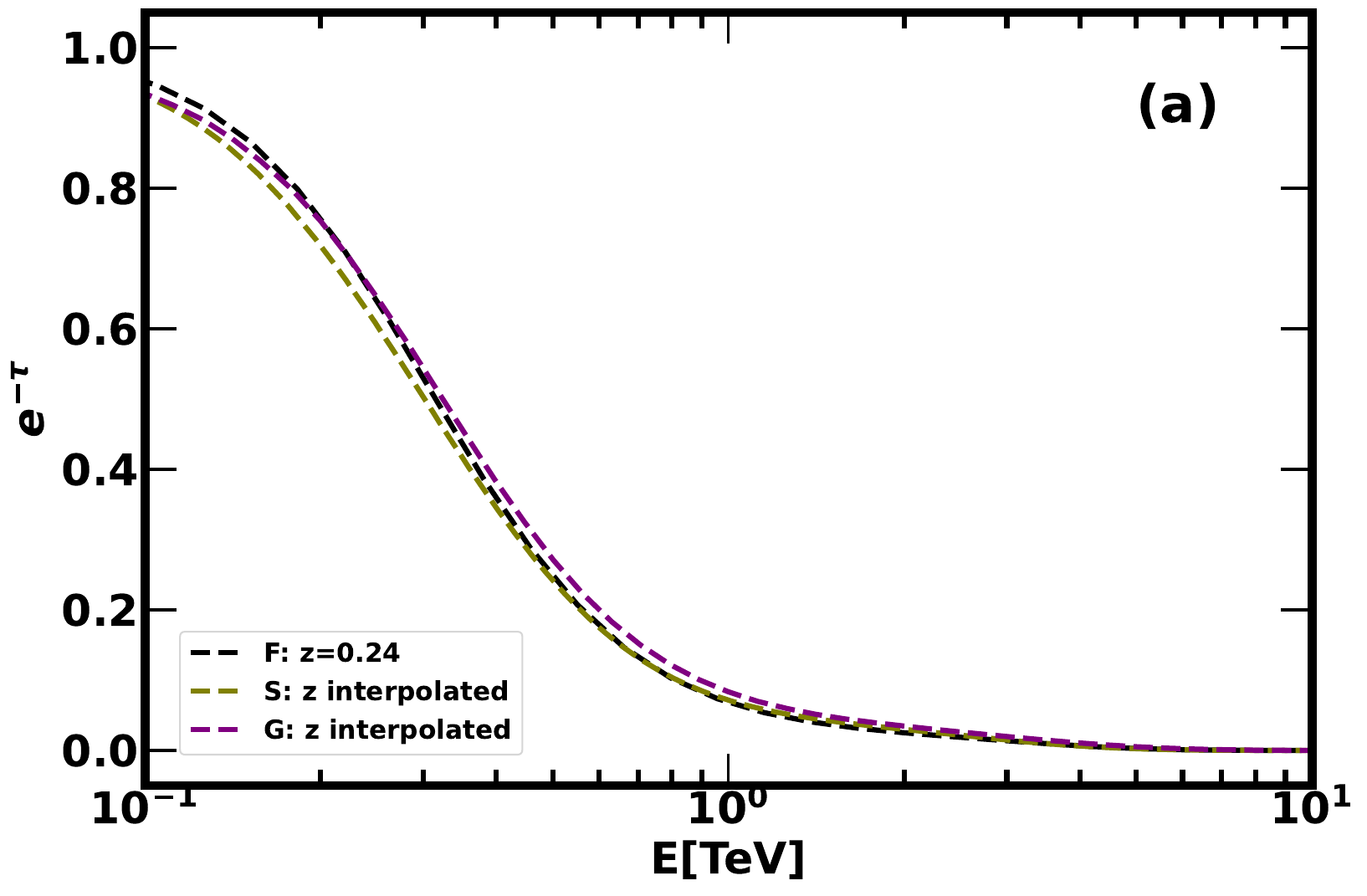}
    \includegraphics[width=0.4\textwidth]{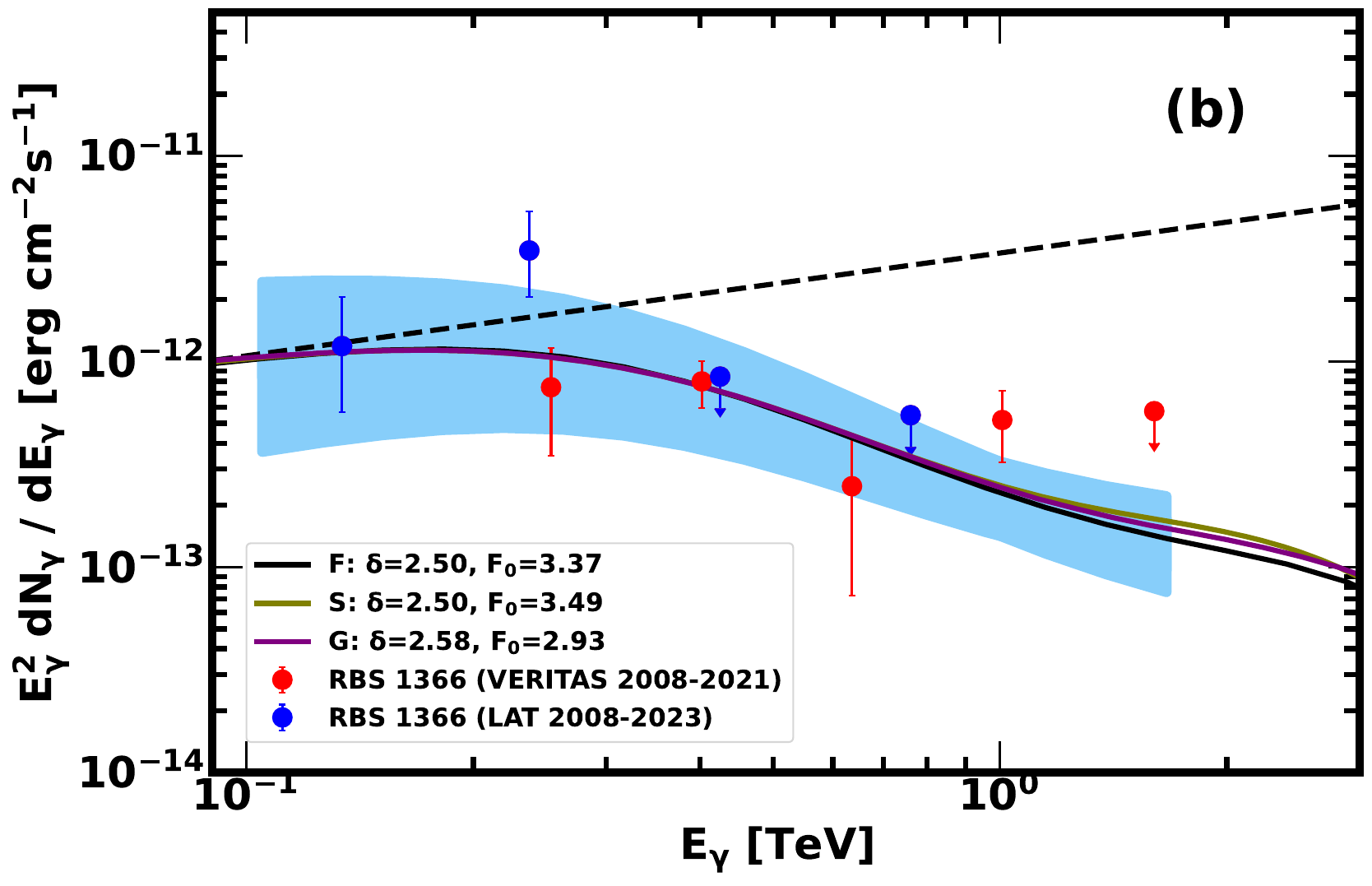}
  \caption{Left: We have plotted the survival probability (SP), $e^{-\tau_{\gamma\gamma}(E_\gamma,z)}$, for $z=0.24$ using the three EBL models S, F and G and compare them. Right: The VHE observations by Fermi-LAT and VERITAS to RBS 1366~\citep{2023arXiv230912230R} are fitted including the EBL models to the  photohadronic model and shown for comparison. 
    The blue butterfly region corresponds to $1\sigma$  confidence level (CL) for the best fit values of the parameters ($\delta, F_0$) using EBL-F, while the dotted line represents the intrinsic flux derived from it. In the rest of the figures below (from Figure \ref{fig: 1ES 0229+200} to Figure \ref{fig: H 2356-309-2006}) the SP is plotted for the three EBL models for comparison 
    and we follow the same definitions and units.
    } 
    \label{fig: RBS 1366}
  \end{center}
\end{figure}

\begin{figure}[!tb]
  \begin{center}
    \includegraphics[width=0.4\textwidth]{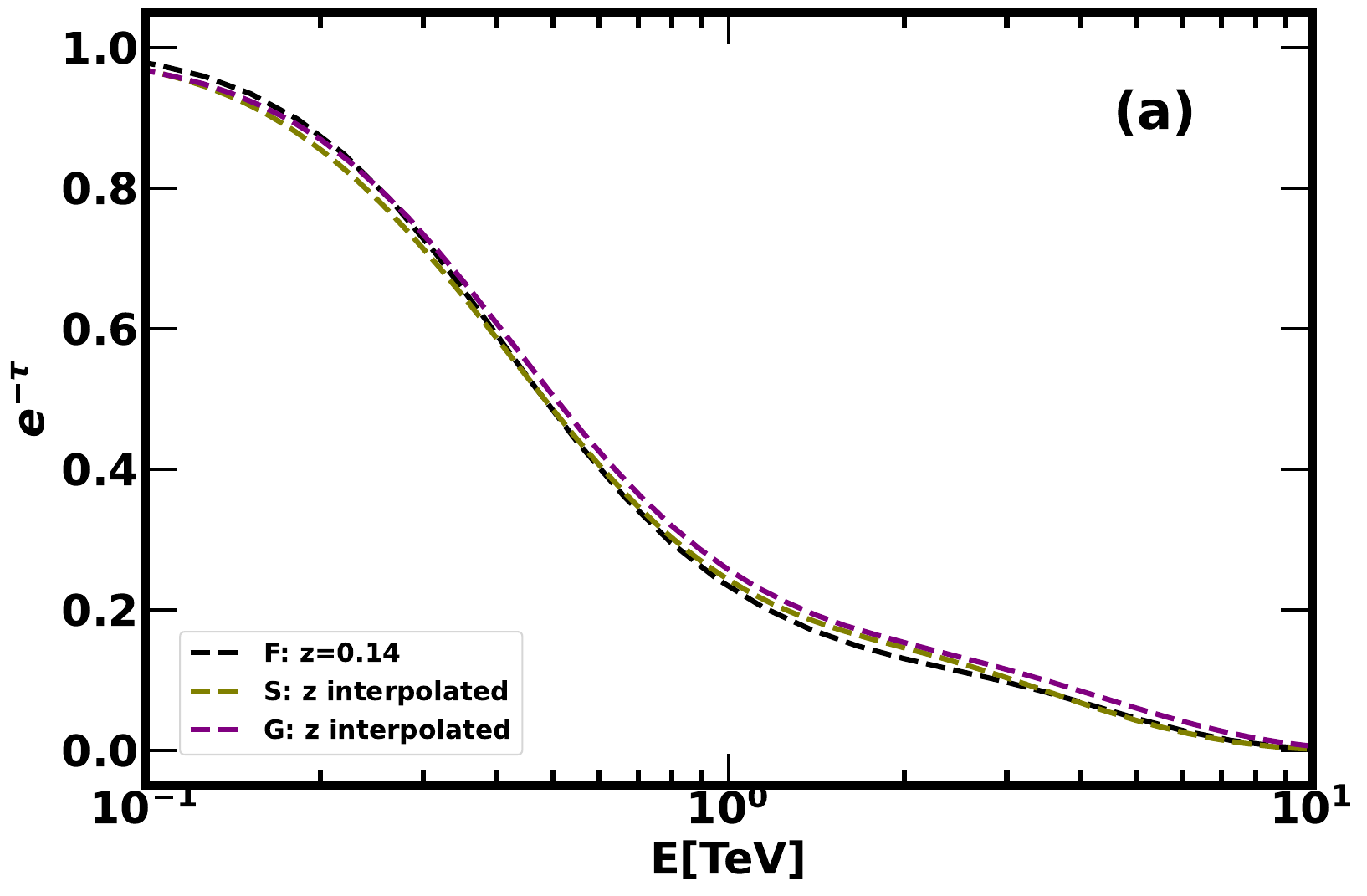}
    \includegraphics[width=0.4\textwidth]{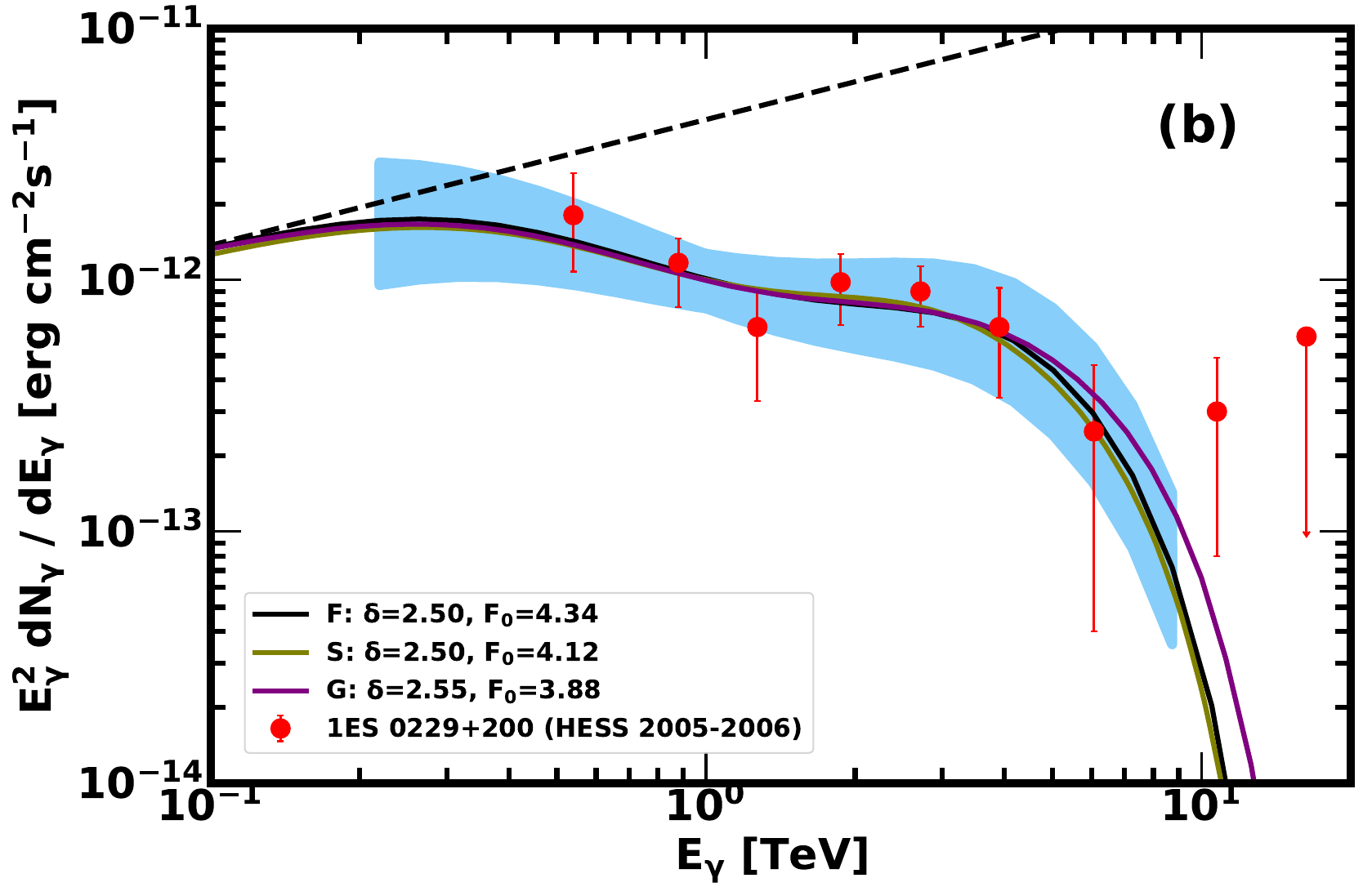}
    \includegraphics[width=0.4\textwidth]{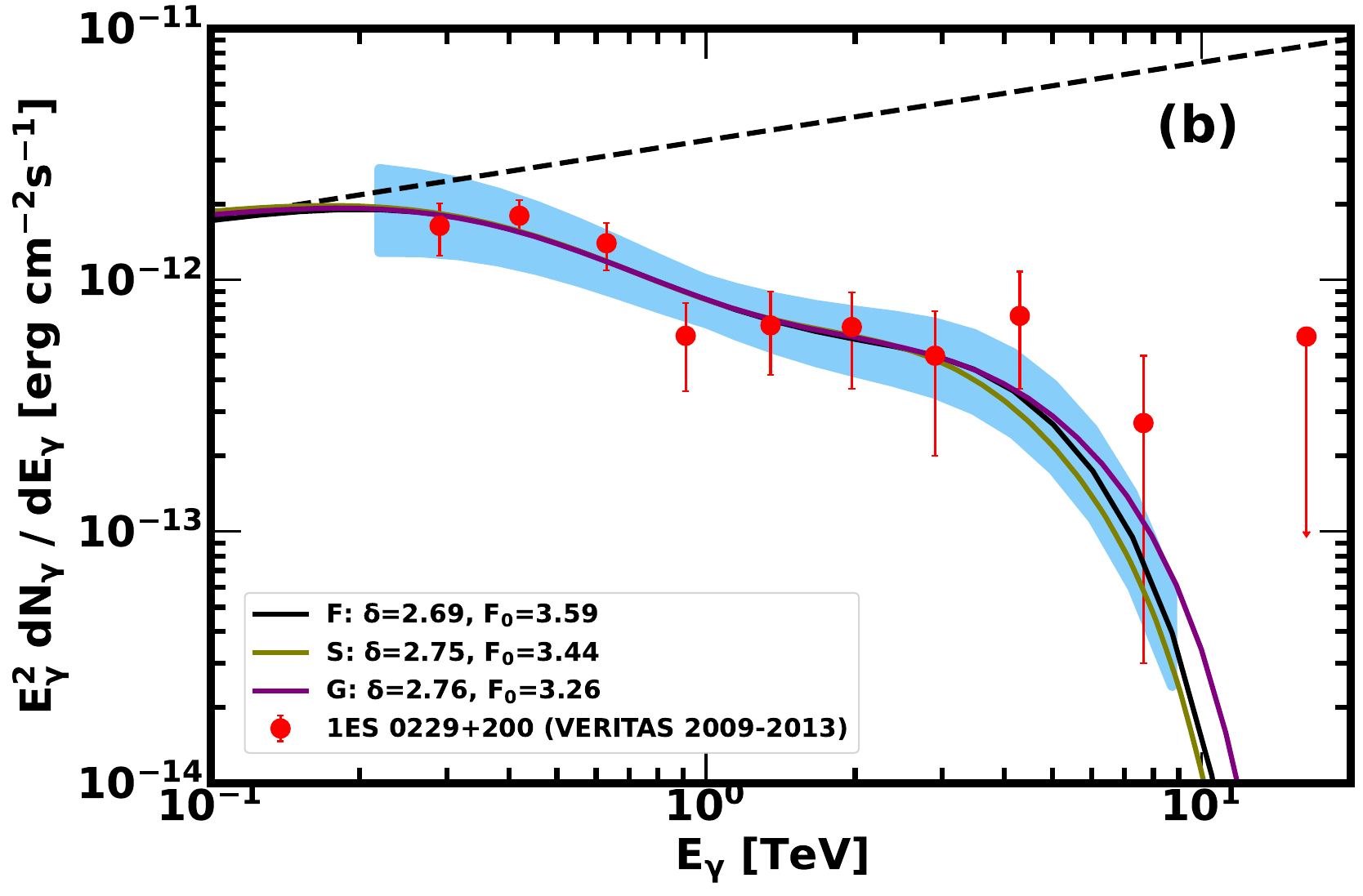}
    \includegraphics[width=0.4\textwidth]{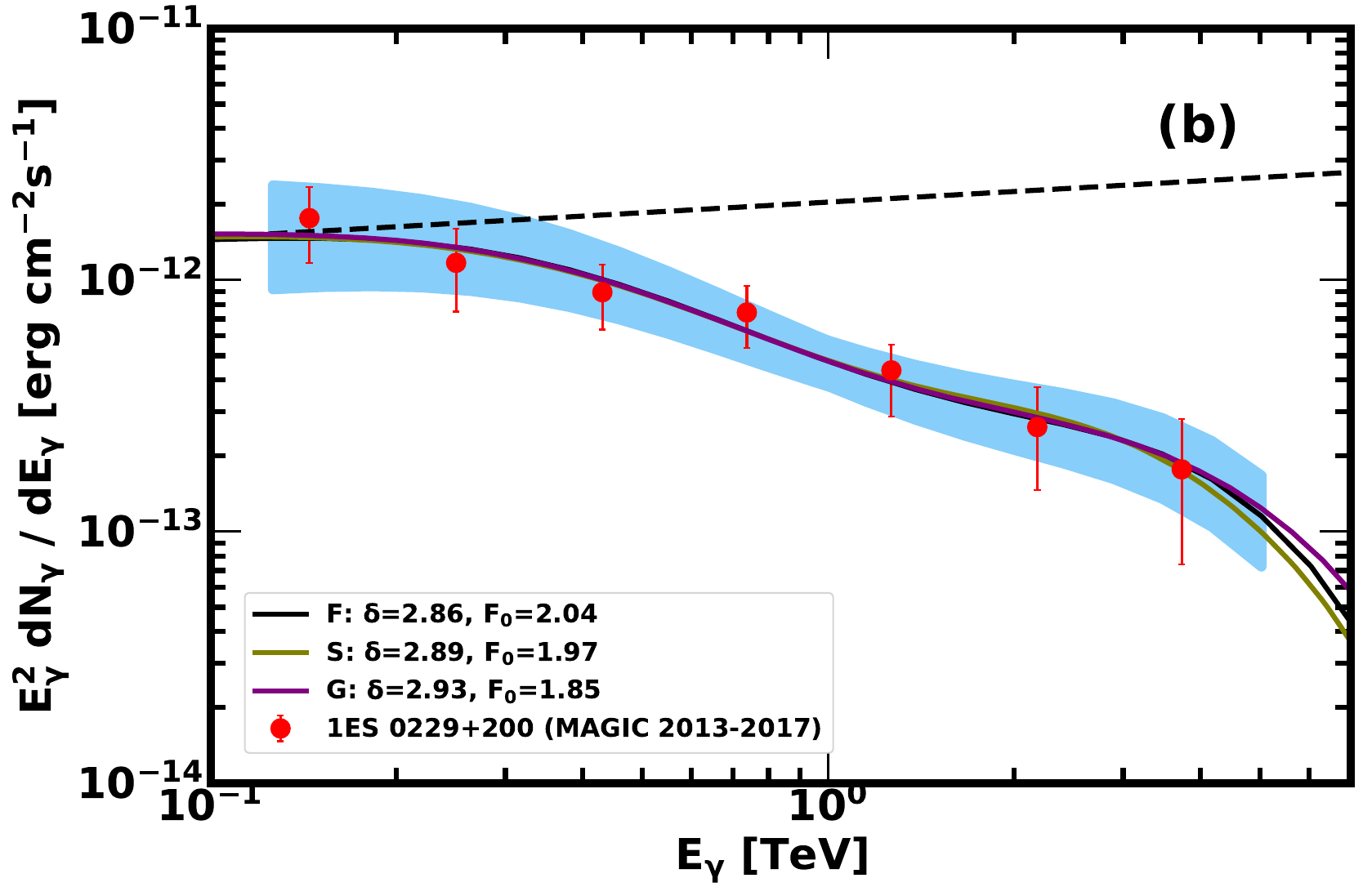}
    \caption{
    The VHE spectra of 1ES 0229+200 observed by HESS (2005-2006)~\citep{2007AA...475L...9A}, by VERITAS (2009-2013)~\citep{2014ApJ...782...13A}  and by MAGIC (2013-2017)~\citep{2023AA...670A.145A} are fitted using the photohadronic model by including the EBL correction from EBL F, S and G.
    }  
    \label{fig: 1ES 0229+200}
  \end{center}
\end{figure}

\begin{figure}[!tb]
  \begin{center}
    \includegraphics[width=0.4\textwidth]{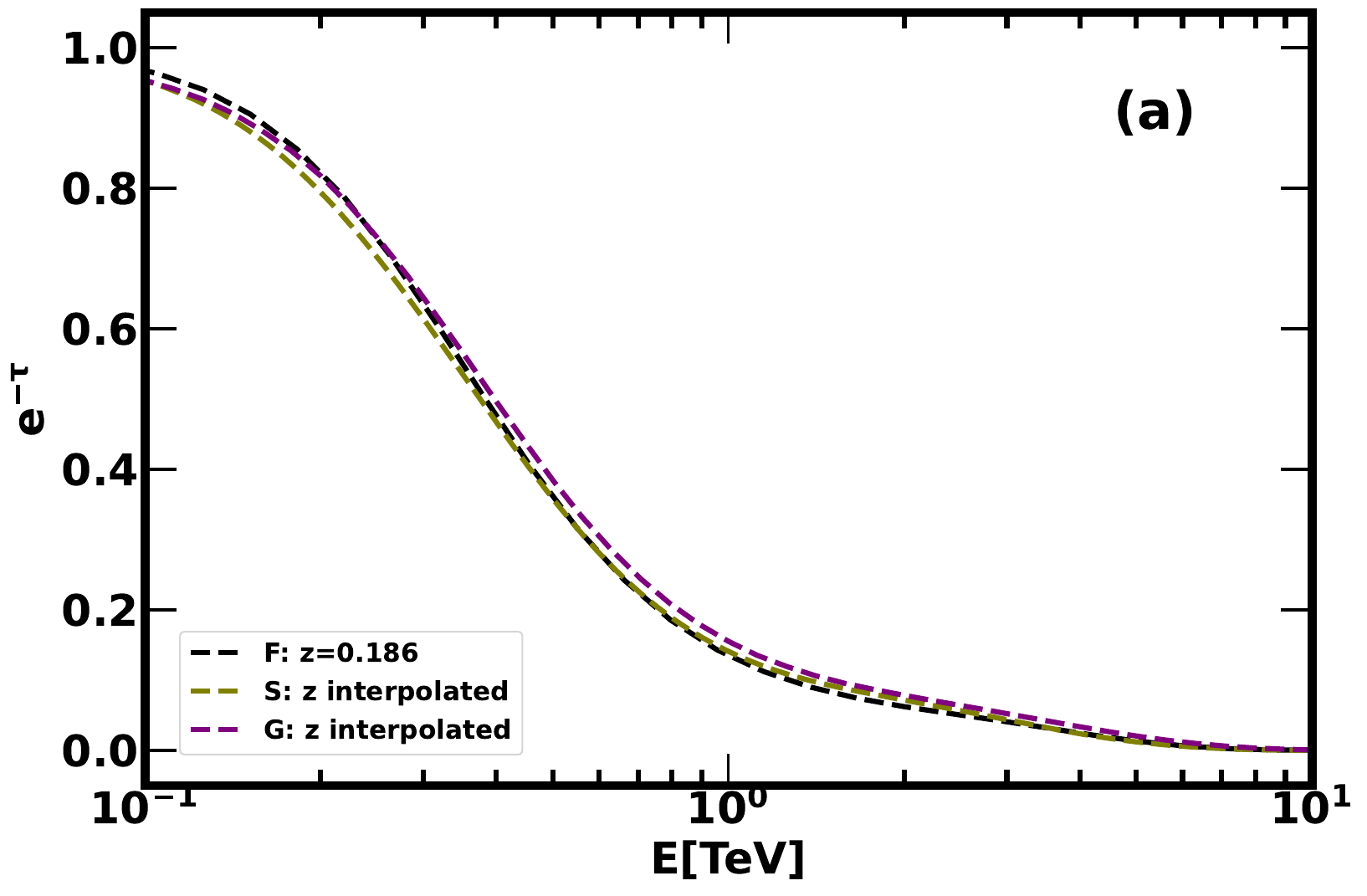}
    \includegraphics[width=0.4\textwidth]{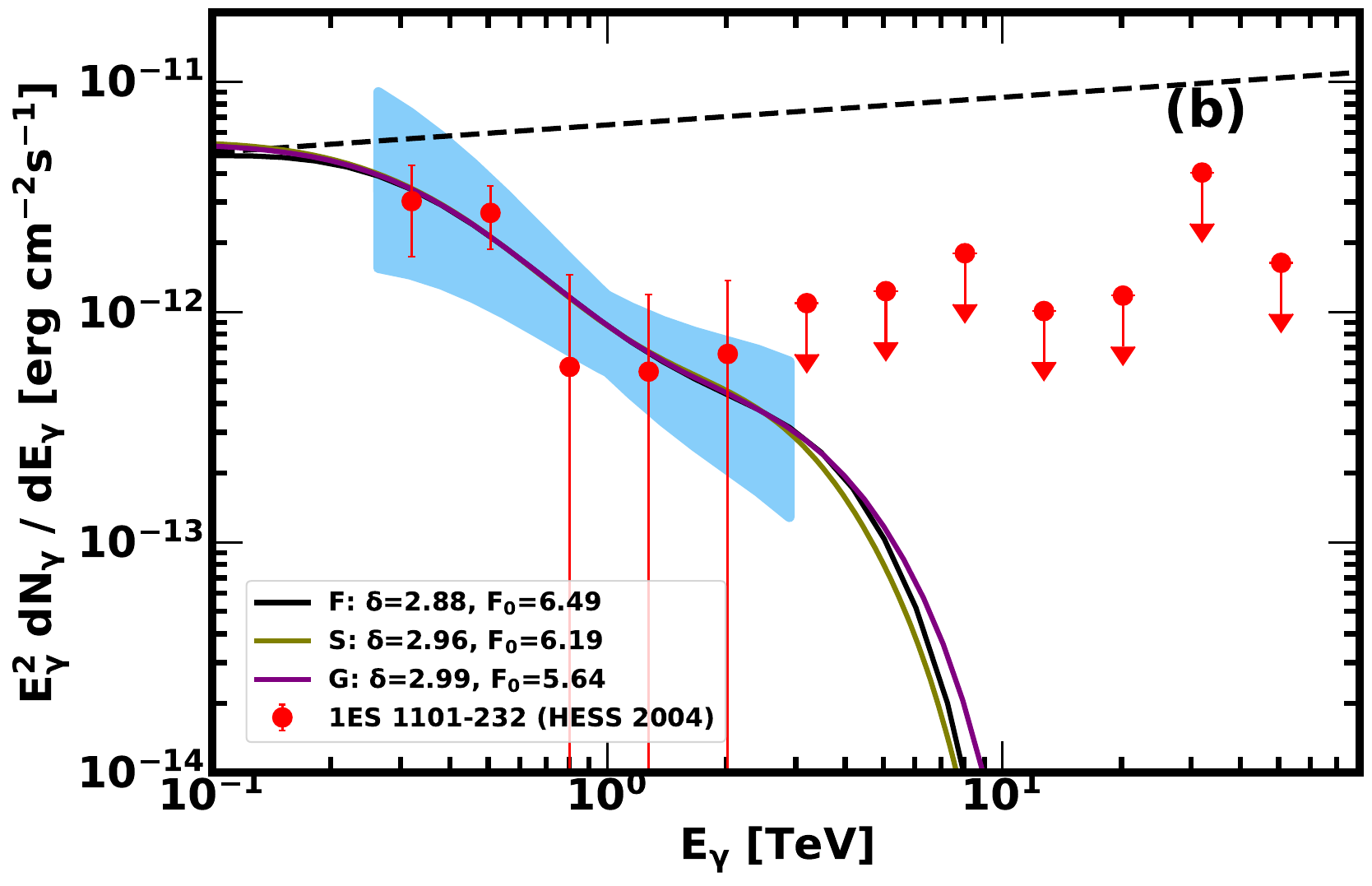}
    \includegraphics[width=0.4\textwidth]{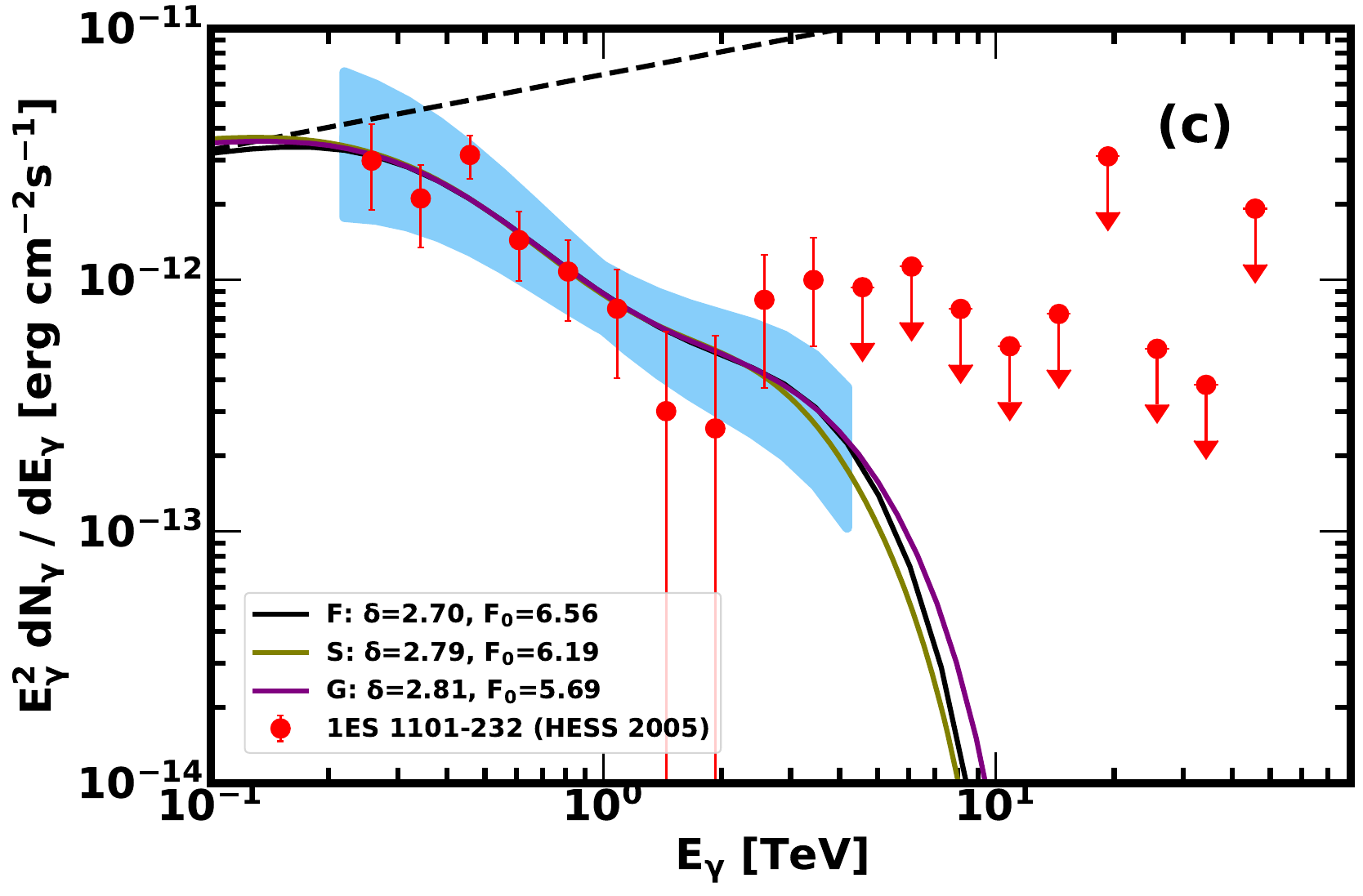}
    \includegraphics[width=0.4\textwidth]{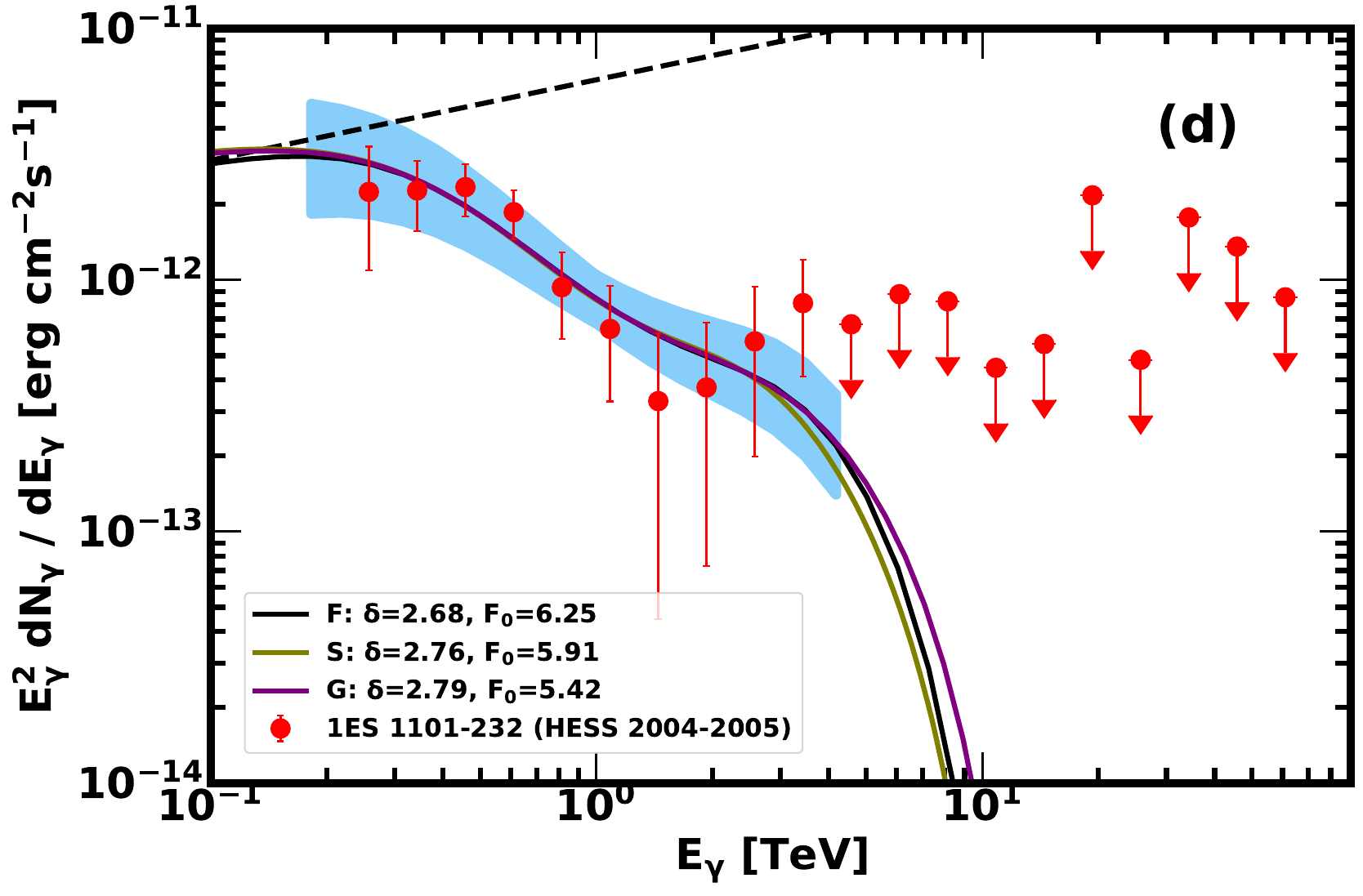}
    \caption{
    The VHE spectra of 1ES 1101-232 observed by HESS in the period 2004 to 2005~\citep{2007AA...470..475A} are fitted including EBL contribution from three EBL models to the photohadronic model.}  
    \label{fig: 1ES 1101-232}
  \end{center}
\end{figure}

\begin{figure}[!tb]
  \begin{center}
    \includegraphics[width=0.4\textwidth]{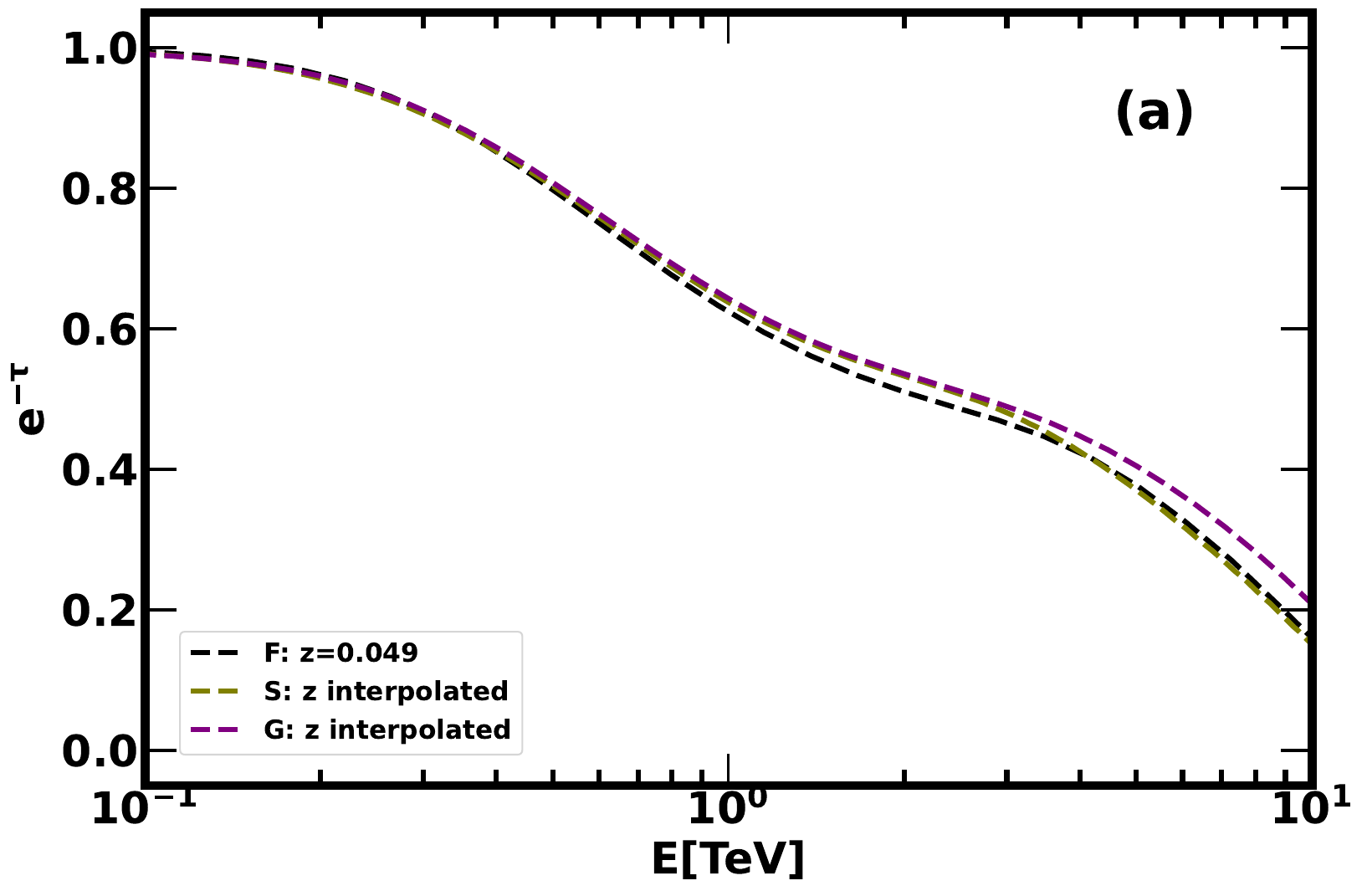}
    \includegraphics[width=0.4\textwidth]{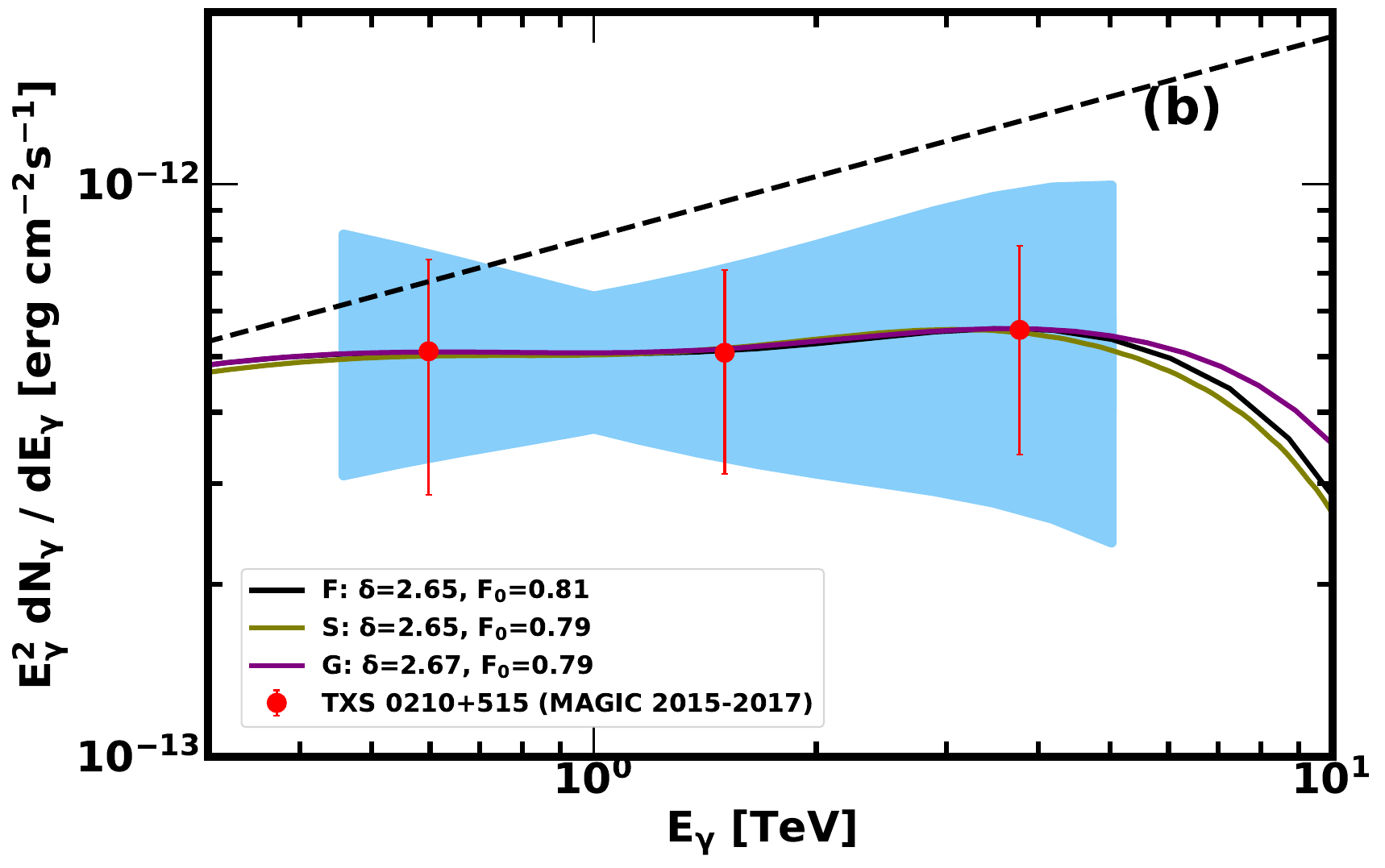}
    \caption{The average VHE spectrum of TXS 0210+515 observed by MAGIC during 2015 to 2017~\citep{2020ApJS..247...16A} is fitted by including the EBL models S, F and G to the photohadronic model. In the observed energy range all the EBL models give almost the same result.}  
    \label{fig: TXS 0210+515}
  \end{center}
\end{figure}

\begin{figure}[!tb]
  \begin{center}
    \includegraphics[width=0.4\textwidth]{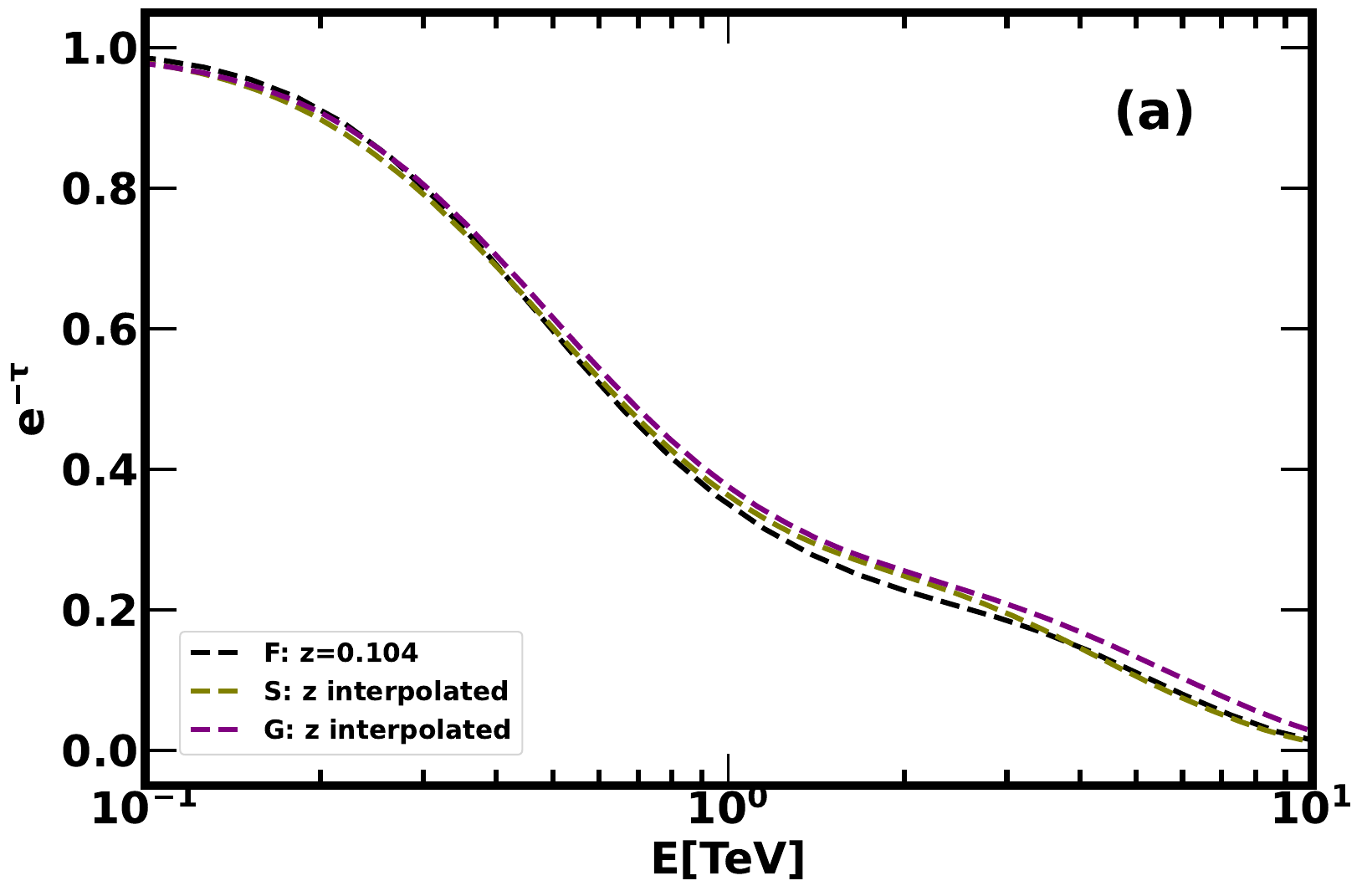}
    \includegraphics[width=0.4\textwidth]{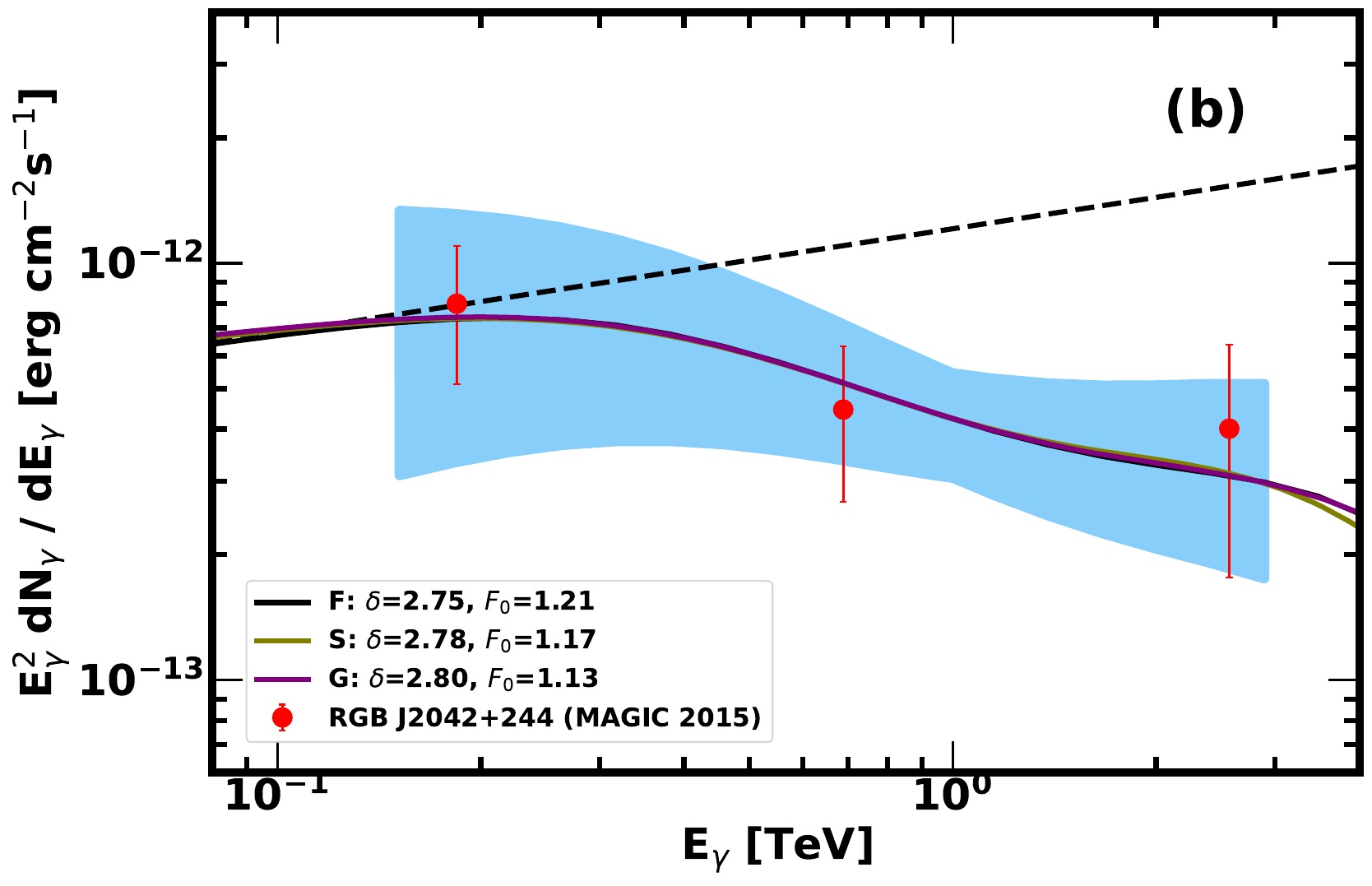}
    \caption{The VHE spectrum of RGB J2042+244 observed by MAGIC in 2015~\citep{2020ApJS..247...16A} is interpreted in the context of the photohadronic model by including the EBL correction from three EBL models discussed above.}  
    \label{fig: RGB J2042+244}
  \end{center}
\end{figure}

\begin{figure}[!tb]
  \begin{center}
    \includegraphics[width=0.4\textwidth]{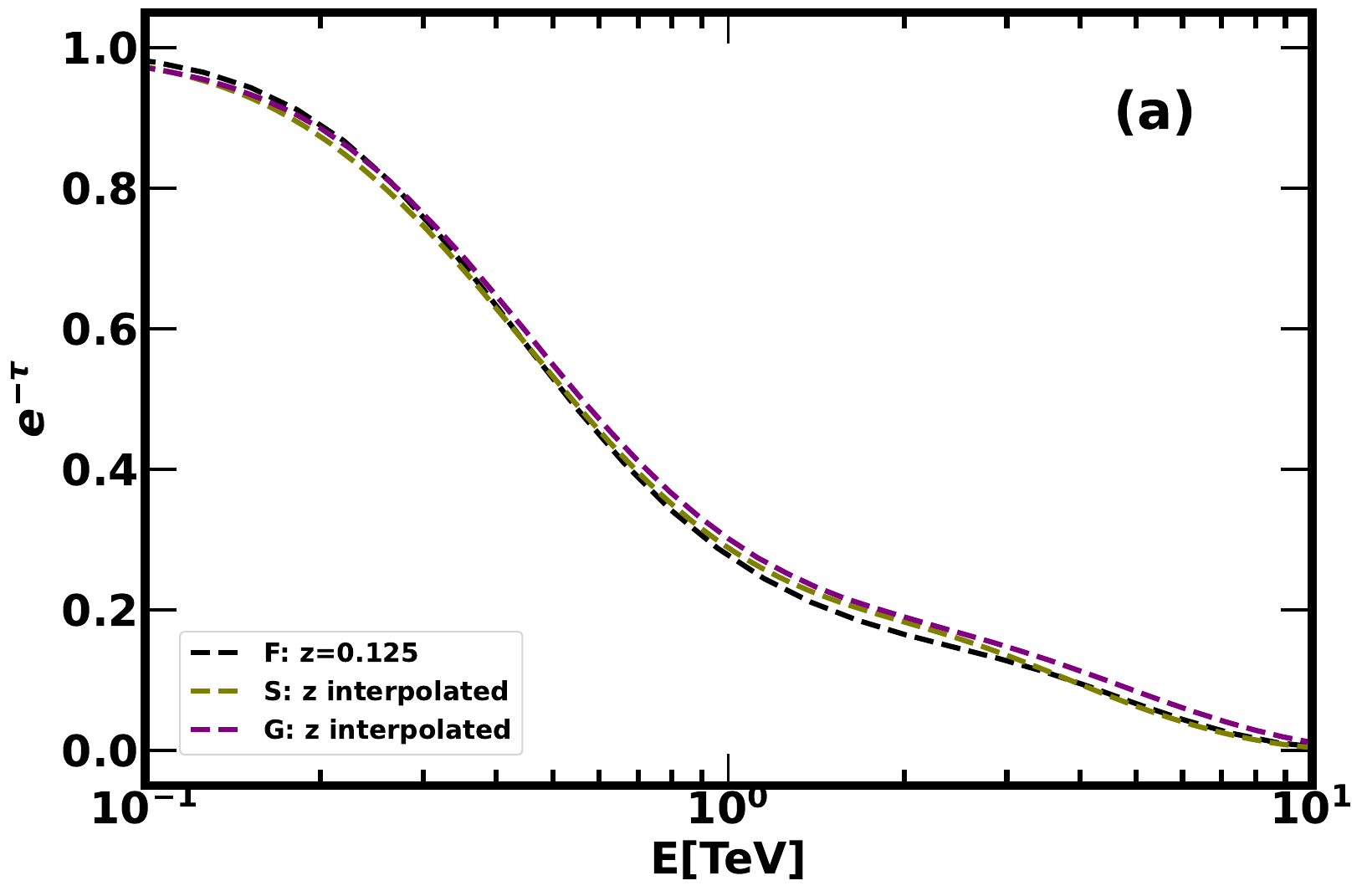}
    \includegraphics[width=0.4\textwidth]{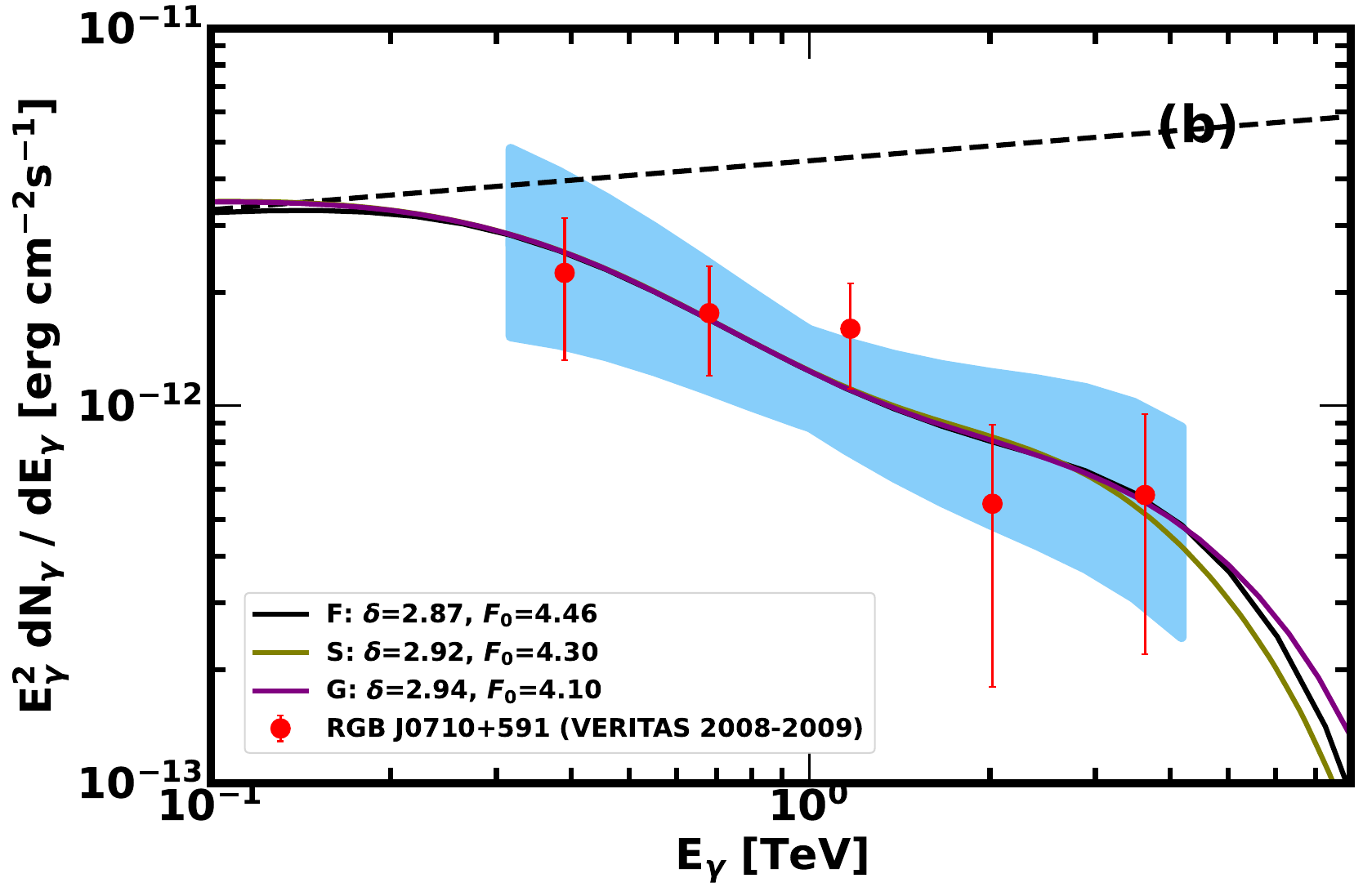}
    \caption{RGB J0710+591 was observed in VHE by VERITAS during 2008-2009~\citep{2010ApJ...715L..49A} and its spectrum is fitted by the photohadronic model with the inclusion of the EBL correction from three EBL models discussed above.}  
    \label{fig: RGB J0710+591}
  \end{center}
\end{figure}

\begin{figure}[!tb]
  \begin{center}
    \includegraphics[width=0.4\textwidth]{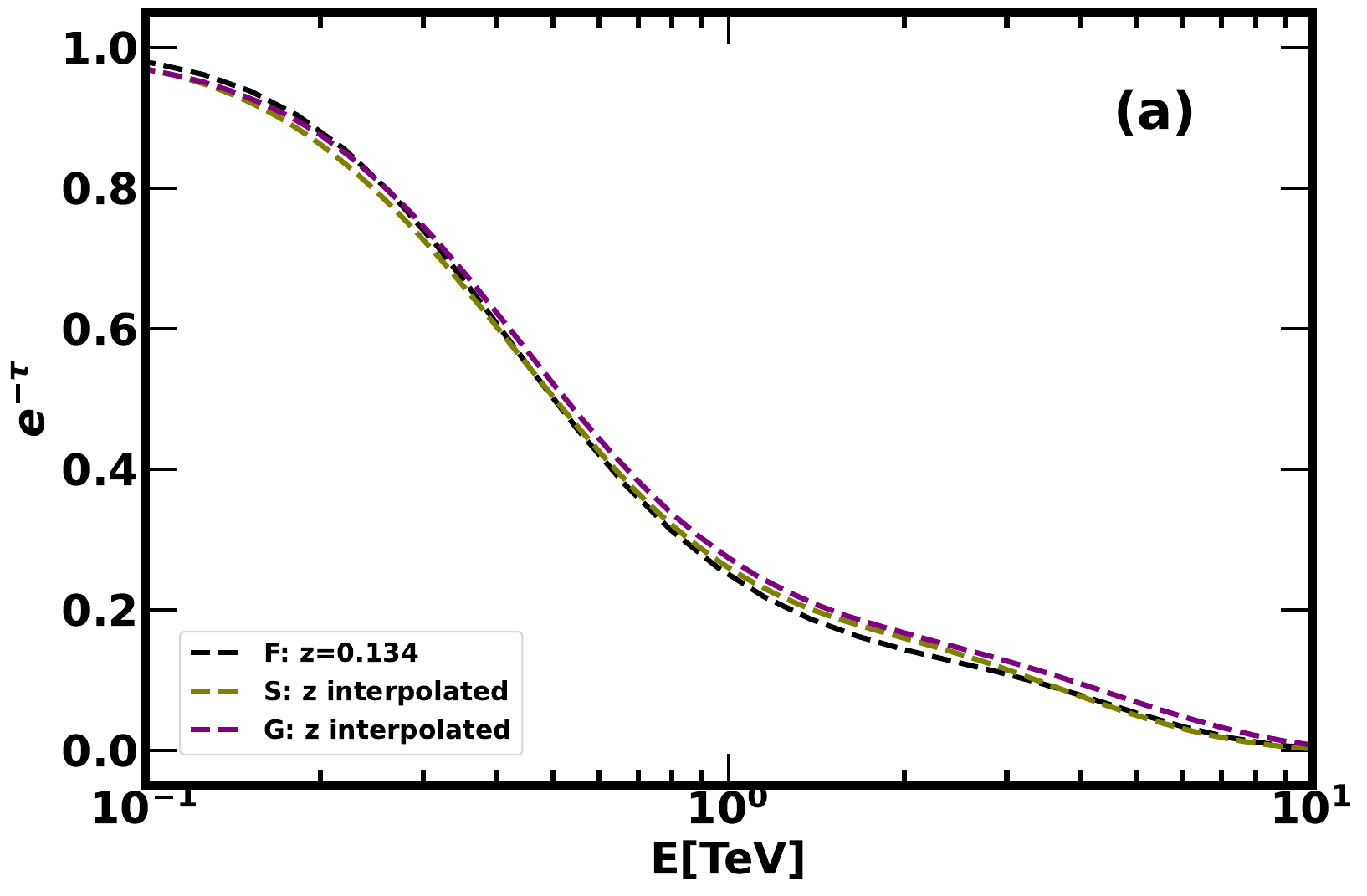}
    \includegraphics[width=0.4\textwidth]{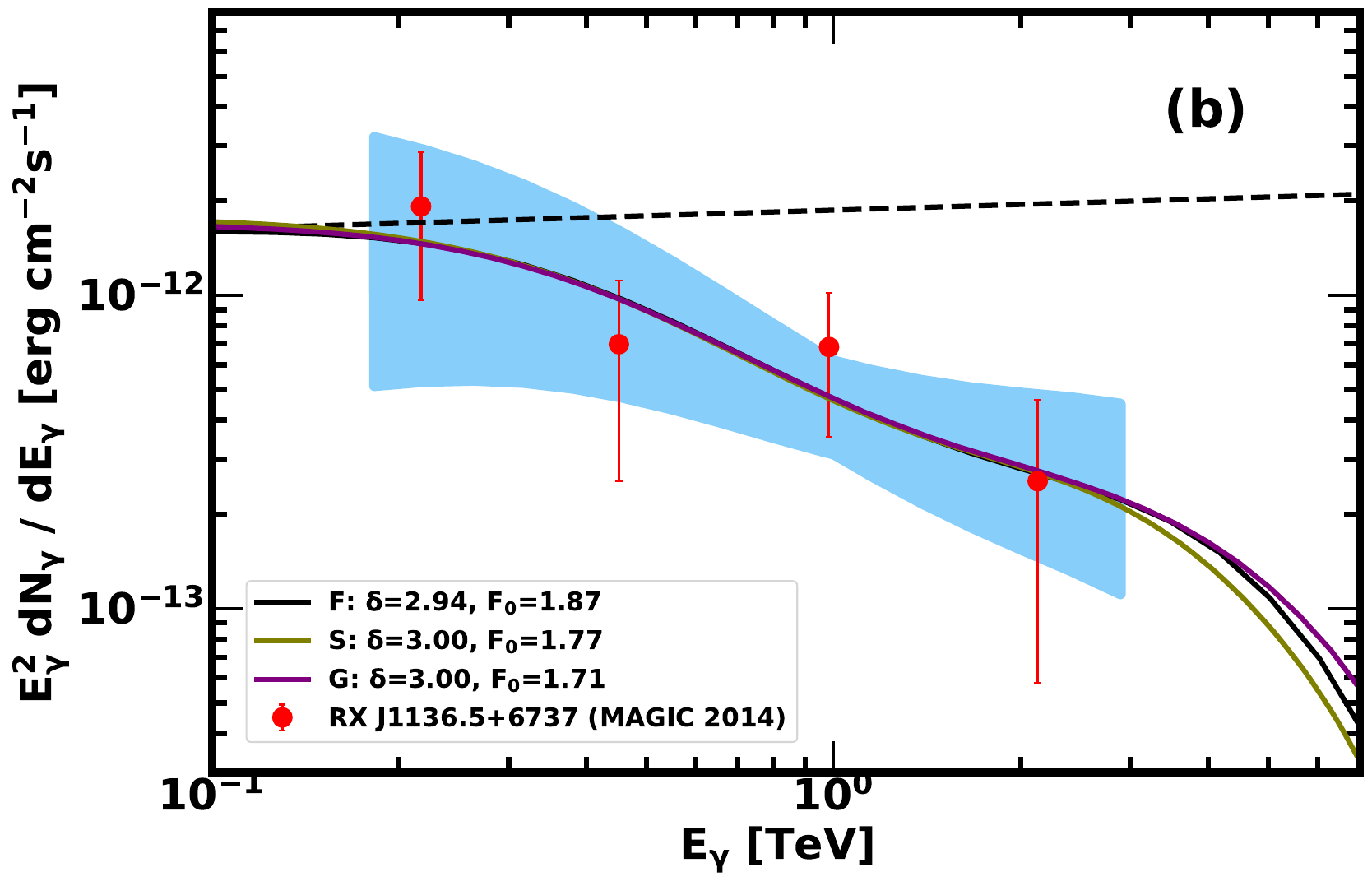}
    \caption{The VHE spectrum of RX J1136.5+6737 observed by MAGIC in 2014~\citep{2015ICRC...34..698H} is interpreted in terms of the photohadronic model and EBL correction to it by three EBL models S, F and G.}  
    \label{fig: RX J1136.5+6737}
  \end{center}
\end{figure}

\begin{figure}[!tb]
  \begin{center}
    \includegraphics[width=0.4\textwidth]{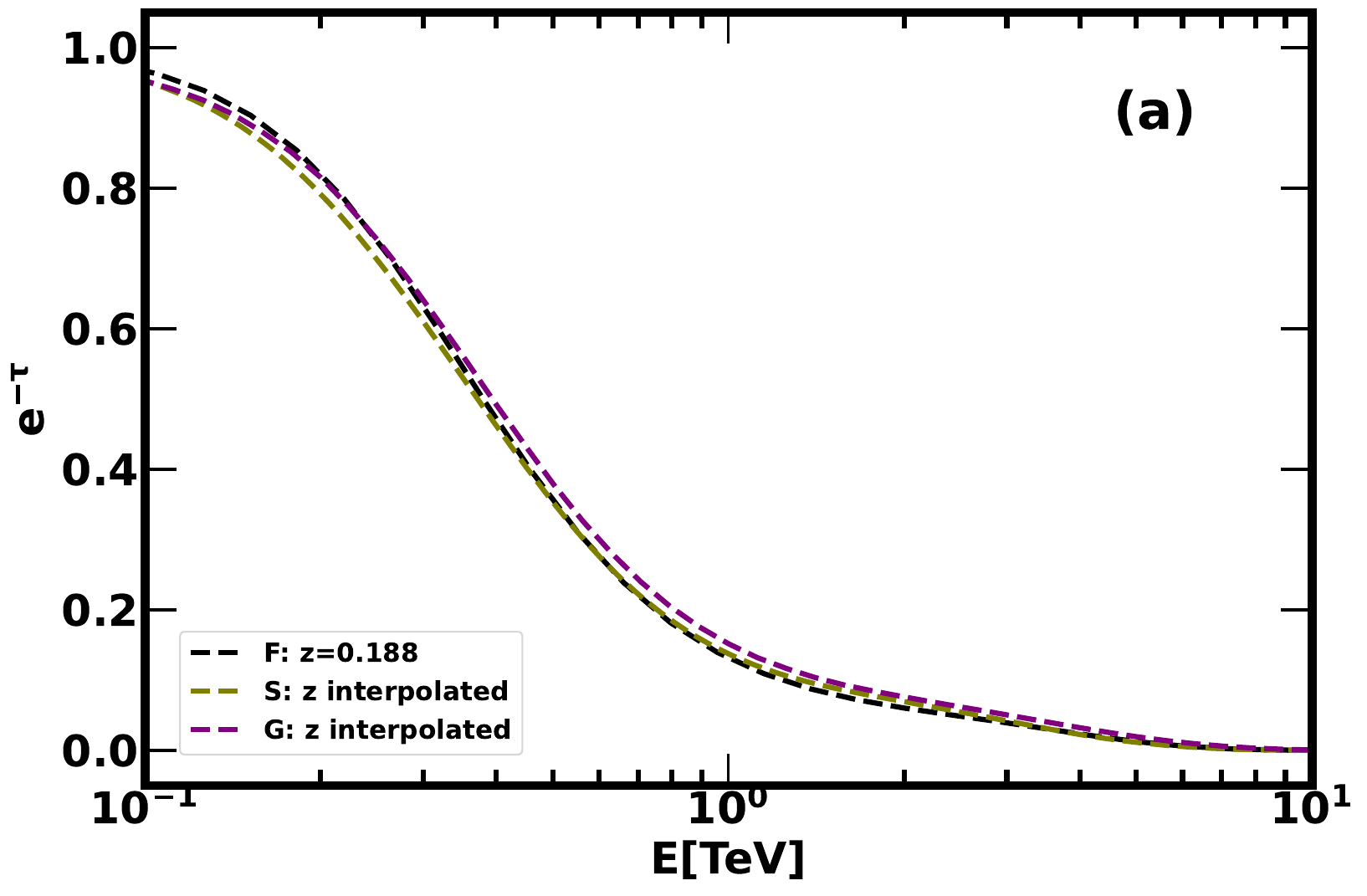}
    \includegraphics[width=0.4\textwidth]{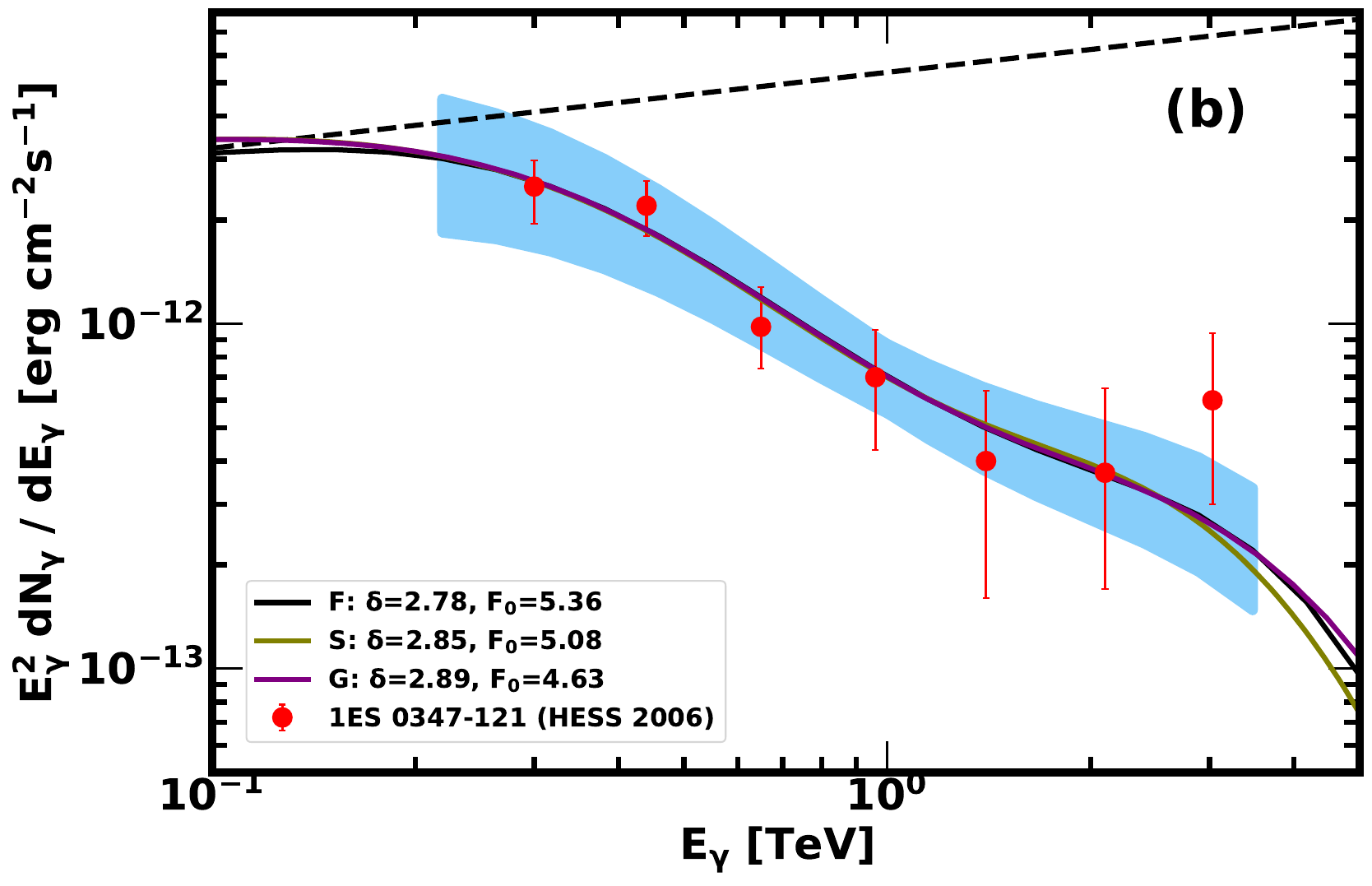}
    \caption{Photohadronic + EBL correction (using EBL model S, F, and G) interpretation to the VHE spectrum of 1ES 0347-121 observed by HESS in 2006~\citep{2007AA...473L..25A}. }  
    \label{fig: 1ES 0347-121}
  \end{center}
\end{figure}

\begin{figure}[!tb]
  \begin{center}
    \includegraphics[width=0.4\textwidth]{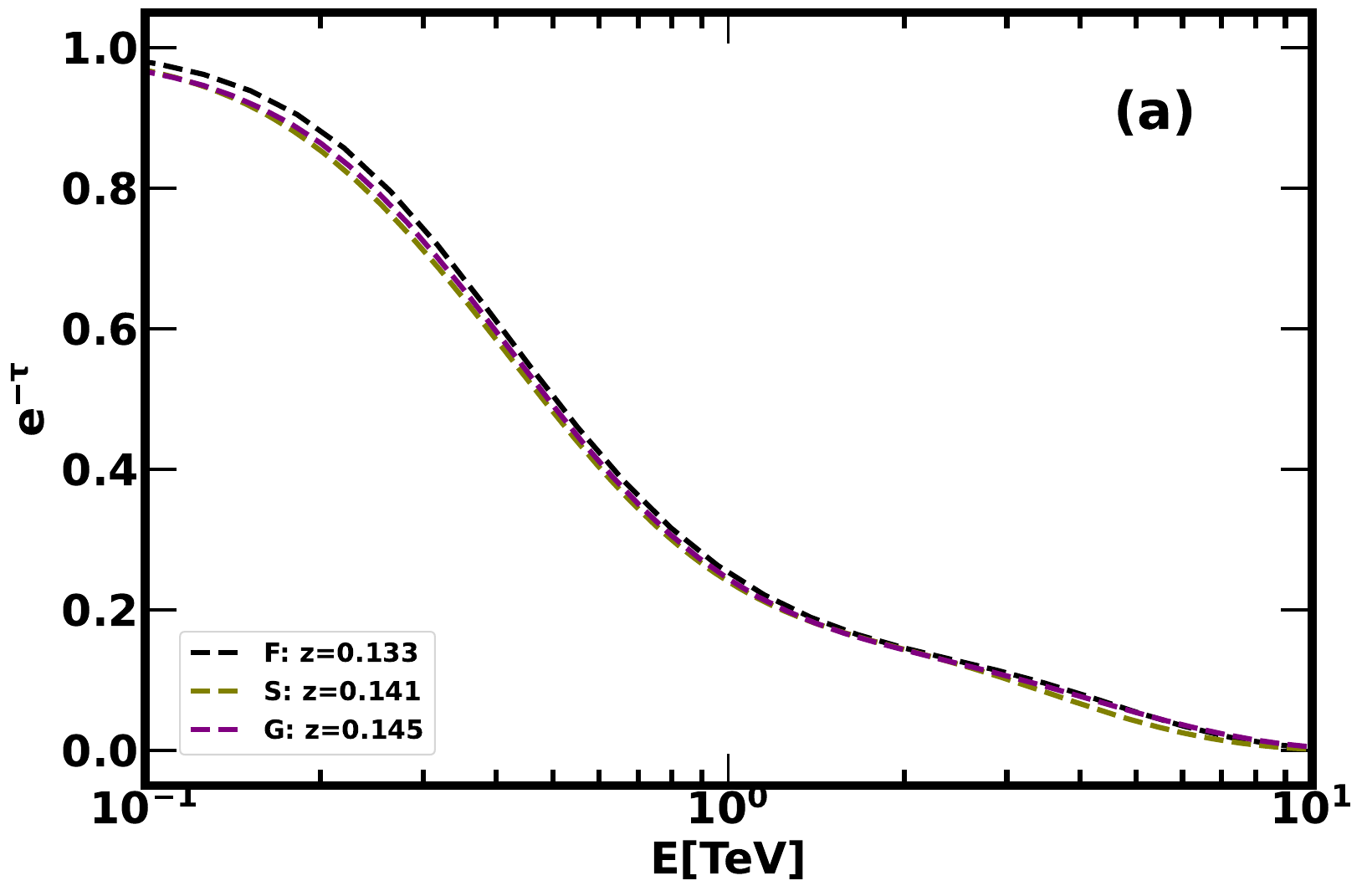}
    \includegraphics[width=0.4\textwidth]{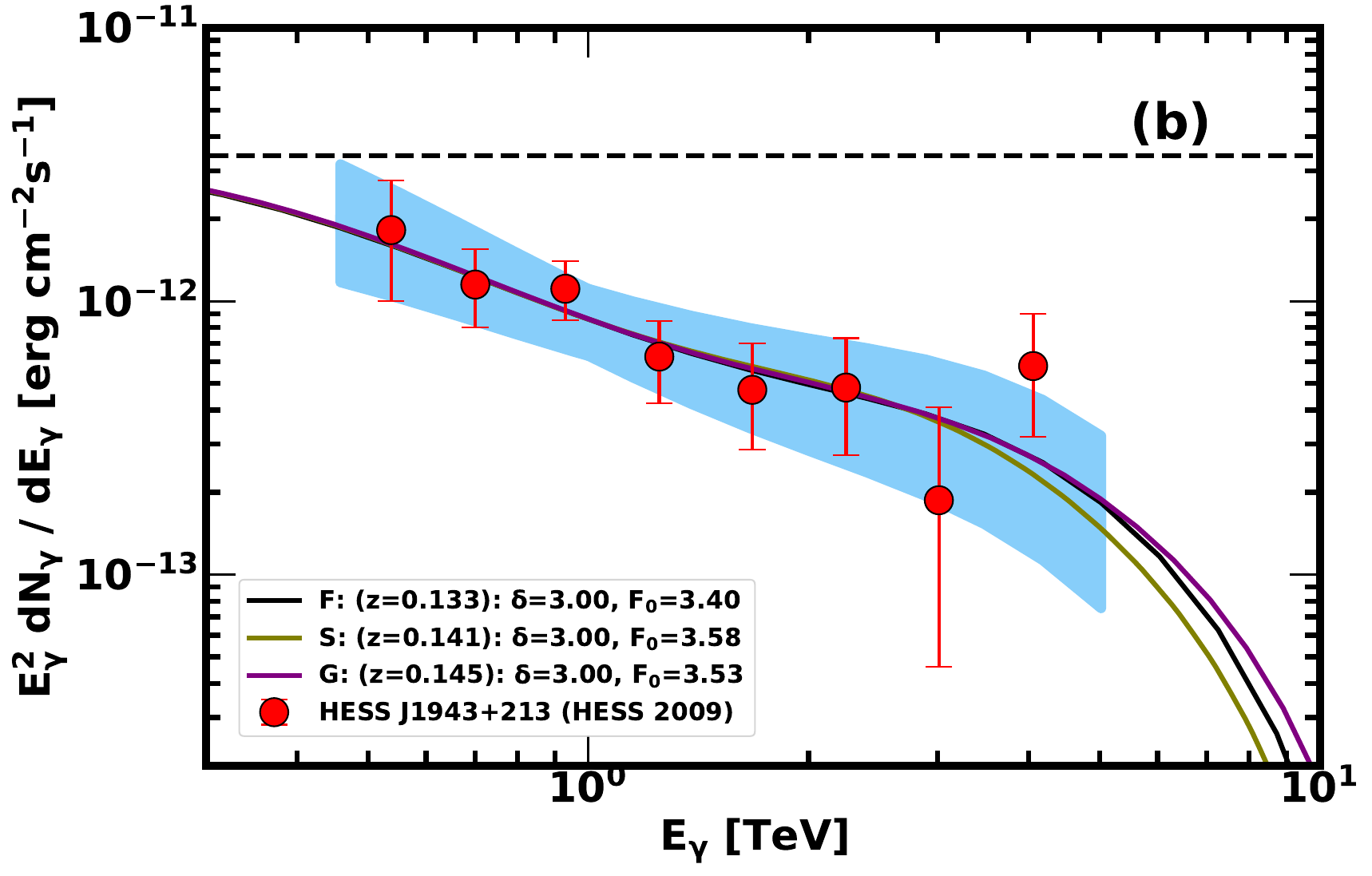}
    \includegraphics[width=0.4\textwidth]{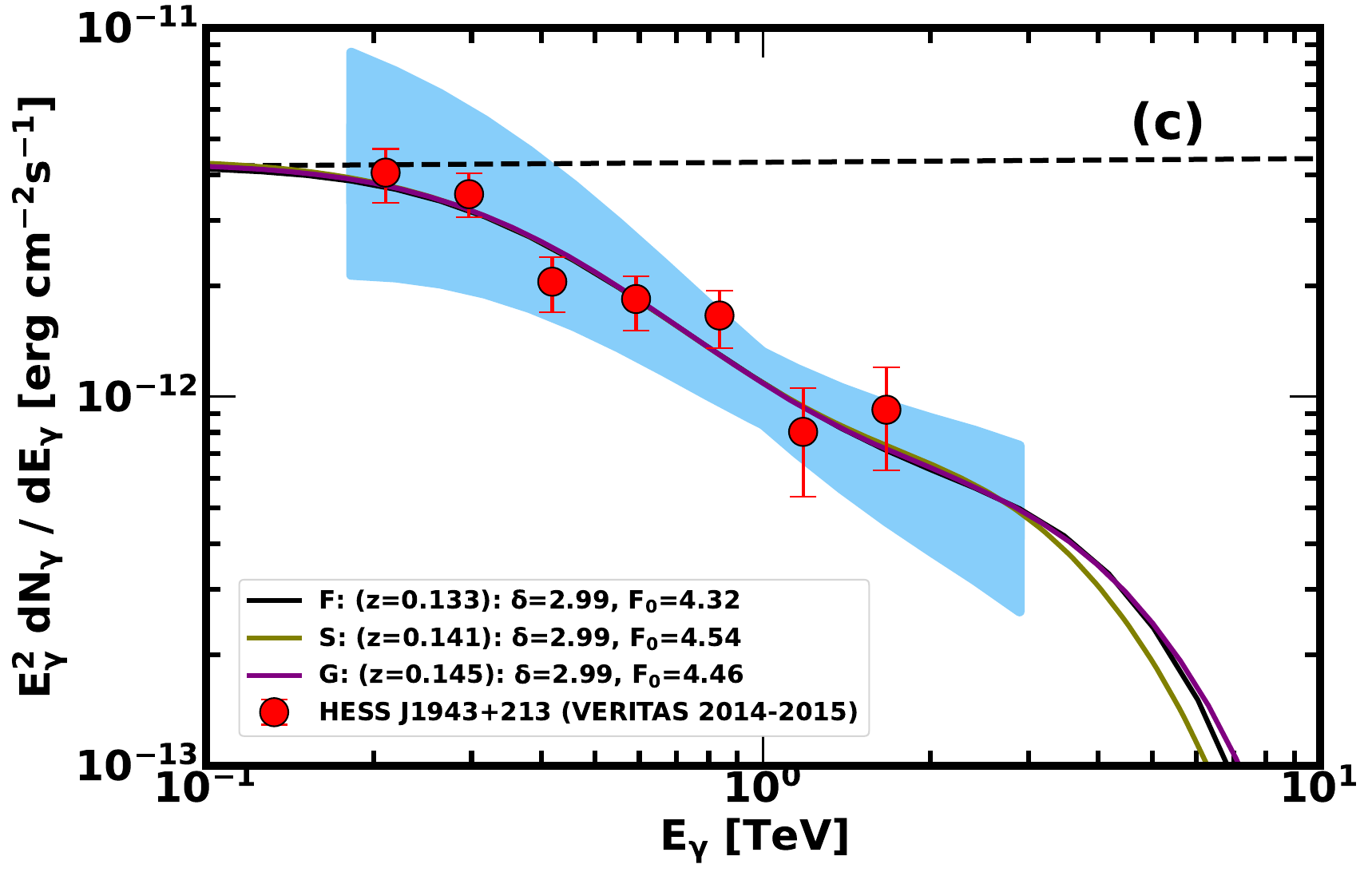}
    \caption{HESS J1943+213 was observed in VHE by HESS in 2009 and VERITAS during 2014 to 2015~\citep{2011AA...529A..49H,2018ApJ...862...41A} are 
   fitted by the photohadronic model with the inclusion of the EBL correction from three EBL models discussed above. In Figure \ref{fig: HESS J1943+213} (a), the best fit value of $z$ by each EBL model is plotted for comparison. In (b) and (c) we use these values of $z$ and fitted the VHE spectra. However, the $1\sigma$ CL is shown only for EBL-F.}
    \label{fig: HESS J1943+213}
  \end{center}
\end{figure}

\begin{figure}[!tb]
  \begin{center}
    \includegraphics[width=0.4\textwidth]{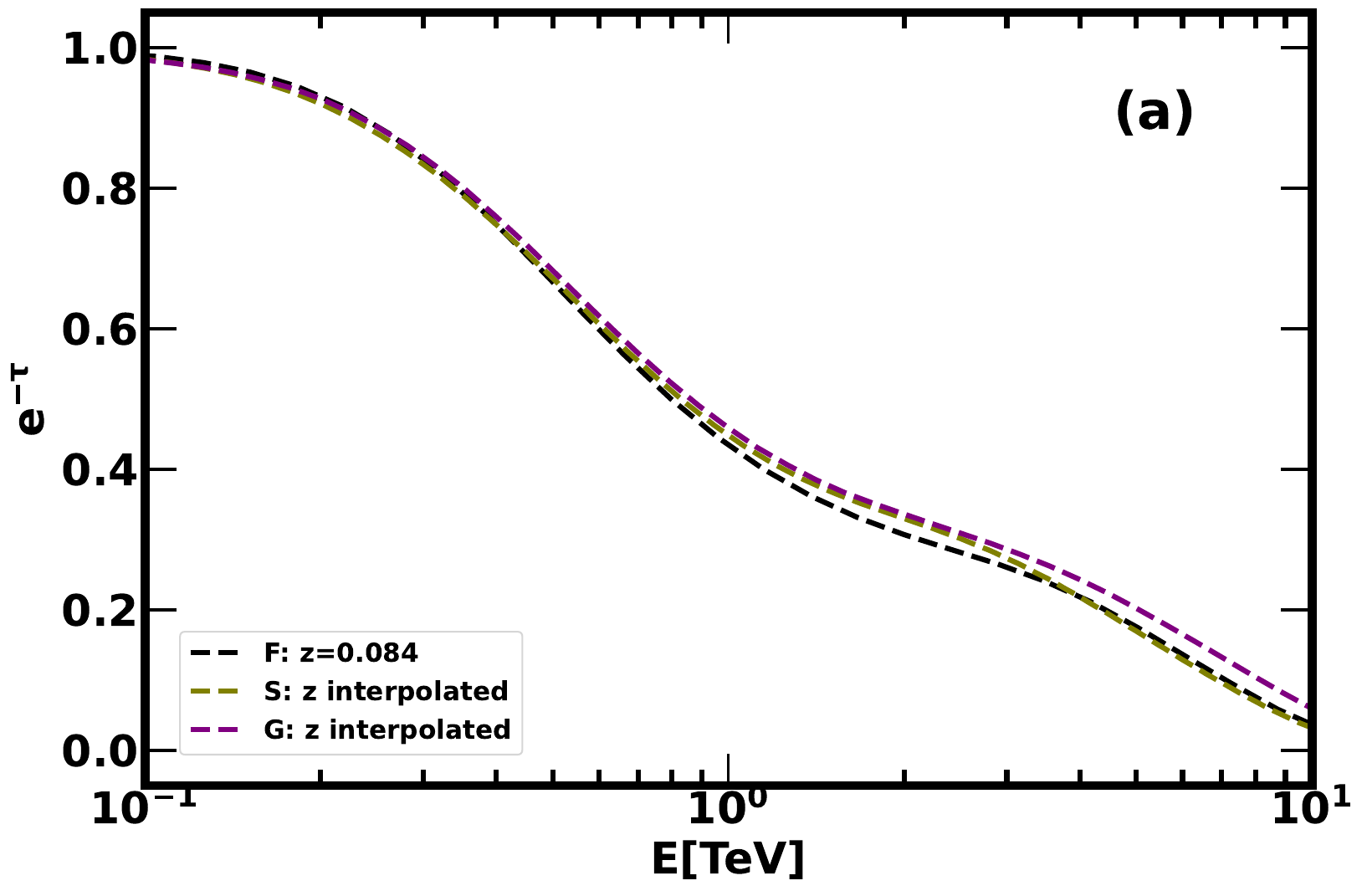}
    \includegraphics[width=0.4\textwidth]{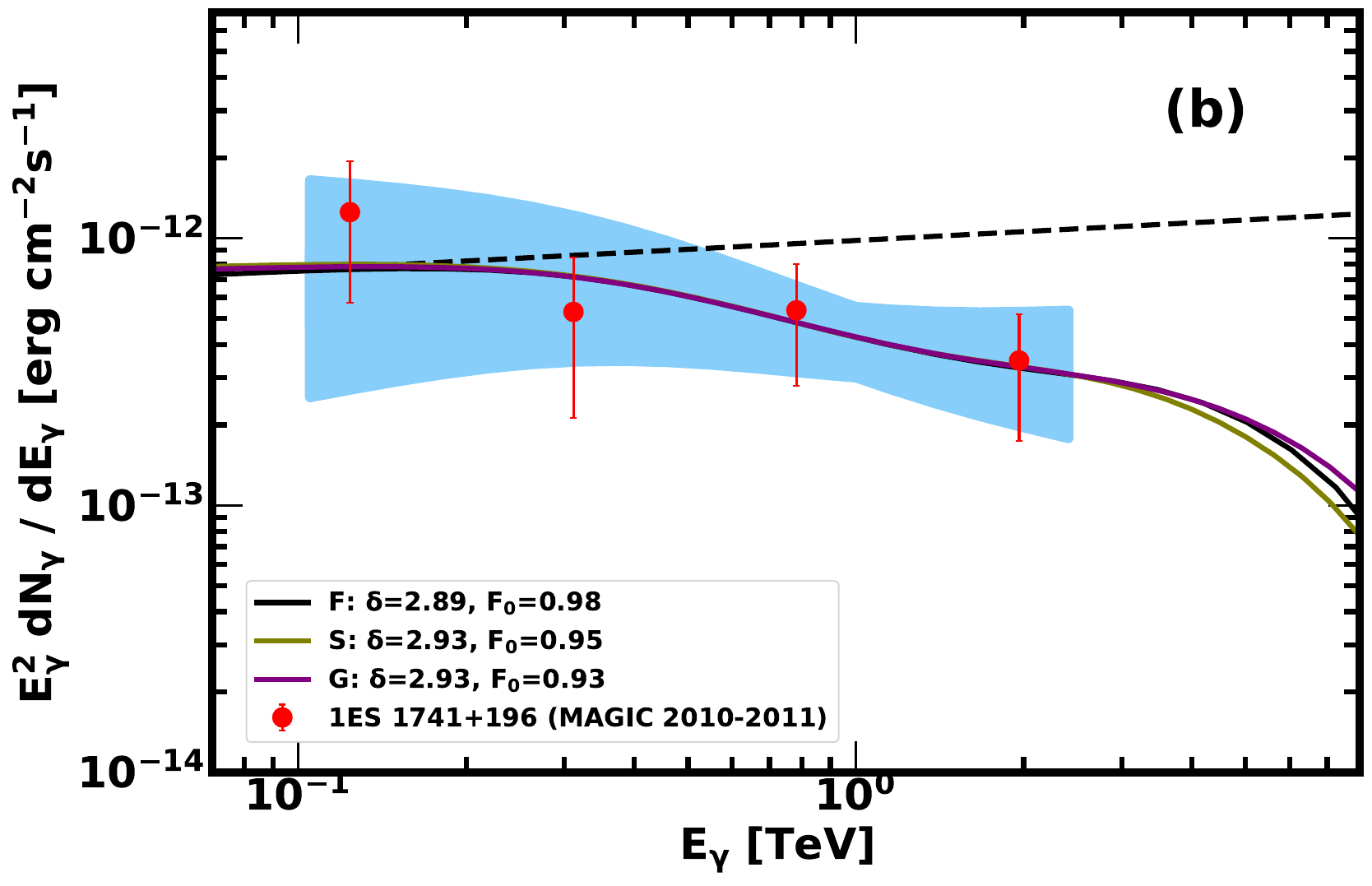}
    \includegraphics[width=0.4\textwidth]{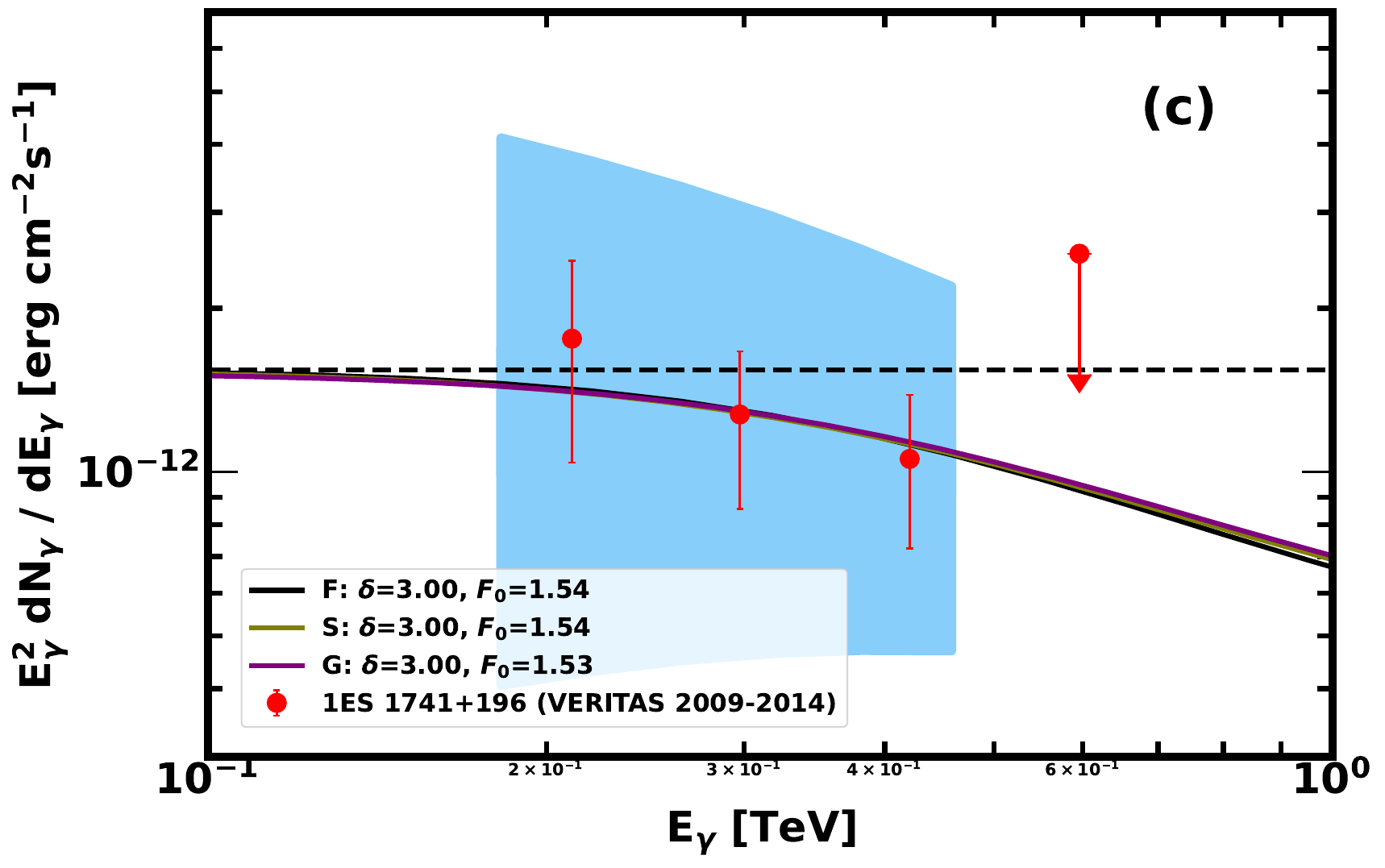}
    \caption{As usual, the VHE spectrum of 1ES 1741+196 observed by MAGIC during 2010-2011 is fitted using the photohadronic model by including the EBL effect from three EBL models. Simillarly, the average spectrum of 1ES 1741+196 observed during 2009-2014 by VERITAS is fitted including the EBL models to the photohdronic model and compared in Figure \ref{fig: 1ES 1741+196} (c)~\citep{2017MNRAS.468.1534A,2016MNRAS.459.2550A}.}  
    \label{fig: 1ES 1741+196}
  \end{center}
\end{figure}

\begin{figure}[!tb]
  \begin{center}
    \includegraphics[width=0.4\textwidth]{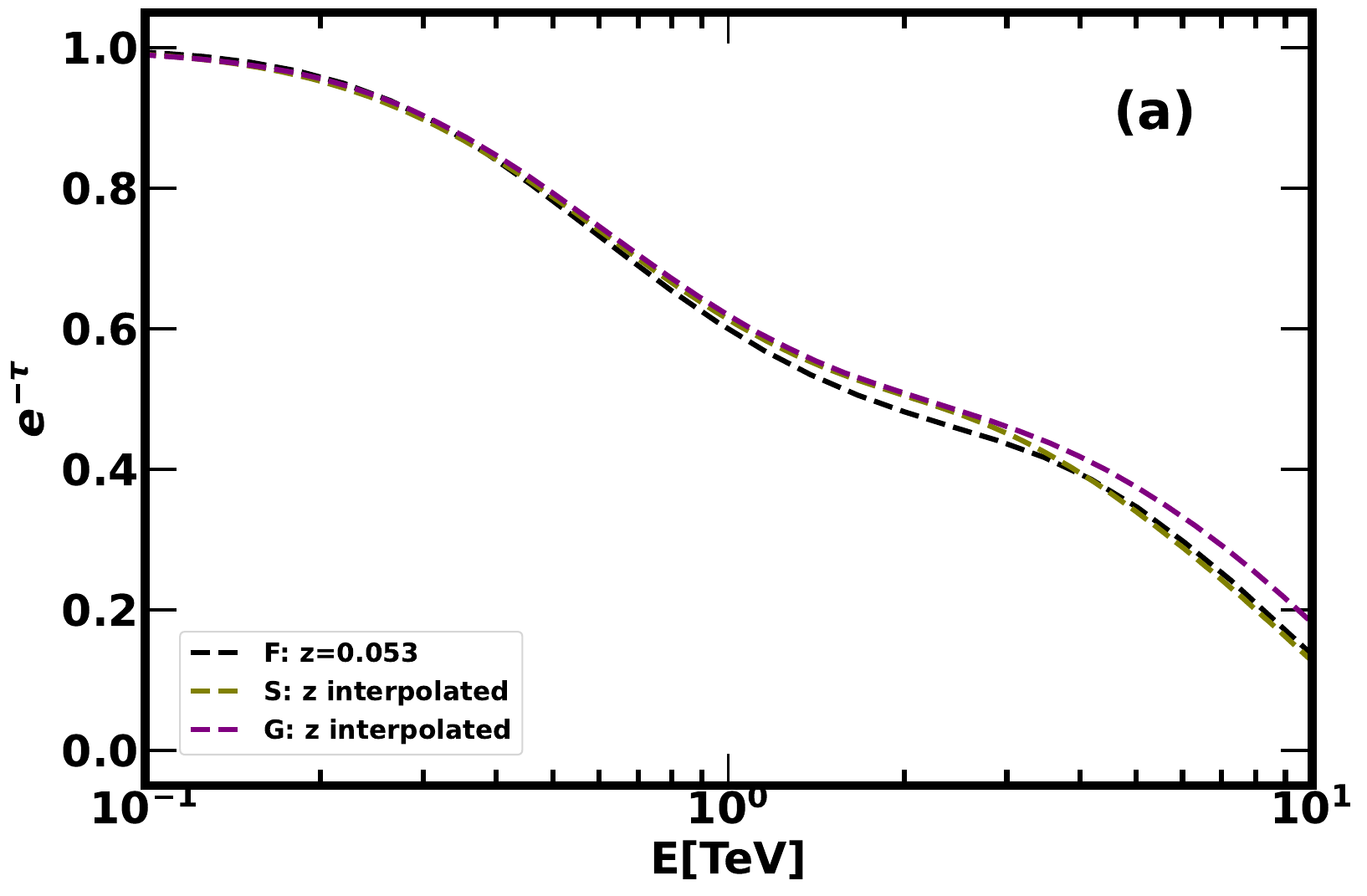}
    \includegraphics[width=0.4\textwidth]{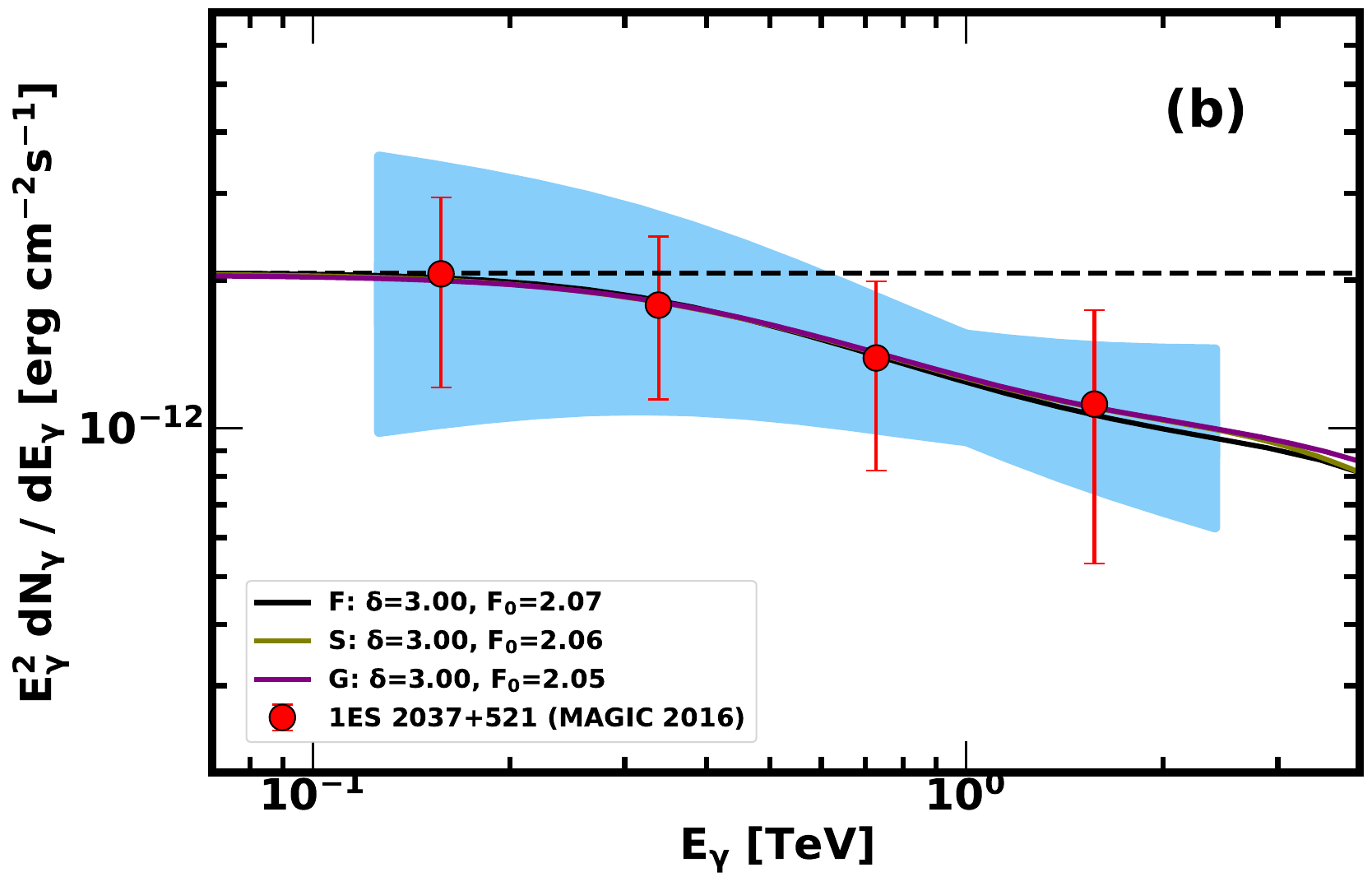}
    \caption{The VHE spectrum of 1ES 2037+521 observed by MAGIC in 2016~\citep{2020ApJS..247...16A} is fitted using photohadronic model and EBL models.}  
    \label{fig: 1ES 2037+521}
  \end{center}
\end{figure}

\begin{figure}[!tb]
  \begin{center}
    \includegraphics[width=0.4\textwidth]{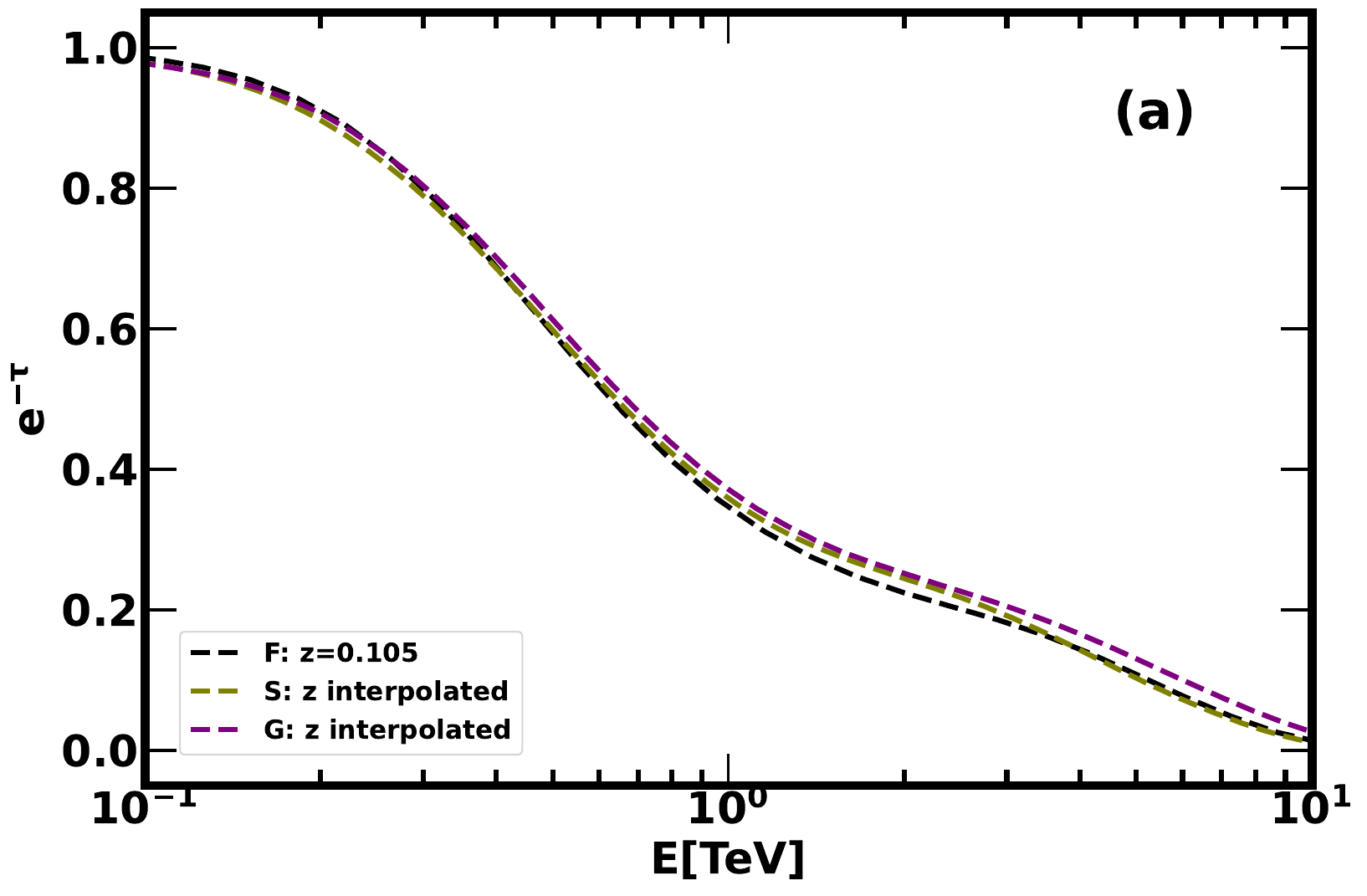}
    \includegraphics[width=0.4\textwidth]{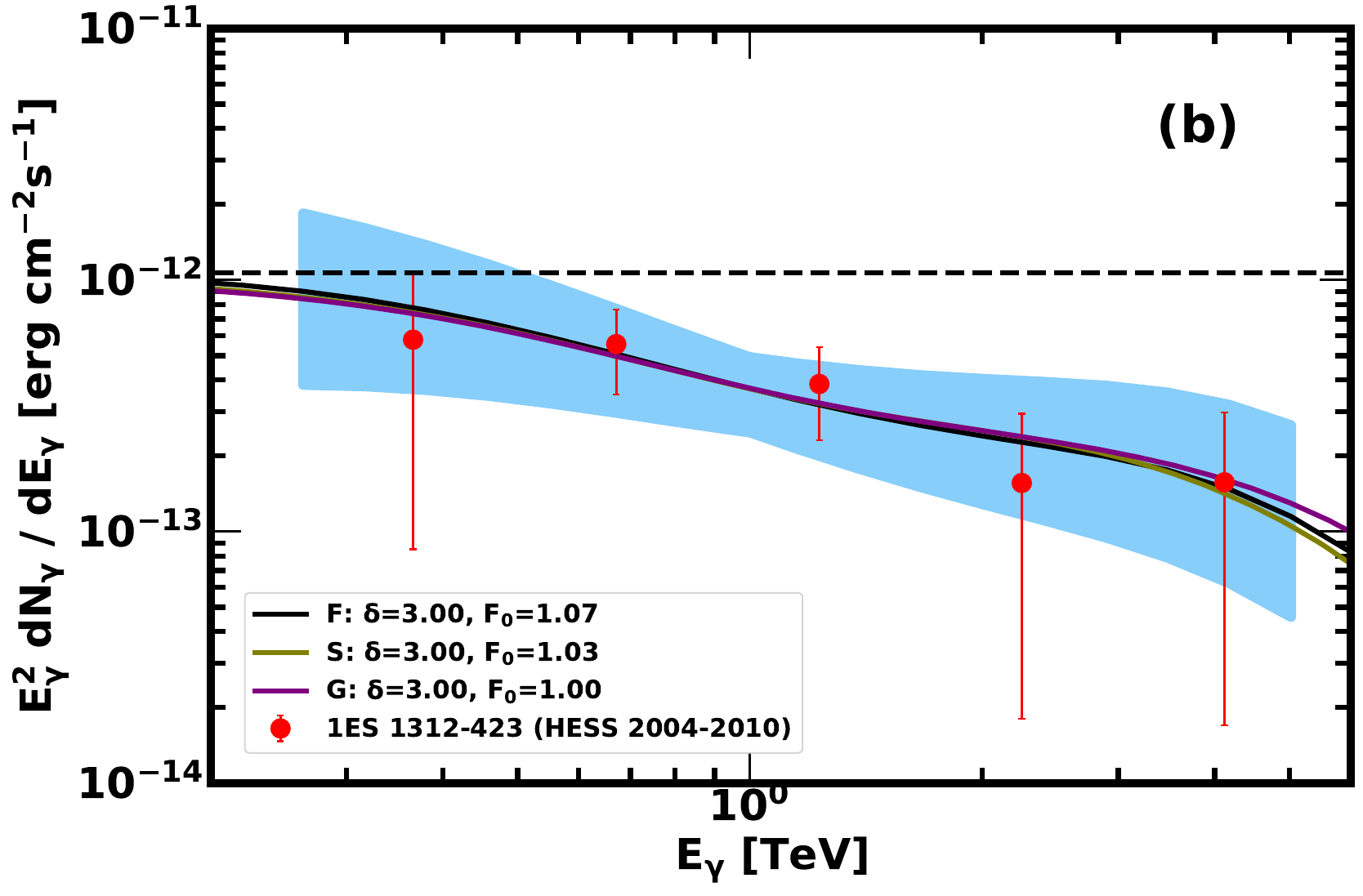}
    \caption{The average VHE spectrum of 1ES 1312-423 observed by HESS during 2004-2016 is fitted~\citep{2013MNRAS.434.1889H}.}  
    \label{fig:1ES 1312-423}.
  \end{center}
\end{figure}

\begin{figure}[!tb]
  \begin{center}
    \includegraphics[width=0.4\textwidth]{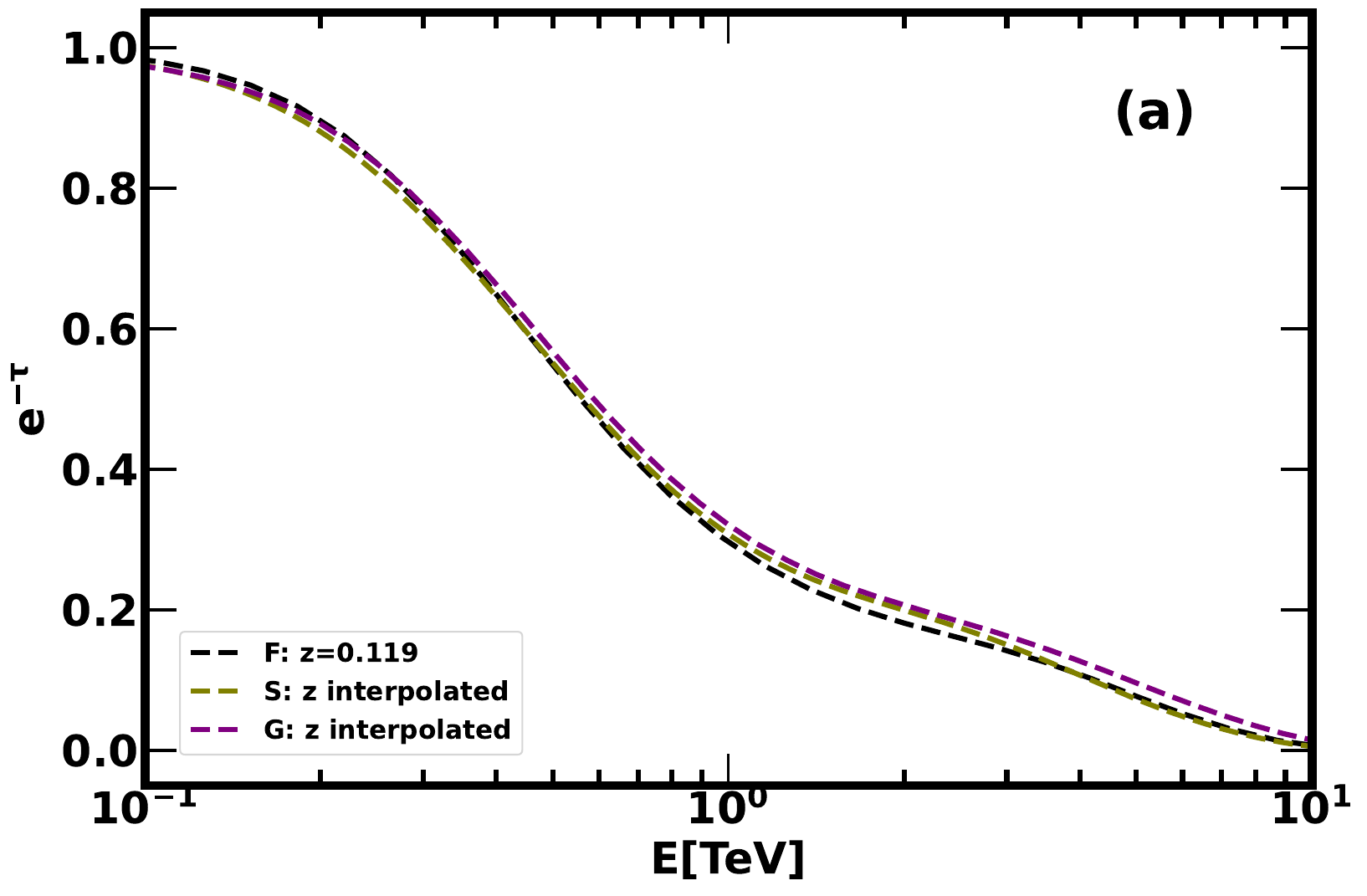}
    \includegraphics[width=0.4\textwidth]{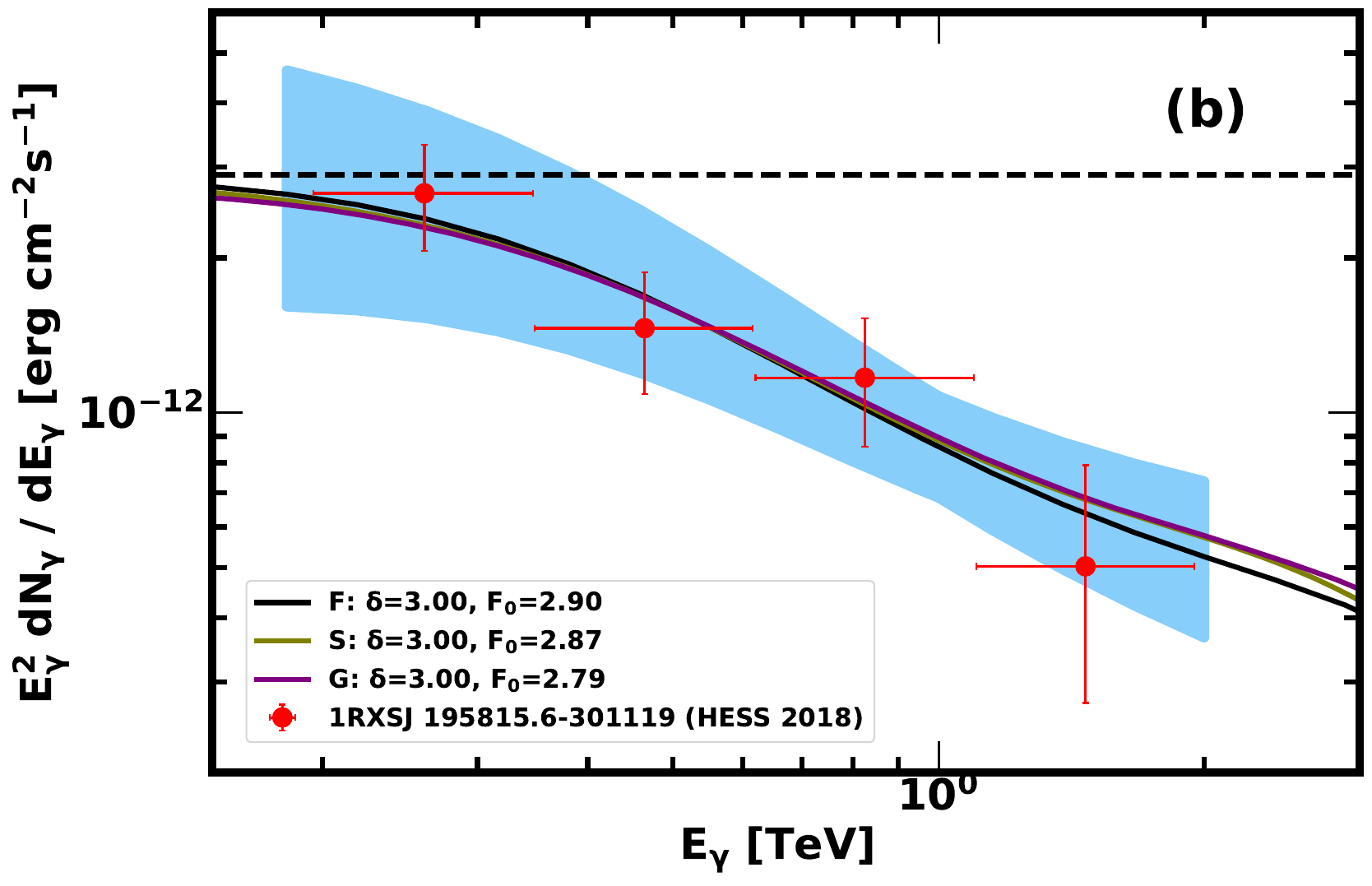}
    \caption{The VHE spectrum of 1RXSJ 195815.6-301119 observed by HESS in 2018 is fitted and results of the EBL models are compared~\citep{2022icrc.confE.823B}.}  
    \label{fig: 1RXSJ 195815.6-301119}
  \end{center}
\end{figure}

\begin{figure}[!tb]
  \begin{center}
    \includegraphics[width=0.4\textwidth]{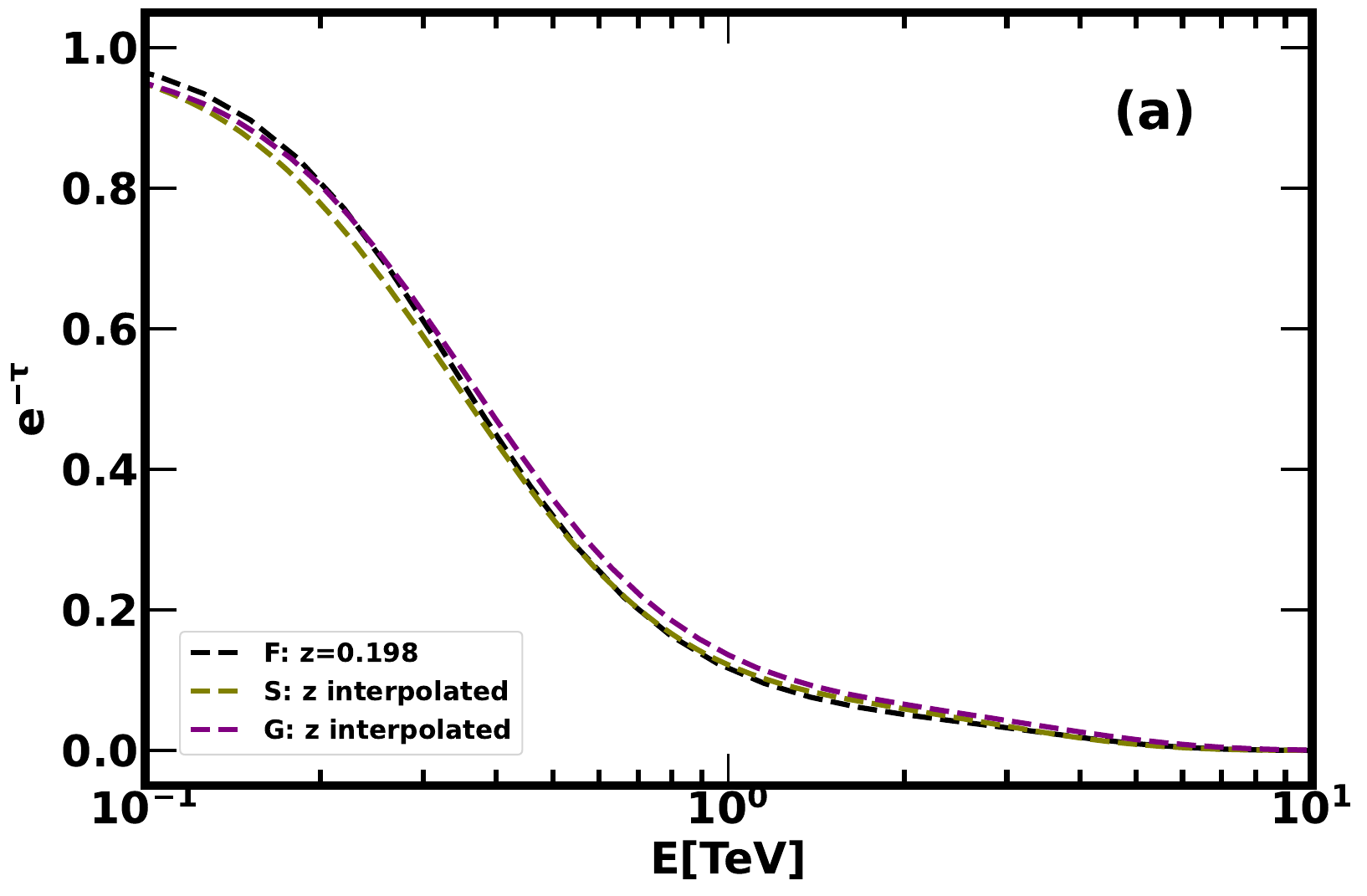}
    \includegraphics[width=0.4\textwidth]{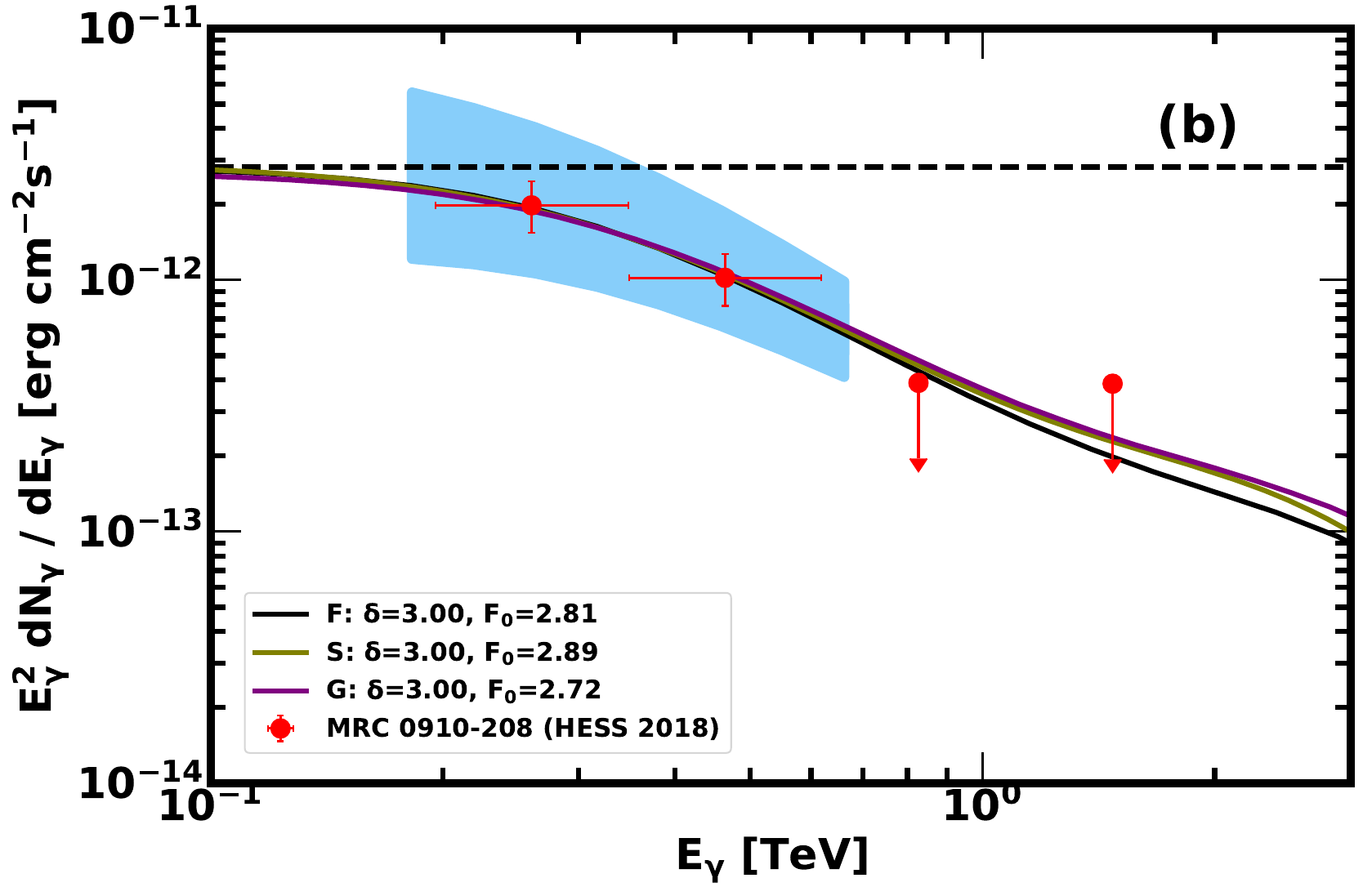}
    \includegraphics[width=0.4\textwidth]{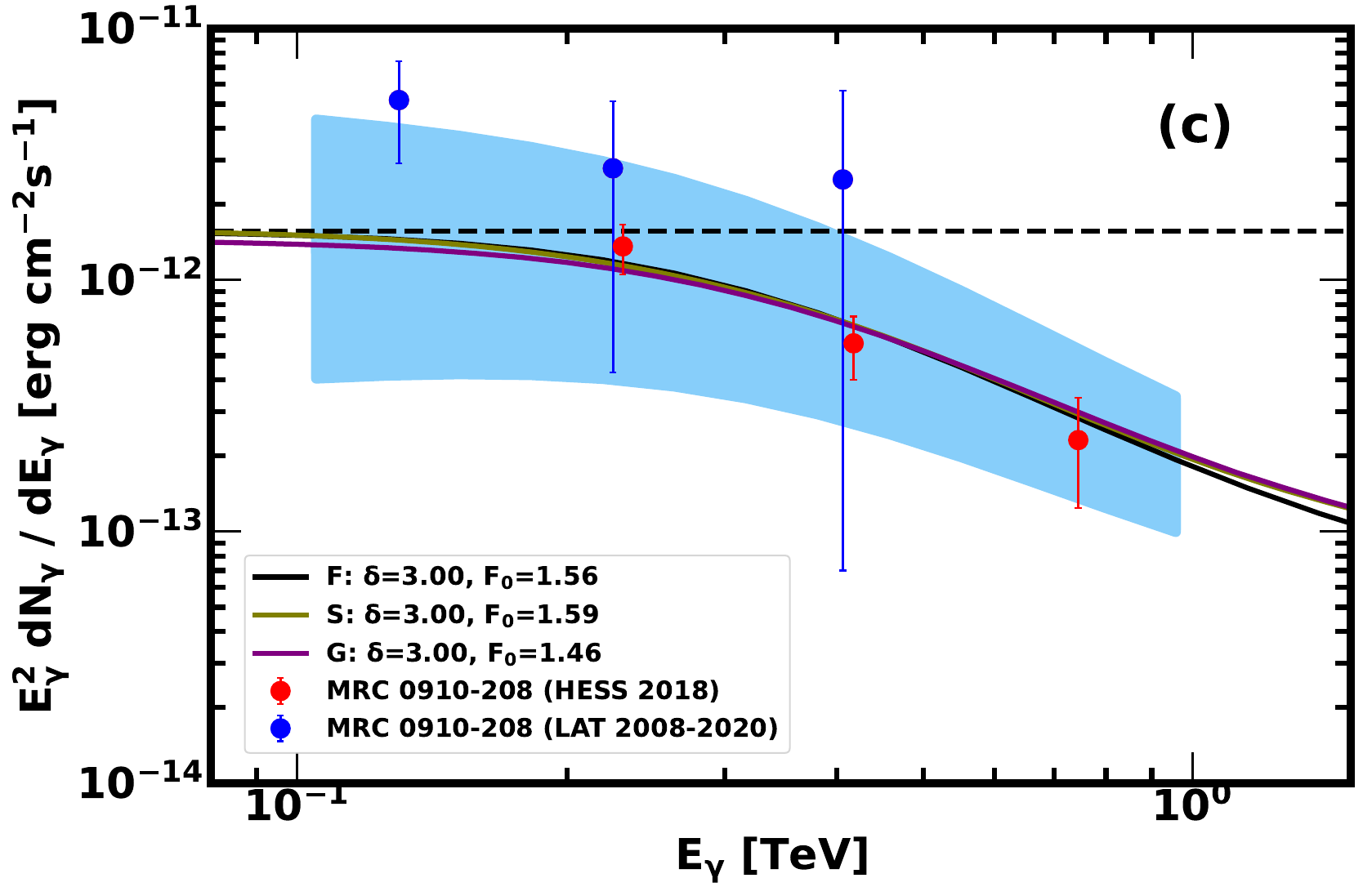}
    \caption{The VHE spectra of MRC 0910-208 observed by HESS in 2018 and the average spectrum observed by LAT+HESS during 20028-2020 are fitted and the results of the EBL models are compared~\citep{2022icrc.confE.823B}.}  
    \label{fig: MRC 0910-208}
  \end{center}
\end{figure}

\begin{figure}[!tb]
  \begin{center}
    \includegraphics[width=0.4\textwidth]{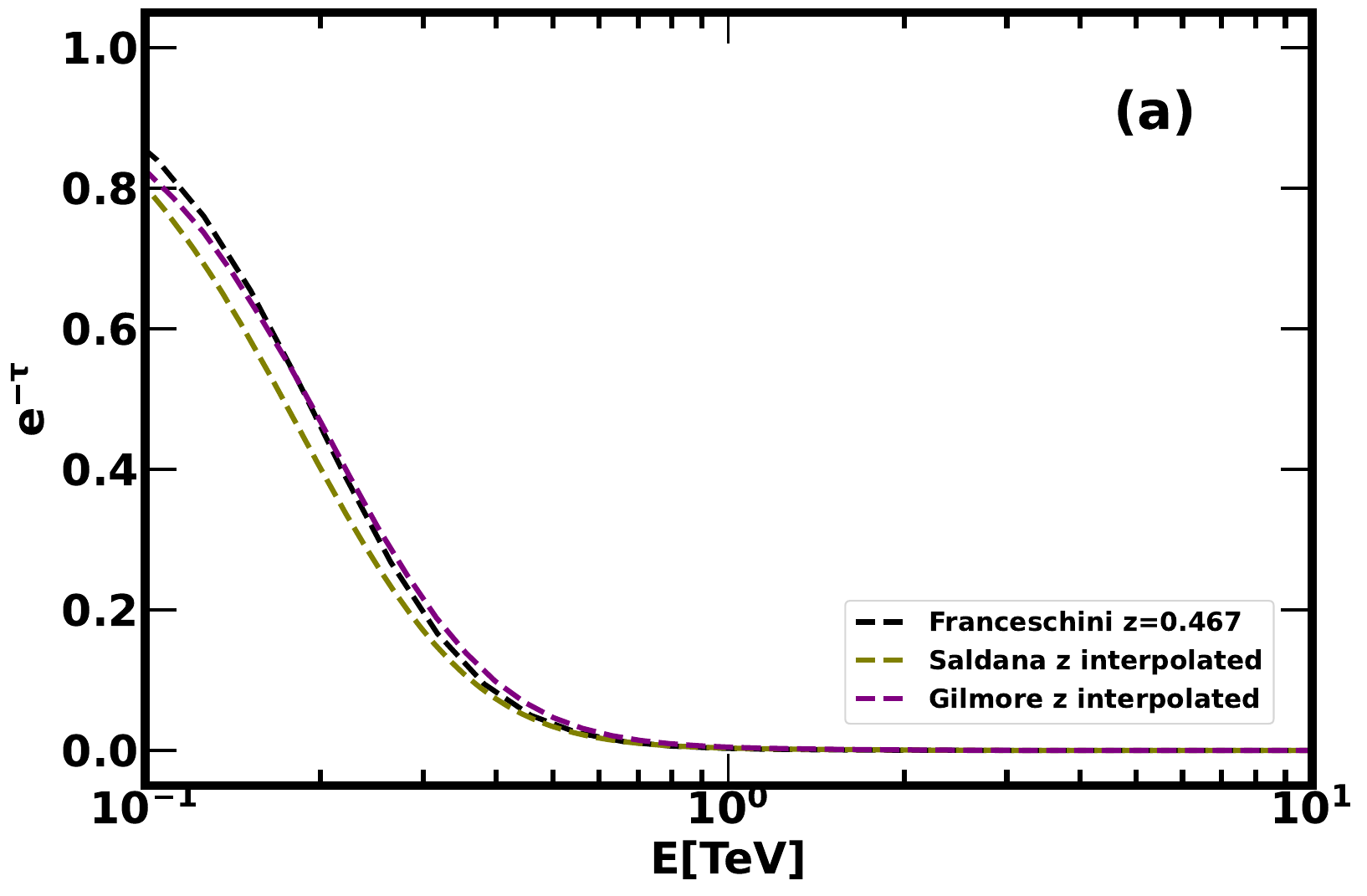}
    \includegraphics[width=0.4\textwidth]{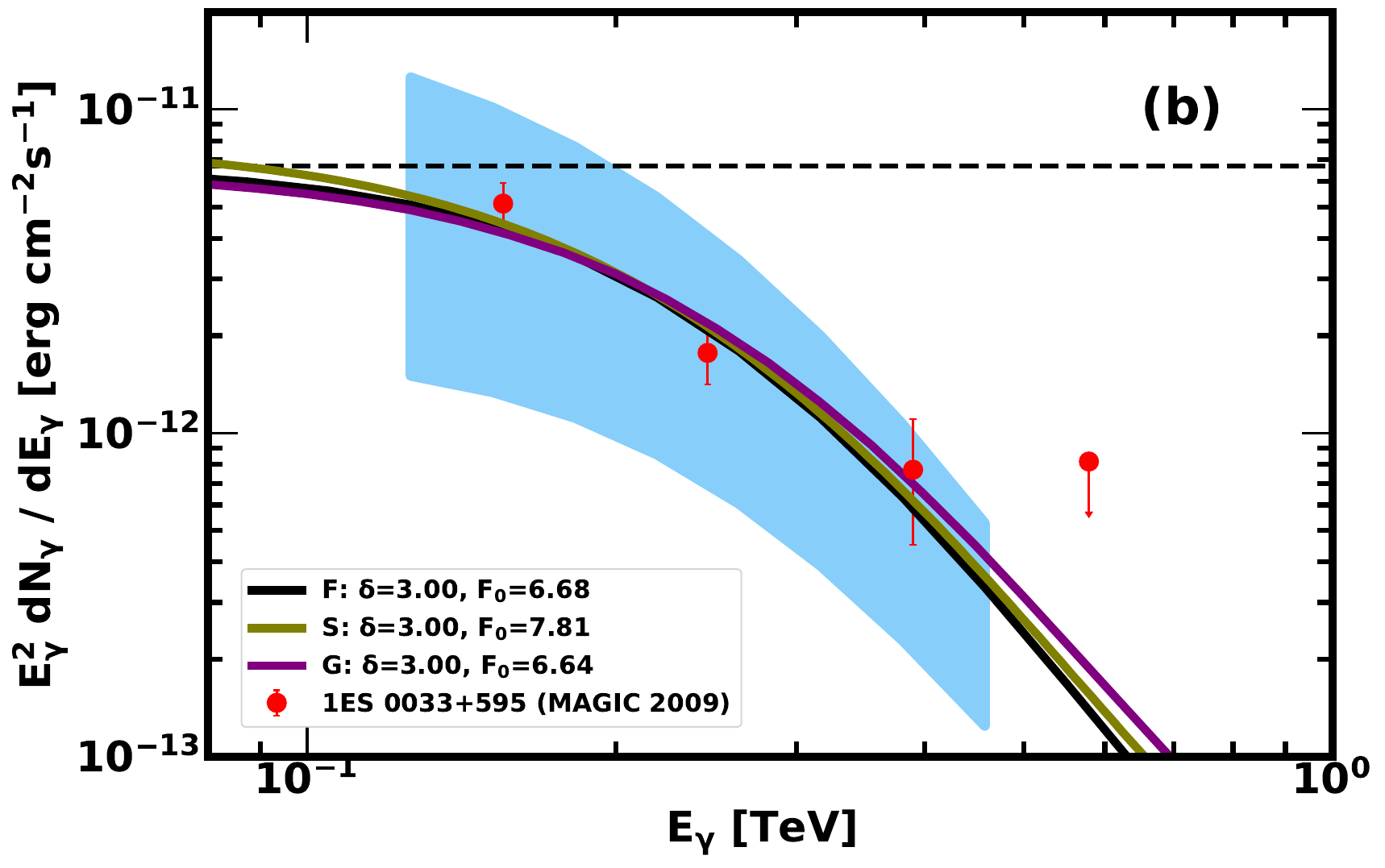}
    \caption{The VHE spectrum of 1ES 0033+595 observed by MAGIC in 2009~\citep{2015MNRAS.446..217A} is interpreted in terms of the photohadronic model and EBL correction to it by three EBL models S, F and G.}
    \label{fig: 1ES 0033+595}
  \end{center}
\end{figure}

\begin{figure}[!tb]
  \begin{center}
    \includegraphics[width=0.4\textwidth]{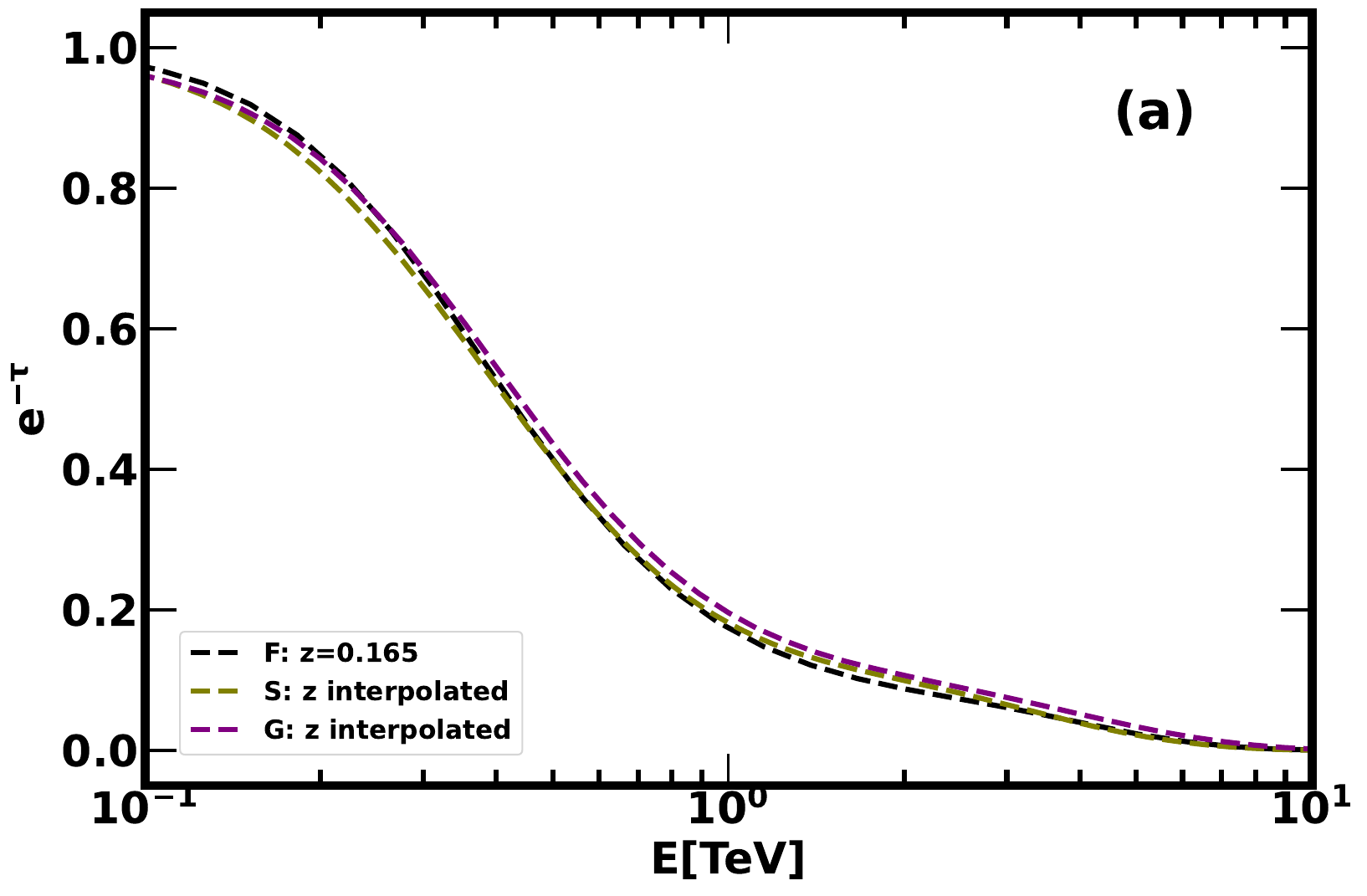}
    \includegraphics[width=0.4\textwidth]{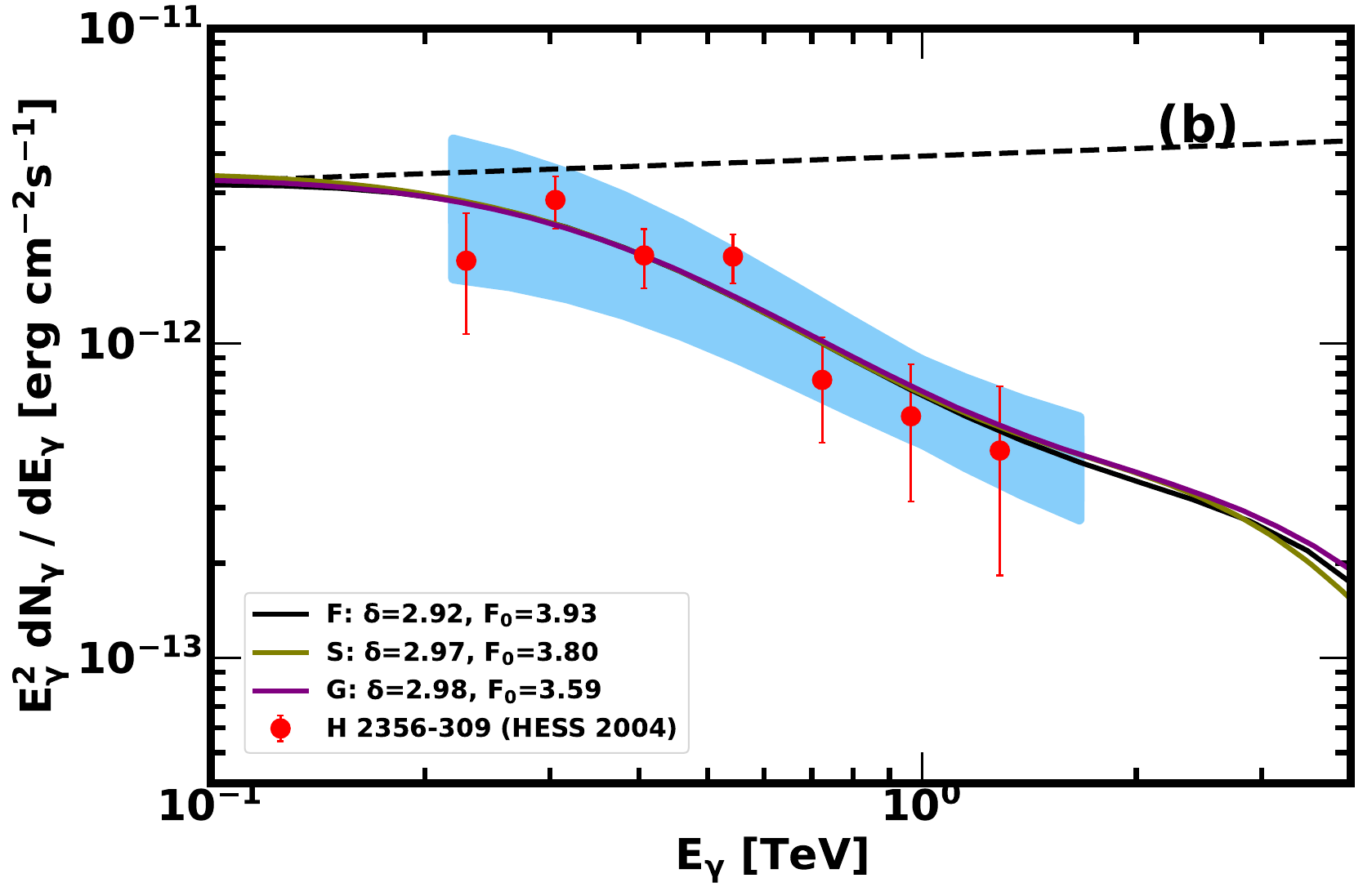}
    \includegraphics[width=0.4\textwidth]{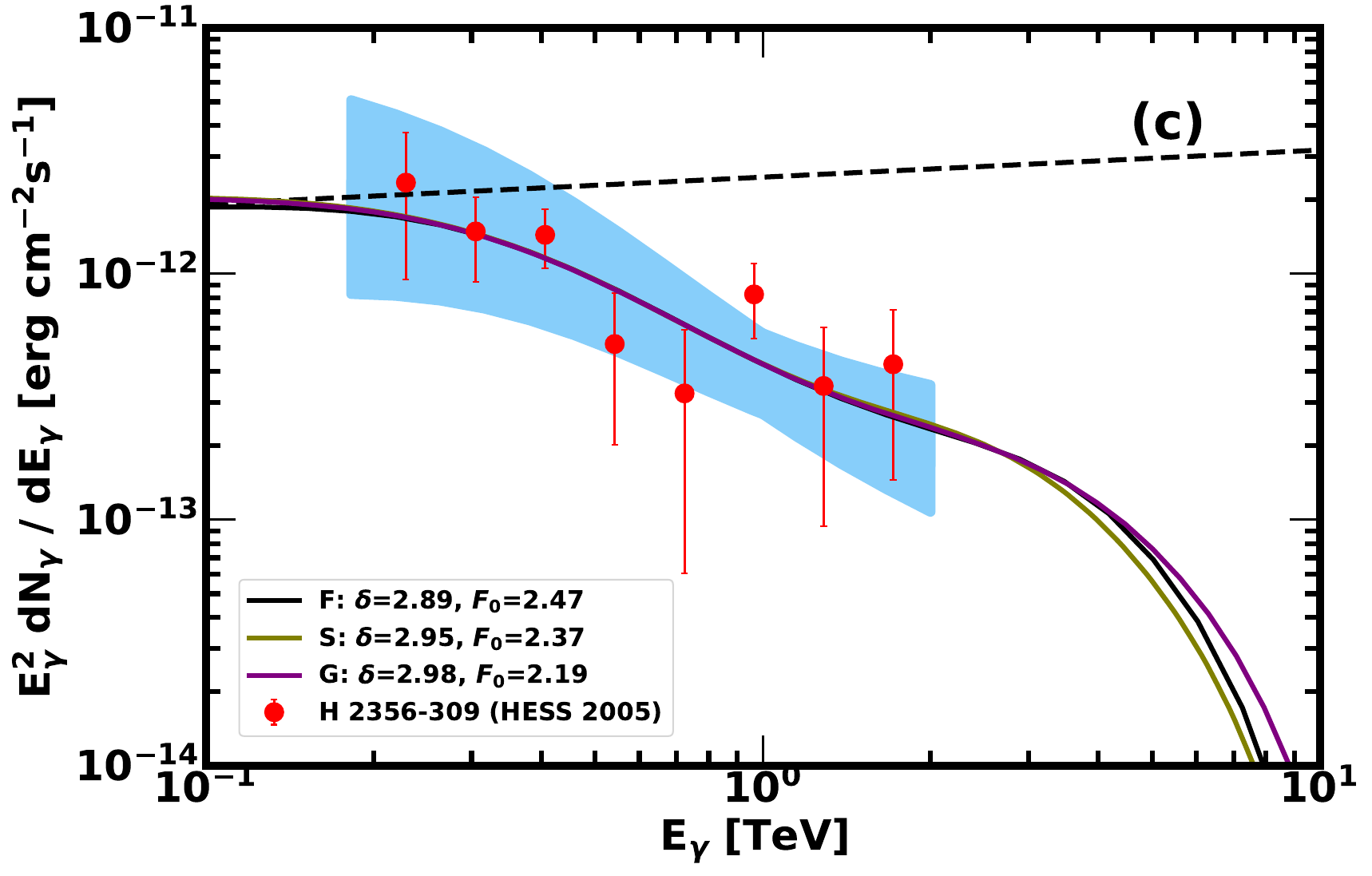}
    \includegraphics[width=0.4\textwidth]{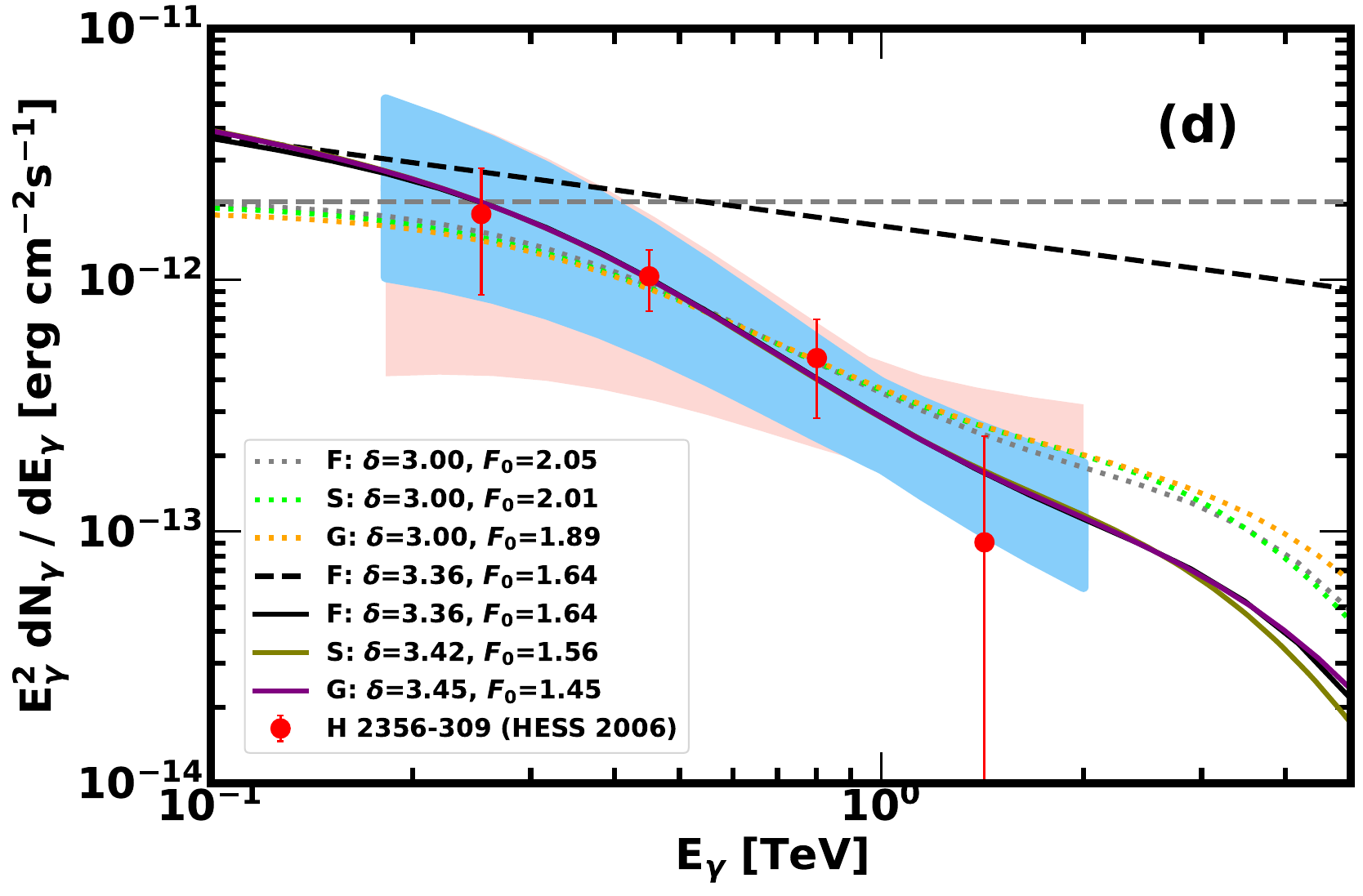}
    \caption{The VHE spectra of H 2356-309 observed by HESS during 2004 to 2006~\citep{2010AA...516A..56H} fitted using the photohadronic model and EBL correction to it. The spectra of 2004 and 2005 are consistent with high emission state. The spectrum in 2006 is peculiar. The best fit to it is obtained for $\delta$ in the range $3.36\le \delta \le 3.45$ (depending on which EBL model is used). For comparison we have also fitted this spectrum with $\delta=3.0$ in Figure (d) and the light green butterfly region corresponds to $1\sigma$ CL to the parameters $\delta$ and $F_0$ for EBL-F. The $\delta > 3.0$ falls faster and fit better than $\delta=3.0$.}  
    \label{fig: H 2356-309-2006}
  \end{center}
\end{figure}

\section{Results and Analysis}

We studied 15 sources and their 22 VHE spectra observed in different epochs. Our selection criteria for the EHBL sources are based on the following conditions: (1) The synchrotron peak frequency of the source $\nu^p_s$ should be always $> 10^{17}$ Hz and contemporaneously observed in VHE gamma-rays. (2) The source should have relatively steady flux over several months to years time scale \citep{2007AA...475L...9A,2014ApJ...782...13A}.  
Normally, the X-rays and the VHE gamma-rays should be observed simultaneous to constrain the synchrotron peak position. However, due to the low fluxes, both in X-rays and in VHE gamma-rays we may not be able to observe simultaneously. Thus, a gap of about a week may not substantially change the spectrum. The above criteria ensure that the sources we are studying have stable spectra in VHE gamma-rays and they belong to permanent EHBL class of sources~\citep{Biteau_2020}.

The Imaging Atmospheric Cerenkov Telescope (IACT) collaborations use reliable EBL models~\citep{Franceschini:2008tp,Dominguez:2010bv,2021MNRAS.507.5144S} to analyze the VHE spectra of extragalactic sources with varying redshifts. Here we use the following EBL models, ~\cite{Franceschini:2008tp} (F), ~\cite{2021MNRAS.507.5144S} (S) and ~\cite{2012MNRAS.422.3189G} (G) for the EBL correction to the VHE spectra and to compare them. For many sources, the optical depth $\tau_{\gamma\gamma}$ for the exact value of the redshift is not given in EBL models S, and G. So, in these models, we have interpolated $z$ to the nearest value and are shown in the respective figures. For all 16 sources under study, we compare the survival probability (SP), 
$exp(-\tau_{\gamma\gamma})$, given in Eq. (\ref{eq:flux}) for each source using the above three EBL models (F, S and G). It is observed that they are consistent with each other, although there can be a slight difference in their values for $E_{\gamma}\, > 1$ TeV. The VHE spectra of the EHBLs are fitted using these three EBL models in the photohadronic scenario and compare them. Finally, for all the VHE spectra, their corresponding 1$\sigma$ confidence intervals for the parameters $\delta$ and $F_0$ are shown in the figures (blue butterfly region) only for the EBL model of ~\cite{Franceschini:2008tp} (EBL-F).

The Tables \ref{Table 1} - \ref{Table 3} contain the results of the statistical analysis to all the flaring epochs and in these tables we have also included the statistical and systematic errors. For the statistical errors we have taken the errors at 68\% according to the $\chi^2$ minimization \citep{2003sppp.conf..250B}. For the systemtic error, we have taken into account the observational errors of each observation for each telescope (e.g. \cite{2012APh....35..435A}). These errors are included in the butterfly regions for each observation.For the multi‑instrument global fits of HESS J1943+213, we include an additional $5\%$ cross‑normalisation systematic error when deriving the redshift constraints.

\begin{table}[!ht]
\centering    
\caption{The EHBL objects studied are listed in the first column by their names, the second column shows their redshift $z$. The third column indicates the observing instrument(s) to the source, and the fourth column specifies the observation time for each VHE flaring period. In the fifth and the sixth columns, the best fit values of $\delta$ and $F_0$ values are given along with their respective confidence intervals at $1\sigma$ for the three EBL models F, S and G respectively. In the last column, the references to each observation are given. $F_0$ is given in units of $10^{-12}\,\mathrm{erg}\, \mathrm{cm^{-2}}\, \mathrm{s^{-1}}$, and the errors correspond to the statistical and the systematic, respectively. In Table \ref{Table 2} same convention is followed.
} 
\label{Table 1}
\setlength{\tabcolsep}{6pt} 
\renewcommand{\arraystretch}{1.3}
    \begin{tabular}{ccccccc}
    \hline
   Object & $z$ & Telescope & Period &  $\delta$ & $F_0$ & Reference \\ \hline
     RBS 1366 & 0.2365 & LAT+VERITAS & 2008-2021\, & (F)2.50$^{+0.15+0.17} _{-0.19-0.16}$ \, & $3.37 ^{+0.60+1.26} _{-0.59-1.15} $\, & \cite{2023arXiv230912230R} \\

     &&&&(S)$2.50 ^{+0.16+0.13} _{-0.19-0.17}$&$3.49^{+0.62+1.15}_{-0.62-1.19}$& \\ 
     &&&&(G)$2.58 ^{+0.15+0.13} _{-0.18-0.17}$&$2.93^{+0.52+0.98}_{-0.50-0.99}$& \\ \hline
        1ES 0229+200 & 0.1396 & HESS & 2005-2006\, & (F)2.50$^{+0.18+0.11} _{-0.20-0.01}$\, & 4.34$^{+0.55+0.87} _{-0.59-0.93}$\, & \cite{2007AA...475L...9A}\\
       &&&&(S)$2.50 ^ {+0.19+0.15} _ {-0.21-0.01} $&$4.12 ^ {+0.52+0.83} _ {-0.55-0.87}$& \\ 
     &&&&(G)$2.55 ^ {+0.21+0.05} _ {-0.16-0.04}$&$3.88 ^ {+0.50+0.79} _ {-0.52-0.81}$& \\ 
         &  & VERITAS & 2009-2013\, & (F)2.69$^{+0.11+0.02} _{-0.12-0.03}$\, & 3.59$^{+0.32+0.62 } _{-0.32-0.65 }$\, & \cite{2014ApJ...782...13A}  \\
         &&&&(S)$2.75 ^ {+0.11+0.01} _ {-0.12-0.03}$&$3.44 ^ {+0.30+0.60} _ {-0.31-0.62}$& \\ 
     &&&&(G)$2.76 ^ {+0.11+0.01} _ {-0.12-0.03}$&$3.26 ^ {+0.29+0.57} _ {-0.29-0.60}$& \\
        ~& ~ & MAGIC & 2013-2017\, & (F)$2.86 ^ {+0.11+0.03 } _ {-0.14-0.03}$\, & 2.04$^{+0.27+0.32 } _{-0.27-0.32 }$ \,& \cite{2023AA...670A.145A} \\ 
        &&&&(S)$2.89 ^{+0.12+0.03} _ {-0.13-0.04}$&$1.97 ^{+0.26+0.30} _{-0.26-0.32}$& \\ 
     &&&&(G)$2.93 ^{+0.12+0.02} _{-0.14-0.03} $&$1.85 ^ {+0.24+0.28} _ {-0.25-0.29}$& \\ \hline
    1ES 1101-232 & 0.186 & HESS & 2004\, & (F)2.88$^{+0.25+0.32} _{-0.33-0.08}$\, & 6.49$^{+1.54+1.63 } _{-1.55+1.85}$ \, & \cite{2007AA...470..475A}  \\ 

     &  &  &  & (S)$ 2.96 ^ {+0.24+0.27} _ {-0.33-0.05}$ & $6.19 ^ {+1.46+1.42} _ {-1.49-1.73}$   &   \\
     &  &  &  & (G)$2.99 ^ {+0.23+0.17} _ {-0.34-0.06}$ & $5.64 ^ {+1.43+1.14} _ {-1.29-1.58}$  &   \\

       ~ & ~ & HESS & 2005\, &(F)2.70$^{+0.14+0.27}_{-0.17-0.06}$\, & 6.56$^{+0.81+1.69} _{-0.79-1.70}$\, & \cite{2007AA...470..475A}  \\
     &  &  &  & (S)$2.79 ^ {+0.14+0.22} _ {-0.17-0.04}$ &   $6.19 ^ {+0.75+1.42} _ {-0.76-1.56}$&   \\
     &  &  &  & (G)$2.81 ^ {+0.14+0.13} _ {-0.17-0.05}$ &$5.69 ^ {+0.69+1.15} _ {-0.70-1.44}$   &   \\

      ~ & ~ & HESS & 2004-2005\,  & (F)2.68$^{+0.14+0.19}_{-0.17-0.04}$\, & 6.25$^{+0.75+1.46} _{-0.75-1.57}$\, & \cite{2007AA...470..475A} \\
     &  &  &  & (S)$2.76 ^ {+0.14+0.15} _ {-0.17-0.03} $ & $5.91 ^ {+0.72+1.25} _ {-0.71-1.45}$  &   \\
     &  &  &  & (G)$2.79 ^ {+0.14+0.07} _ {-0.17-0.04}$ & $5.42 ^ {+0.65+1.09} _ {-0.65-1.35}$  &   \\

      \hline
     TXS 0210+515  & 0.049 & MAGIC & 2015-2017\, & (F)$2.65 ^ {+0.32+0.01} _ {-0.24-0.01}$& $0.81 ^ {+0.19+0.09} _ {-0.19-0.09}$ \,& \cite{2020ApJS..247...16A} \\
     &  &  &  & (S)$2.65 ^ {+0.32+0.01} _ {-0.24-0.02}$ & $0.79 ^ {+0.19+0.09} _ {-0.19-0.09}$  &   \\
     &  &  &  & (G)$2.67 ^ {+0.32+0.01} _ {-0.24-0.01}$ & $0.79 ^ {+0.19+0.09} _ {-0.18-0.09}$  &   \\

     \hline
   RGB J2042+244 & 0.104 & MAGIC & 2015 \,& (F)2.75$^{+0.19+0.04} _{-0.27-0.05}$\, & 1.21$^{+0.31+0.15} _{-0.29-0.17}$\, & \cite{2020ApJS..247...16A}  \\
     &  &  &  & (S)$ 2.78 ^ {+0.19+0.04} _ {-0.27-0.04}$ &$1.17 ^ {+0.29+0.16} _ {-0.29-0.16}$   &   \\
     &  &  &  & (G)$2.80 ^ {+0.19+0.03} _ {-0.28-0.05}$ & $1.13 ^ {+0.29+0.15} _ {-0.28-0.16}$  &   \\

   \hline
   RGB J0710+591 & 0.125 & VERITAS & 2008-2009\, & (F)2.87$^{+0.24+0.01}_{-0.24-0.01}$\, & 4.46$^{+0.81+1.00}_{-0.79-0.99}$\, & \cite{2010ApJ...715L..49A}  \\
     &  &  &  & (S)$2.92 ^ {+0.24+0.01} _ {-0.24-0.01}$ & $4.30 ^ {+0.76+0.96} _ {-0.77-0.96}$  &   \\
     &  &  &  & (G)$2.94 ^ {+0.24+0.01} _ {-0.24-0.01}$ & $4.10 ^ {+0.73+0.92} _ {-0.74-0.92}$  &  \\
   \hline
   RX J1136.5+6737& 0.1342 & MAGIC & 2014\, & (F)2.94$^{+0.27+0.01} _{-0.40-0.01}$\, & 1.87$^{+0.53+0.29} _{-0.55-0.30}$\, & \cite{2015ICRC...34..698H}  \\
    &  &  &  & (S)$3.00 ^ {+0.27+0.01} _ {-0.39-0.01}$ & $1.77 ^ {+0.51+0.28} _ {-0.52-0.28}$  &   \\
     &  &  &  & (G)$3.00 ^ {+0.27+0.01} _ {-0.39-0.02} $ &$1.71 ^ {+0.48+0.26} _ {-0.50-0.29}$   &   \\
\hline   
1ES 0347-121 & 0.188 & HESS & 2006\,  & (F)2.78$^{+0.11+0.02}_{-0.15-0.06}$\, & 5.36$^{+0.58+0.02}_{-0.62-1.12}$\, & \cite{2007AA...473L..25A}  \\
        &  &  &  & (S)$2.85 ^ {+0.11+0.01} _ {-0.14-0.05}$ & $5.08 ^ {+0.56+0.96} _ {-0.58-1.03}$  &   \\
     &  &  &  & (G)$2.89 ^ {+0.11+0.02} _ {-0.15-0.05}$ &$4.63 ^ {+0.51+0.89} _ {-0.53-0.95}$   &   \\
\hline    
    \end{tabular}
    \label{Table1}  
\end{table}

\begin{table}[!ht]
\centering    
\caption{The redshift of HESS J1943+213 is unknown. Including the EBL models F, S and G to the photohadronic model we have fitted the VHE spectra of it and best fit value of each EBL model is shown here along with the $1\sigma$ CL to their respective parameters $(\delta, F_0)$. We follow the same convention as Table \ref{Table 1} here.
 }
\label{Table 2}
\setlength{\tabcolsep}{6pt} 
\renewcommand{\arraystretch}{1.3}
    \begin{tabular}{ccccccc}
    \hline
   Object & $z$ & Telescope & Period &  $\delta$ & $F_0$ & Reference \\ \hline

    HESS J1943+213 & 0.133 & HESS & 2009\, & (F)$3.00 ^ {+0.35+0.08} _ {-0.19-0.02}$\, & $3.40 ^ {+0.47+0.87} _ {-0.44-0.79}$ & \cite{2011AA...529A..49H} \\
    & 0.141  &  &  & (S)$3.00 ^ {+0.36+0.07} _ {-0.20-0.03}$ & $3.58 ^ {+0.48+0.90} _ {-0.48-0.87}$  &   \\
     & 0.145 &  &  & (G){$3.00 ^ {+0.35+0.03} _ {-0.19-0.03}$} & $3.53 ^ {+0.51+0.90} _ {-0.45-0.86}$  &   \\

     ~ & 0.133 & VERITAS & 2014-2015\, & 
     (F)$2.99^{+0.07+0.03}_{-0.08-0.07}$ & 4.32$^{+0.28+1.06}_{-0.33-1.12}$\, & \cite{2018ApJ...862...41A}  \\ 
    & 0.141  &  &  & (S)$2.99 ^ {+0.06+0.03} _ {-0.08-0.06}$ & $4.54 ^ {+0.31+1.10} _ {-0.30-1.17}$ &   \\
     & 0.145 & &  & (G)$2.99 ^ {+0.06+0.04} _ {-0.09-0.06}$  & $4.46 ^ {+0.29+1.09} _ {-0.30-1.15}$  &   \\

        \hline
    1ES 1741+196 & 0.084 & MAGIC & 2010-2011\, & (F)2.89$^{+0.23+0.01}_{-0.33-0.01}$\, & 0.98$^{+0.26+0.13}_{-0.26-0.12}$ \,& \cite{2017MNRAS.468.1534A}  \\
    &  &  &  & (S)$2.93 ^ {+0.23+0.01} _ {-0.34-0.02}$ & $0.95 ^ {+0.25-0.11} _ {-0.26-0.12}$   &   \\
     &  &  &  & (G)$2.93 ^ {+0.23+0.01} _ {-0.33-0.01}$  & $0.93 ^ {+0.25+0.11} _ {-0.25-0.11}$   &   \\

      & ~ & VERITAS & 2009-2014\, & (F)3.00$^{+0.17+0.31}_{-0.18-0.45}$\, & 1.54$^{+0.30+0.47}_{-0.31-0.78}$\, & \cite{2016MNRAS.459.2550A}  \\
    &  &  &  & (S)$3.00 ^ {+0.17+0.32} _ {-0.18-0.45}$ & $1.55 ^ {+0.30+0.47} _ {-0.31-0.78}$   &   \\
     &  &  &  & (G)$3.00 ^ {+0.17+0.34} _ {-0.18-0.38}  $ & $1.53 ^ {+0.29+0.49} _ {-0.30-0.70}$   &   \\

       \hline       
    1ES 2037+521 & 0.053 & MAGIC & 2016 & (F)3.00$^{+0.16+0.02}_{-0.22-0.01}$ & 2.07$^{+0.45+0.25}_{-0.43-0.25}$ & \cite{2020ApJS..247...16A}  \\ 
    &  &  &  & (S)$3.00 ^ {+0.17+0.01} _ {-0.21-0.03}$  & $2.06 ^ {+0.44+0.24} _ {-0.44-0.27}$   &   \\
     &  &  &  & (G)$3.00 ^ {+0.17+0.01} _ {-0.21-0.03}$ & $2.05 ^ {+0.43+0.24} _ {-0.43-0.27}$   &   \\
     \hline  
    1ES 1312-423 & 0.105 & HESS & 2004-2010\, & (F)3.00$^{+0.34+0.01} _{-0.32-0.01}$\, & 1.07$^{+0.26+0.24} _{-0.26-0.24}$ \,& \cite{2013MNRAS.434.1889H}  \\ 
     &  &  &  & (S)$3.00 ^ {+0.38+0.02} _ {-0.30-0.04}$ & $1.03 ^ {+0.26+0.24} _ {-0.25-0.24}$   &   \\
     &  &  &  & (G)$3.00 ^ {+0.40+0.05} _ {-0.27-0.06}$ &$1.00 ^ {+0.24+0.23} _ {-0.25-0.23}$   &   \\ \hline
    1RXS J195815.6-301119 & 0.119 & HESS &  2018\, & (F)3.00$^{+0.18+0.07}_{-0.14-0.08}$\, & 2.90$^{+0.42+0.60}_{-0.42-0.59}$\, & \cite{2022icrc.confE.823B}  \\ 
         &  &  &  & (S)$3.00 ^ {+0.21+0.12} _ {-0.13-0.19}$ & $2.87 ^ {+0.42+0.58} _ {-0.41-0.89}$   &   \\
     &  &  &  & (G)$3.00 ^ {+0.22+0.14} _ {-0.12-0.20}$ & $2.78 ^ {+0.41+0.56} _ {-0.39-0.87}$   &   \\ \hline
    MRC 0910-208 & 0.19802 & LAT+HESS & 2008-2020\, & (F)$3.00 ^ {+0.21+0.48} _ {-0.10-0.67}$   & $1.56 ^ {+0.24+0.51} _ {-0.26-1.00}$  & \cite{2022icrc.confE.823B} \\
     &  &  &  & (S)$3.00 ^{ +0.22+0.53} _ {-0.10-0.79}$ & $1.59 ^ {+0.24+0.56} _ {-0.27-1.11}$   &   \\
     &  &  &  & (G)$3.00 ^ {+0.25+0.59} _ {-0.10-0.79} $ & $1.46 ^ {+0.24+0.54} _ {-0.13-1.00}$   &   \\

    &  & HESS & 2018\, & (F)3.00$^{+0.16+0.15}_{-0.15-0.25}$\, & 2.81$^{+0.49+0.56}_{-0.45-1.18}$\, & \cite{2022icrc.confE.823B}  \\ 
    &  &  &  & (S)$3.00 ^ {+0.15+0.01} _ {-0.15-0.29} $ & $2.89 ^ {+0.50+0.78} _ {-0.46-1.27}$   &   \\
     &  &  &  & (G)$3.00 ^ {+0.15+0.04} _ {-0.15-0.33}$ &$2.72 ^ {+0.45+0.62} _ {-0.45-1.27}$   &   \\ \hline
    1ES 0033+595 & 0.467 & MAGIC & 2009\, & (F)3.00$^{+0.08+0.16} _{-0.08-0.34}$\, & 6.68$^{+1.08+0.80} _{-0.69-3.75}$\, & \cite{2015MNRAS.446..217A}  \\ 
     &  &  &  & (S)$3.00 ^ {+0.08+0.08} _ {-0.08-0.74}$ & $7.81 ^ {+0.98+1.47} _ {-0.98-6.20}$   &   \\
     &  &  &  & (G)$3.00 ^ {+0.09+0.29} _ {-0.08-0.57} $ & $6.64 ^ {+0.84+1.30} _ {-0.85-4.69}$   &   \\
\hline

    \end{tabular}
    \label{Table2}  
\end{table}

\begin{table}[!ht]
\centering
\caption{Redshift constrain of HESS J1943+213 including different EBL models to the photohadronic model are given. In the first column, the EBL models are listed and in the second column, the best redshift derived from the respective EBL model is given. From third to fifth column the $1\sigma$, $2\sigma$, and $3\sigma$ CL are given for each EBL model, where the errors include statistical and systematic errors applied at quadratures.}
    \label{Table 3}
    \setlength{\tabcolsep}{6pt} 
\renewcommand{\arraystretch}{1.3}
\centering
    \begin{tabular}{ccccc}
    \hline
        EBL & z & 1 $\sigma$ & 2$ \sigma$ & 3 $\sigma$ \\ \hline
        Franceschini (F) & 0.133 & 0.116$\leq$ z $\leq$ 0.152 & 0.107$\leq$ z $\leq$0.164 & 0.102$\leq$ z $\leq$0.170 \\ \hline
        Saldana (S) & 0.141 & 0.123$\leq$ z $\leq$0.160 & 0.113$\leq$ z $\leq$0.173 & 0.109$\leq$ z $\leq$0.179 \\ \hline
        Gilmore (G) & 0.145 & 0.126$\leq$ z $\leq$0.165 & 0.116$\leq$ z $\leq$0.178 & 0.111$\leq$ z $\leq$0.185\\ \hline
    \end{tabular}
    \label{Table3} 
\end{table}

\begin{table}[!ht]
\centering    
\caption{H 2356-309 with redshift $z=0.165$ was observed by HESS in three different periods 2004, 2005 and 2006~\citep{2010AA...516A..56H}. The spectra are fitted using the photohadronic model where three EBL models are included to account for the the EBL correction to the spectra. In the table the first column indicates the observing instrument(s) to the source, the second column specifies the observation period for each VHE flaring state. In the third and the fourth columns, the best fit values of $\delta$ and $F_0$ values are given along with their respective confidence intervals at $1\sigma$. $F_0$ is given in units of $10^{-12}\,\mathrm{erg}\, \mathrm{cm^{-2}}\, \mathrm{s^{-1}}$. }
\label{Table 4}
\setlength{\tabcolsep}{6pt} 
\renewcommand{\arraystretch}{1.5}
\centering
    \begin{tabular}{cccc}
    \hline
Telescope & Period &  $\delta$ & $F_0$ \\ 
\hline
 HESS & 2004 & (F) $2.92 ^ {+0.11+0.06} _ {-0.11-0.09}$ & $3.93 ^ {+0.42+1.06} _ {-0.38-1.11}$\\
&  & (S) $2.97 ^ {+0.10+0.04} _ {-0.11-0.08} $ & $3.80 ^ {+0.40+0.99} _ {-0.37-1.06}$ \\
&  &(G) $2.98 ^ {+0.11+0.02} _ {-0.11-0.11}$ & $3.59 ^ {+0.35+0.89} _ {-0.37-1.05}$\\
\hline
        
HESS & 2005 & (F) $2.89 ^ {+0.16+0.08} _ {-0.27-0.35}$ & $2.47 ^ {+0.40+0.72} _ {-0.58-0.72}$\\
&  &(S) $2.95 ^ {+0.22+0.03} _ {-0.22-0.07}$ &$2.37 ^ {+0.40+0.61} _ {-0.40-0.66}$ \\
&  & (G) $2.98 ^ {+0.19+0.03} _ {-0.25-0.05}$ & $2.19 ^ {+0.37+0.58} _ {-0.37-0.48}$\\
\hline

HESS & 2006 & (F) $3.36 ^ {+0.21+0.03} _ {-0.28-0.07}$ & $1.64 ^ {+0.34+0.51} _ {-0.35-0.51}$ \\
&  & (S) $3.42 ^ {+0.21+0.02} _ {-0.27-0.05}$ & $1.56 ^ {+0.33+0.47} _ {-0.33-0.48}$ \\
&  & (G) $3.45 ^ {+0.21+0.14} _ {-0.27-0.49}$ & $1.45 ^ {+0.30+0.62} _ {-0.31-0.62}$ \\
      \hline
    \end{tabular}
    \label{Table4}  
\end{table}

\subsection{RBS 1366}
RBS 1366 is a new source at a redshift of $z=0.2365$ \citep{2023arXiv230912230R} observed in VHE gamma-rays by both {\it Fermi}-Large Area Telescope (LAT) and VERITAS telescopes between 2008 and 2021 blue{(for $\sim 56.8$ h)}~\citep{2023arXiv230912230R}. 
The synchrotron peak of this object is always found to be above 
$1$ keV~\citep{2019MNRAS.486.1741F,2023arXiv230912230R,2025arXiv250311543B}. The light curves of LAT and VERITAS were consistent with steady emission and the average VHE spectrum observed during 2008-2021 is fitted using the photohadronic model by including the EBL correction from F, S and G EBL models and the best fit is obtained for $\delta = 2.50^{+0.15} _{-0.19}$ (F), $\delta = 2.50^{+0.16} _{-0.19}$ (S), and $\delta = 2.58^{+0.15} _{-0.18}$ (G) respectively. These values of $\delta$ correspond to 
very high-emission state ($2.5\leq \delta \leq 2.6$) with a 
hard intrinsic spectrum which behaves as $F_{in} \propto  E_{\gamma}^{\lambda} $, where  $0.42 \le \lambda \le 0.5$ (depending on the EBL model).
For comparison, we have plotted SP for $z=0.24$ using the EBL models (F, S and G) in Figure \ref{fig: RBS 1366} (a) and they are very similar. In Figure \ref{fig: RBS 1366} (b), the VHE spectrum is fitted using the F, S and G EBL models and all are practically the same. The 1$\sigma$ confidence intervals for 
the parameters $\delta$ and $F_0$ for the EBL-F are shown in Figure \ref{fig: RBS 1366} (b).
Unfortunately, there are no other observations of this EHBL. 

\subsection{1ES 0229+200}

The BL Lac source 1ES 0229+200 is at a redshift of $z=0.1396$ \citep{2005ApJ...631..762W} and since 2002 its synchrotron peak is observed to have $\nu^p_s > 1$ keV \citep{2002babs.conf...21C}. It is observed in multiwavelength and at least three VHE flaring epochs 2005-2006, 2009-2013 and 2013-2017 have been observed by HESS, VERITAS and MAGIC collaborations respectively \citep{2007AA...475L...9A,2014ApJ...782...13A,2023AA...670A.145A}. Using the EBL models F, S and G, we have plotted the SP for VHE photons coming from 1ES 0229+200 in Figure \ref{fig: 1ES 0229+200} (a). All these models give similar results. 
Its VHE spectrum observed during 2005-2006 by HESS~\citep{2007AA...475L...9A} is fitted using the photohadronic model by including the EBL corrections from the models F, S and G in Figure \ref{fig: 1ES 0229+200} (b). The spectrum is fitted very well for the spectral index $\delta$ in the range $2.5\le \delta\le 2.6$. 
This shows that the intrinsic spectrum is hard and grows rapidly, like the spectrum of RBS 1366. For $E_{\gamma} > 5$ TeV, the fit with EBL-G falls slightly slower then the other two. The blue butterfly region corresponding to 1$\sigma$ confidence intervals for 
the parameters $\delta$ and $F_0$ for the EBL-F is shown in Figure \ref{fig: 1ES 0229+200} (b).

The broadband SED of 1ES 0229+200 during the VHE flaring of 2005-2006 is given in Fig. 4 of ~\cite{2007AA...475L...9A} and the $\epsilon_{\gamma}$ in the low-energy SSC tail is much above $10^{20}$~Hz. Taking $\epsilon_{\gamma}\sim 5\times 10^{20}$ Hz ($\sim 2.1$ MeV) and the maximum observed photon energy $E_{\gamma}=10.75$~TeV, the bulk Lorentz factor from Eq. (\ref{eq:Epegmaa}) is $\Gamma\simeq 30$ or larger. This is much larger than the bulk Lorentz factor for a HBL \citep{2019MNRAS.490.2284M,2023A&A...670A..49M}.

The VHE flaring of 1ES 0229+200 observed during 2009-2013 by VERITAS \citep{2014ApJ...782...13A} and by MAGIC during 2013-2017~\citep{2023AA...670A.145A} are 
shown in Figure \ref{fig: 1ES 0229+200} (c) and \ref{fig: 1ES 0229+200} (d) respectively. These spectra are fitted very well using the photohadronic model by including the EBL corection from models F, S and G. For all the EBL models, the $\delta$ value is in the range $2.6 < \delta < 3.0$ which corresponds to high-emission state.

\subsection{1ES 1101-232}

1ES 1101-232 has a redshift of $z=0.186$~\citep{1994ApJS...93..125F}. It was observed by HESS for four nights in April and six nights in June in the year 2004 \citep{2007AA...470..475A}. Again in March 2005, HESS observed for 31.6 h, a total live time of 43 h during 2004 and 2005~\citep{2007AA...470..475A}. It was observed that the VHE gamma-ray flux observed during 2004 and 2005 remained constant throughout these observation periods. Also, no hint for spectral variability was found. The SP of 1ES 1101-232 in the energy range $0.1\, \mathrm{TeV} < E_{\gamma} < 10\, \mathrm{TeV} $ is shown in Figure \ref{fig: 1ES 1101-232} (a) for the three EBL models discussed above. These are compartible with each other. Again, these EBL models are included in the photohadronic model to fit the VHE spectra of June 2004, March 2005 and the combined data of 2004 and 2005, which are shown in Figures \ref{fig: 1ES 1101-232} (b), \ref{fig: 1ES 1101-232} (c) and \ref{fig: 1ES 1101-232} (d) respectively. Best fits to these spectra are obtained for $2.6 < \delta < 3.0$ for all the EBL models. This shows that the flaring epochs are in high-emission states and the intrinsic flux during the above observation periods are hard.

\subsection{TXS 0210+515}

This EHBL is located at a redshift of z = 0.049 ~\citep{2011NewA...16..503M,2017A&A...598A..17C}.  The MAGIC telescopes observed this source during 2015-2017 and its intrinsic spectrum was found to be hard~\citep{2020ApJS..247...16A}. We fitted this spectrum using the photohadronic model. For EBL correction to the observed spectrum we used the EBL models F, S and G (comparison of the SP is shown in Figure \ref{fig: TXS 0210+515} (a)). The VHE spectrum is in high-emission state (hard intrinsic spectrum) and all the EBL models fit very well to the observed spectrum as shown in Figure \ref{fig: TXS 0210+515} (b). In the observed energy range, all the EBL models coincide with each other. We have also shown the $1\sigma$ confidence intervals for the parameters $\delta$ and $F_0$ for EBL-F in Figure \ref{fig: TXS 0210+515} (b).

\subsection{RGB J2042+244}

In the aim of increasing the population of EHBL sources, the MAGIC telescopes observed 10 potential sources from 2010 to 2017 for 265 h\citep{2020ApJS..247...16A}. For the SED characterization of these sources, simultaneous $\it{Swift}$-XRT observations were carried out \citep{2020ApJS..247...16A}. During this observation period, a hint of the VHE gamma-ray signal was detected from RGB J2042+244 which is at a redshift of $z=0.104$ \citep{2013ApJ...764..135S}. This source was found to have hard intrinsic spectrum and its synchrotron peak frequency was around $\nu^p_s\sim 10^{17.5}$ Hz, necessary conditions for the source to be EHBL \citep{2020ApJS..247...16A}. All the EBL models have almost the same SP (Figure \ref{fig: RGB J2042+244} (a)). The observed spectrum is fitted well with all the EBL models, Figure \ref{fig: RGB J2042+244} (b) and it is in high-emission state.

\subsection{RGB J0710+591}
The EHBL RGB J0710+591 was first detected in VHE gamma-rays by VERITAS during 2008-2009 observation~\citep{2010ApJ...715L..49A}. This was complemented by contemporaneous observation by various telescopes in multiwavelength. No variability in the flux was observed during this period \citep{2010ApJ...715L..49A}. The EHBL RGB J0710+591 is situated at a redshift of $z=0.125$~\citep{1991ApJ...378...77G}. The EBL models F, S and G have the same SP for RGB J0710+591 as shown in Figure \ref{fig: RGB J0710+591} (a). It is also observed that the VHE gamma-ray spectrum is fitted very well by the photohadronic model including the EBL correction from the three EBL models discussed (Figure \ref{fig: RGB J0710+591} (b)).

\subsection{RX J1136.5+6737}

The BL Lac object RX J1136.5+6737 was observed by MAGIC telescopes for about 35 night between January 19 and May 28, 2014 and discovered VHE gamma-rays with $>5\sigma$ significance \citep{2015ICRC...34..698H}. Simultaneous multiwavelength observation in lower wavelenghts were also carried out. No significant variability in the VHE flux was observed. The $\it{Swift}$ observation shows that its synchrotron peak is near $10^{17.90\pm 0.29}$ Hz. This source is at a redshift of $z=0.1342$ \citep{2014ATel.6062....1M}. The SP curves for the EBL models F, S and G are practically the same (Figure \ref{fig: RX J1136.5+6737} (a)). The inclusion of EBL-F to the photohadronic model predict the VHE event in the high-emission state. However, inclusion of EBL-S and EBL-G to the photohadronic model predict the VHE flaring state to be in low emission state and these are shown in  Figure \ref{fig: RX J1136.5+6737} (b) for comparison.

\subsection{1ES 0347-121}

1ES 0347-121 is classified as a BL Lac object and it is located at a redshift of $z=0.188$~\citep{2005ApJ...631..762W}. The HESS telescopes observed 1ES 0347-121 a total of 25.4 h live time between August and
December 2006 in the energy range $0.25\, \mathrm{TeV} < E_{\gamma} < 3\,\mathrm{TeV}$~\citep{2007AA...473L..25A}. Contemporaneous X-ray and UV/optical observations from the \textit{Swift} satellite were also made. No significant flux 
variability was detected on time-scales of days or months \citep{2007AA...473L..25A}. As the source is at a redshift of 0.188, the observed VHE spectrum is expected to be strongly attenuated by the EBL. The survival probability of the VHE photons from this source are shown in Figure \ref{fig: 1ES 0347-121} (a) for different EBL models  discussed above and all of them give similr results. Including these EBL models in the photohadronic model, we have fitted the VHE spectrum of the source 1ES 0347-121 very well which are shown in Figure \ref{fig: 1ES 0347-121} (b), all these EBL models predict the spectrum in the high-emission state and the intrinsic spectrum is hard.

\subsection{HESS J1943+213}
The redshift of the EHBL HESS J1943+213 is unknown and it is observed in VHE by HESS in 2009~\citep{2011AA...529A..49H}  and by VERITAS during 2014-2015~\citep{2018ApJ...862...41A}. Using the photohadronic model, the VHE spectra of HESS and VERITAS are fitted simultaneously through a global $\chi^2$ minimization method blue{(i.e., minimising the sum of $\chi^2$ from both data sets with a common redshift parameter)}  using the EBL models F, G and S. For a given EBL model and for each assumed redshift $z$, the parameters ($F_0$, $\delta$) are optimized independently for every spectrum, and the process is repeated by varying $z$ to obtain the global best-fit value. The confidence level intervals of the redshift are subsequently derived at $1\sigma$, $2\sigma$ and $3\sigma$ for the three EBL models. In Table \ref{Table2}, the best fit value of $z$ for each EBL model and their corresponding ($F_0, \delta$) values with $1\sigma$ confidence interval are shown. However, in Table \ref{Table3} the best fit value of $z$ and $1\sigma$, $2\sigma$ and $3\sigma$ confidence intervals to the redshift for the EBL models  are shown. By taking the best fit values of $z$ from each EBL model we have plotted the SP for HESS J1943+213 as shown in Figure \ref{fig: HESS J1943+213} (a). Similarly, by taking the same best fit values of $z$ for the EBL models we have fitted the observed spectra of HESS and VERITAS and shown them in Figure \ref{fig: HESS J1943+213} (b) and Figure \ref{fig: HESS J1943+213} (c) respectively. The HESS 2009 data correspond to a low-emisson state with a flat intrinsic spectrum whereas, the VERITAS 2014-15 spectrum is in high-emission state with $\delta=2.99$. The redshift calculated using the three EBL models (F, S and G) as shown in Table \ref{Table3} give tighter constraint to $z$ compared to the existing results. For example, by using the infrared counterpart \cite{2014A&A...571A..41P} get $0.03< z < 0.45$. Additionally, by using \textit{Fermi} data,  
\cite{2018ApJ...862...41A} obtain a best-fit value of $z = 0.20$, with an upper limit of $z < 0.23$.

\subsection{1ES 1741+196}

The MAGIC telescopes observed the BL Lac 1ES 1741+196 from 2010 April 10 to 2011 May 26 for 53 nights a total of about 57 h and during this reported the first observation of VHE gamma-ray from this source. The $\it{Swift}$ also observed this source from 2010 July 30 to 2011 January 21~\citep{2017MNRAS.468.1534A}. This object has a redshift of $z=0.084$ \citep{1999A&A...348..113H}.

The VERITAS telescopes also observed 1ES 1741+196 between 2009 April 19 and 2014 June 26 for about six years and the time averaged spectrum above 180 GeV can be fitted with a power low \citep{2016MNRAS.459.2550A}. The multiwavelength observation of the source suggests it an EHBL. Also, the observation of 1ES 1741+196 is consistent with no varibility in VHE during the $\sim 6$ yr~\citep{2016MNRAS.459.2550A}.

We have fitted the VHE spectra of both MAGIC and VERITAS observations which are shown in Figure {\ref{fig: 1ES 1741+196}}. In Figure {\ref{fig: 1ES 1741+196}} (a) we have compared the SP of VHE photons for $z=0.084$. It shows that all the models have similar prediction. In Figure {\ref{fig: 1ES 1741+196}} (b) and Figure {\ref{fig: 1ES 1741+196}} (c), we have fitted the VHE spectra observed by MAGIC and VERITAS using the different EBL models respectively. The photohadronic model explains well the spectra. However, the MAGIC spectrum of 2010-2011 is in high-emission state, wheras, the VERITAS $\sim 6$ yr average spectrum is consistent with low emission state, corresponding to $\delta=3.0$. 

\subsection{1ES 2037+521}
1ES 2037+521 is cataloged as BL Lac object at a redshift of $z=0.053$ \citep{2003A&A...400...95N}. It was observed in VHE by MAGIC telescopes during 2026 September 26 to 30 for 28.1 h. Also, contemporaneous observation was made by $\it{Swift}$ indicating that the synchrotron peak is above 4 keV~\citep{2020ApJS..247...16A}. Using the photohadronic model and three EBL models for EBL correction, we analyzed the VHE spectrum of 1ES 2037+521 observed by MAGIC telescopes. It is observed that all the EBL models are consistent with each other and the spectrum is well fitted by a spectral index $\delta=3.0$
 (Figures \ref{fig: 1ES 2037+521} (a), (b)). This shows that the emission is in low emission state and the intrinsic spectrum is flat, does not depend on photon energy. Also, the intrinsic spectrum is soft, which shows that VHE spectrum of EHBL is not necessarily always hard. 

\subsection{1ES 1312-423}

The BL Lac object 1ES 1312-423 is at a redshift of $z=0.105$~\citep{2000AJ....120.1626R}, discovered in the field-of-view of Centaurus A in VHE  by HESS telescopes with 6.8 $\sigma$ significance in 168 h of observation between 2004 and 2010~\citep{2013MNRAS.434.1889H,2010tsra.confE.167B}. It is one of the faintest extragalactic sources ever detected in the VHE gamma-rays. No significant variability was observed during this long observation period by HESS. Its observed VHE spectrum is fitted including the EBL corrections from the EBL models F, S and G (as given in Figure \ref{fig:1ES 1312-423} (a)) to the photohadronic model. All the EBL models predict the VHE spectrum to be in low-emission state which has $\delta=3.0$ as shown in Figure \ref{fig:1ES 1312-423} (b). Like the spectrum of 1ES 2035+521, this source also is in low-emission state with its intrinsic spectrum flat and soft. 

\subsection{1RXS J195815.6-301119 and MRC 0910-208}

These two sources, 1RXS J195815.6-301119 ($z=0.119$) and MRC 0910-208 ($z=0.19802$) \citep{2009MNRAS.399..683J} were selected for observation by HESS due to their synchrotron peak above $10^{17}$ Hz and were observed in 2018 May (1RXS J195815.6-301119) and September (MRC 0910-208) for 7.3 h and for 7.3 h and 17.2 h respectively \citep{2022icrc.confE.823B}. Also, 
to extract average source spectra above 100 MeV, a Fermi-LAT analysis was performed over a time range of 11.5 years (August 8, 2008 - January 4, 2020). These objects were also observed in X-ray \citep{2022icrc.confE.823B}.

We analyzed their spectra by taking EBL-F, S and G models. The SP and the spectral fit of 1RXS J195815.6-301119 are shown in Figure \ref{fig: 1RXSJ 195815.6-301119} (a) and (b) respectively. The spectrum is in low-emission state and the intrinsic spectrum is soft.
Similarly, the EBL contribution by different models for MRC 0910-208 are compared in Figure \ref{fig: MRC 0910-208} (a) and the fit to the HESS and HESS and LAT data are fitted in Figure \ref{fig: MRC 0910-208} (b) and Figure \ref{fig: MRC 0910-208} (c) respectively. These spectra are also consistent with low-emission state $\delta=3.0$ and the intrinsic spectrum is flat.

\subsection{1ES 0033+595}
The BL Lac object 1ES 0033+595 is near the galactic plane and detected first as a hard X-ray source by the Einstein Slew Survey in 1992 \citep{1992ApJS...80..257E}. It has a redshift of $z=0.467$ \citep{2017ApJ...837..144P}. It was observed by the X-ray satellite BeppoSAX in December 1999 \citep{2001A&A...371..512C} and motivated by this observations, MAGIC observed this source in 2006 and in 2008
for about 5 h. However, only a flux upper limit was obtained \citep{2011ApJ...729..115A}. It is classified as an extreme high-frequency peaked
(HBL) object with synchrotron emission peaking near
$10^{19}$ Hz \citep{2006A&A...445..441N}. New observations during August 17 to October 14, 2009, for a total time of 23.5 h in the
commissioning phase of the MAGIC stereoscopic system, VHE gamma-rays were observed \citep{2011ATel.3719....1M,2012AIPC.1505..494U}. The source showed only marginally significant variability throughout the MAGIC observations in 2009. We have fitted this MAGIC observed spectrum using the photohadronic model by including the EBL corrections from the EBL models F, S and G (Figure \ref{fig: 1ES 0033+595} (a). The fitted spectrum by these EBL models is shwon in Figure \ref{fig: 1ES 0033+595} (b). It can be seen that all the EBL models fit well to the observed spectrum and the emission is in low-emission state. Thus the intrinsic spectrum is soft. 

\section{VHE Spectra of H 2356-309}

Depending on the relation between the TeV slopes and the synchrotron peak frequencies, the TeV detected EHBLs can be split into two groups with opposite slopes but with the same synchrotron peak position and completely different TeV behaviors (see Fig. 4 of \cite{2019MNRAS.486.1741F}). The sources with positive TeV slopes are  for example 1ES 0229+200 and 1ES 1101-232 having  hard TeV spectra. On the other hand, the sources with negative TeV slopes are for example Mrk 501, Mrk 421, and 1ES 1959+650 with short flux variability. The sources with negative TeV slopes are occasional members of the EHBL family and belong to the tEHBL subclass.

H 2356-309 (z=0.165 \citep{1991AJ....101..821F}) was observed by HESS between 2004 and 2007 for a total of 175.3 h. During this period, its $\nu^p_s > 1$ keV and no evidence for flux variability on time scales shorter than months and years within the HESS data were observed. The optical sensitivity corrected VHE spectrum of 2004 \citep{2010AA...516A..56H} is fitted using the EBL correction from the three EBL models (shown in Figure \ref{fig: H 2356-309-2006} (a)). It is observed that the spectrum can be fitted very well with $2.6 < \delta < 3.0$, and the spectrum is in high-emission state as shown in Figure \ref{fig: H 2356-309-2006} (b). Similarly, in Figure \ref{fig: H 2356-309-2006} (c) the time averaged VHE spectrum from 2005 (June to September 2005) is fitted with different EBL models and all fit the spectrum very well and the spectrum is in high-emission state. These two spectra of H 2356-309 have positive TeV slope. However, the spectrum of 2006 has a peculiar behavior. Its spectrum is fitted using different $\delta$ values as shown in Figure \ref{fig: H 2356-309-2006} (d). The best fit is obtained for $\delta > 3.0$ ($\delta=$ 3.36, 3.42, and 3.45 for EBL-F, S, and G respectively) and for comparison we have also fitted the spectrum with $\delta=3.0$. Thus, $\delta > 3.0$ implies that the intrinsic spectrum is very soft with a negative TeV slope. In Table \ref{Table4} we have summarized three different observations by HESS and their fit using different EBL models in the photohadronic scenario. For the first time, an EHBL having both positive and negative TeV slopes is observed (see Fig. 4 of \cite{2019MNRAS.486.1741F}). Probably, for this reason 
\cite{2012arXiv1208.0808C} and \cite{2019MNRAS.486.1741F} proposed H 2356-309 as a transitional type EHBL with spectral properties between the HBL-like and the hard-TeV EHBLs. We have studied VHE spectra of several tEHBLs which can be explained with two-zone photohadronic model. In zone-1, and zone-2 the spectral indices are respectively in the ranges $2.6 < \delta_1 < 3.0$ and $3.0 < \delta_2 \leq 4.0$. But these sources are HBLs and occasionally become EHBL with variable spectral behavior \citep{2021ApJ...914..120S,2022MNRAS.515.5235S,2026EPJC...86..345S}. It is noteworthy to mention about the VHE spectra of the tEHBL  1ES 2344+514 observed on August 11 and 12, 2016 \citep{2022MNRAS.515.5235S}. These spectra can be fitted very well with a photon spectra index $\delta \gtrsim 3.1$,  However, 1ES 2344+514 is the first tEHBL whose spectra can be fitted only with zone-2 with $\delta\gtrsim 3.1$~\citep{2022MNRAS.515.5235S}.   
H 2356-309 is having synchrotron peak always above 1 keV and short scale variability is not observed  from it. Thus it cannot be a transitional type EHBL, like Mrk 501 or 1ES 2344+514. So, it is possible that this can be a new subclass of EHBL with both hard and very soft intrinsic spectra in different flaring epochs, a feature not seen before in other EHBLs. 


 \section{Discussion}

 EHBLs are an emerging subclass of BL Lac objects. The sources are very faint, and therefore very challenging to detect. Thus, only a limited number of EHBL sources have been observed in gamma-rays so far. We selected 15 EHBL sources with 22 epochs of observation that have both X-ray and VHE gamma-ray spectra, observed simultaneously or quasi-simultaneously, to analyze their VHE gamma-ray spectra and fit them using the photohadronic model. For a detailed analysis, we use three well known EBL models F, S and G and compare the predicted photon survival probability for the redshift used for all the EHBLs before proceeding for the spectral fit and the results are consistent. We could fit all 22 spectra extremely well for which the spectral index $\delta$ is in the range $2.5 \le \delta \le 3.0$, similar to the one observed in HBLs~\citep{2019ApJ...884L..17S}. We have also estimated the $1\sigma$ confidence intervals of the parameters $\delta$ and $F_0$ of the photohadronic model using the three EBL models which are shown in Table \ref{Table1} and Table \ref{Table2}.

We caution that the present sample is not statistically complete. Most VHE observations of EHBLs do not arise from unbiased monitoring; they are usually triggered by hard X-ray states or long exposures aimed at detecting faint sources.
The sample is therefore likely biased toward harder spectra and longer-lived, steadier emission. Observation windows also differ widely—from a few hours (e.g. 1ES2037+521) to several years (e.g. RBS1366)—so flaring intervals are not directly comparable in intrinsic variability timescale. Within the sample, a plurality of flaring epochs fall in the high-emission state ($2.6 < \delta < 3.0$): 12 of 22 for EBL-F (55\%; binomial (1$\sigma$) interval 35–74\%). The low-emission state ($\delta = 3.0$) accounts for 8/22 (36\%), and the very-high-emission state ($2.5 \le \delta \le 2.6$) for 2/22 (9\%). These fractions mildly depend on the EBL model (e.g. RX J1136.5+6737, Table \ref{Table4}).
Despite these limitations, we performed a simple binomial test of the null hypothesis that high and low states are equally likely (excluding the very-high state). The result is statistically not significant ($p \approx 0.25$). These fractions characterize only the present sample and not the EHBL population at large.

Using the photohadronic model, we fit the VHE spectra of the EHBL HESS J1943+213 very well and constrain the redshift using the three EBL models discussed above. The photohadronoic model has much tighter constraint to the redshift than other existing results~\citep{2014A&A...571A..41P,2018ApJ...862...41A}.

 The source H 2356-309 has all the characteristic features of an EHBL except for the behavior of its VHE spectra, which are peculiar. In our study we have observed that EHBLs have both hard and soft intrinsic spectra. But the spactra of H 2356-309 are 
both hard and very soft. This anomalous behavior is not seen in any other EHBL, except in tEHBLs which have different characteristics. So, it cannot be a transitional type EHBL. The possibility of a new subclass of EHBL of which H 2356-309 is a member cannot be ruled out.
 
We compare and contrast the VHE spectra of HBLs and EHBLs in the context of the photohadronic model. VHE spectra of both objects can be fitted extremely well with $2.5 \leq \delta \leq 3.0$. Previously, it has been shown that for HBLs, most of t he flaring periods are in the low-emission state, (48\%), followed by a high-emission state, (38\%), and then a very high-emission state, (14\%)~\citep{2019ApJ...884L..17S}. In contrast, our analysis shows that plurality of EHBL flaring periods are in high-emission state, followed by a low-emission state and then a very high-emission state. However, for a larger sample of EHBLs, this statistics may change.

Our analysis shows that an EHBL can have hard, soft and very soft intrinsic photon spectrum observed during different flaring epochs.
Another contrasting feature is that the EHBLs have a much larger bulk Lorentz factor $\Gamma$ than the HBLs. The combination of long duration VHE flaring of EHBLs with hard spectra and large bulk Lorentz factor possibly accelerate protons to very high energies. Interaction of these protons with the background photons and/or protons can produce high-energy neutrinos. So, these are potential sources to search for spatial and/or temporal correlation of high-energy neutrinos by neutrino observatories, IceCube~\citep{2004APh....20..507A} and KM3NeT~\citep{2016JPhG...43h4001A}.

In the present study, we have a sample of only 15 EHBLs. The sources, which have syncrotron peak above $10^{17} \text{ Hz}$ but are not detected in VHE by IACTs ~\citep{2019A&A...632A..77C,2019MNRAS.486.1741F,2026ApJ..1002..138A}, could be detected by the Large High Altitude Air Shower Observatory (LHAASO)~\citep{2019arXiv190502773C}, and the next generation Cherenkov Telescope Array Observatory (CTAO). So far, the number of EHBLs detected in VHE is very small and the study shows that the EHBL class is not homogeneous. 

Hence, the possibility of further subclassification cannot be ruled out. However, with a limited number of EHBL sources observed in VHE, a firm conclusion about their properties cannot be drawn at the present time. This gives a strong motivation to look for new EHBL sources and to study existing ones in depth. It is important to mention that given the small and non-complete sample of EHBLs, this should not be read as a population fraction. Thus, a richer sample with many sources may provide better information on the different emission states. 
We expect that in the future in multiwavelength observations with a richer sample, the very high-emission state and the high-emission state should be above their present estimates and the low-emission state should be lower than the present estimate.

\section*{Acknowledgements}

We thank the referee for the suggestions and
constructive remarks that have considerably improved the manuscript. The work of S.S. is partially supported by DGAPA-UNAM (Mexico) Project No. IN105326. D.I.P.S., A.U.P.O., R.D.J.P.-C. and G. S.-C. would like to thank SECIHTI (Mexico) for partial support. Partial support from CSU-Long Beach is gratefully acknowledged. P.F.C. would like to acknowledge support from the Generalitat Valenciana (Spain) through the PROMETEO CIPROM/2023/32 grant and from the Agencia Estatal de Investigación (Spain) through the  PID2022-142407NB-I00 and PID2024-162480OB-I00 grants.
M. E. Iglesias Martínez’s research is partially supported by a postdoctoral research grant “Subvenciones para la contratación de personal investigador en fase postdoctoral APOSTD/2025 (MODALIDAD A)”, funded by the Conselleria de Educación, Cultura, Universidades y Empleo of the Generalitat Valenciana (Reference CIAPOS/2024/238) and co-funded by the European Social Fund Plus (FSE+) through the 2021–2027 Operational Programme of the Comunitat Valenciana.

\bibliography{EHBLref}{}
\bibliographystyle{aasjournal}

\end{document}